\documentclass[a4paper,11pt]{article}

\usepackage{jheppub}
\usepackage[english]{babel}
\usepackage{relsize}
\usepackage{mathrsfs}  

\usepackage{amstext}  
\usepackage{amsfonts}
\usepackage{amsthm}
\usepackage{graphicx}
\usepackage{amsmath}
\usepackage{amsfonts}
\usepackage{mathtools}

\usepackage{array}                
\usepackage{xcolor} 

\usepackage{slashed}

\renewcommand{\refeq}[1]{\mbox{\eqref{#1}}}

\newcommand{\reffi}[1]{\mbox{Fig.~\ref{#1}}}

\newcommand{\refse}[1]{\mbox{Section~\ref{#1}}}
\newcommand{\refses}[2]{\mbox{Sections~\ref{#1}--\ref{#2}}}
\newcommand{\refapp}[1]{\mbox{Appendix~\ref{#1}}}

\newcommand{\ie}{i.e.\ }

\newcommand{\fieldcomp}[2]{{#1}_{\hspace{-.3ex} {\scriptscriptstyle #2}}}

\newcommand{\f}[2]{\frac{#1}{#2}}

\newcommand{\nosss}[1]{#1}

\newcommand{\ben}{\begin{enumerate}}
\newcommand{\een}{\end{enumerate}}
\newcommand{\bit}{\begin{itemize}}
\newcommand{\eit}{\end{itemize}}
\newcommand{\bea}{\begin{eqnarray}}
\newcommand{\eea}{\end{eqnarray}}
\newcommand{\be}{\begin{equation}}
\newcommand{\ee}{\end{equation}}
\newcommand{\ba}{\begin{align}}
\newcommand{\ea}{\end{align}}
\newcommand{\beas}{\begin{eqnarray*}}
\newcommand{\eeas}{\end{eqnarray*}}
\newcommand{\bes}{\begin{equation*}}
\newcommand{\ees}{\end{equation*}}
\newcommand{\bas}{\begin{align*}}
\newcommand{\eas}{\end{align*}}

\newcommand{\eps}{{\varepsilon}}
\newcommand{\als}{\alpha_{\scriptscriptstyle{\rS}}}
\newcommand{\gs}{g_{\scriptscriptstyle{\rS}}}
\newcommand{\Tgen}{\mathrm{I}}
\newcommand{\gen}{\mathrm{gen}}
\newcommand{\e}{e}
\newcommand{\cw}{c_\mathrm{w}}
\newcommand{\sw}{s_\mathrm{w}}
\newcommand{\ms}{\mathrm{MS}}

\newcommand{\msbar}{\overline{\mathrm{MS}}}

\newcommand{\alphas}{\alpha_{\scriptscriptstyle{\rS}}}
\newcommand{\lb}{\left(}
\newcommand{\rb}{\right)}

\newcommand{\dendim}{d}
\newcommand{\numdim}{D}
\renewcommand{\dendim}{D}
\renewcommand{\numdim}{D_{\mathrm{n}}}

\newcommand\loeq{\stackrel{\mathclap{\mbox{{\tiny LO}}}}{=}}

\def\rcarg{\chi}

\newcommand{\msfact}{S}

\def\param{\theta}

\newcommand{\lndev}[1]{#1\frac{\partial}{\partial #1}}

\newcommand{\denbar}{\bar}

\newcommand{\Dbar}[1]{D_{\nosss{#1}}}

\newcommand{\tilq}{\tilde{q}}

\newcommand{\calA}{\mathcal{A}}
\newcommand{\calC}{\mathcal{C}}
\newcommand{\calD}[1]{\mathcal{D}^{(#1)}(\bar q_{#1})}

\newcommand{\calL}{\mathcal{L}}

\newcommand{\calN}{\mathcal{N}}

\newcommand{\calQ}{\mathcal{Q}}
\newcommand{\ntilde}{\tilde\calN}

\newcommand{\calR}{\mathcal{R}}
\newcommand{\calT}{\mathcal{T}}
\newcommand{\calU}{\mathcal{U}}
\newcommand{\calZ}{\mathcal{Z}}
\newcommand{\dcalZ}{\delta \hat{\mathcal{Z}}}
\newcommand{\bfM}{{\textbf{M}}}
\newcommand{\bfm}{{\textbf{m}}}

\newcommand{\bfV}{{\textbf{V}}}

\newcommand{\bfP}{{\textbf{P}}}

\newcommand{\bfK}{\textbf{K}}

\newcommand{\bfR}{\textbf{R}}

\newcommand{\bfT}{\textbf{T}}

\newcommand{\barN}{\bar{\mathcal{N}}}

\newcommand{\calV}{\mathcal{V}}

\newcommand{\Nc}{N}

\newcommand{\nq}{n_{\mathrm{q}}}

\newcommand{\QiL}{Q_{i,\mathrm{L}}}
\newcommand{\QiLbar}{\bar{Q}_{i,\mathrm{L}}}

\newcommand{\uiR}{u_{i,\mathrm{R}}}
\newcommand{\uiRbar}{\bar{u}_{i,\mathrm{R}}}

\newcommand{\diR}{d_{i,\mathrm{R}}}
\newcommand{\diRbar}{\bar{d}_{i,\mathrm{R}}}

\newcommand{\CA}{C_{\mathrm{A}}}
\newcommand{\CF}{C_{\mathrm{F}}}
\newcommand{\TF}{T_{\mathrm{F}}}

\newcommand{\bk}{}

\newcommand{\hYM}{\widehat{\mathrm{YM}}}
\newcommand{\ubk}{{\scriptscriptstyle \mathrm{YM}}}

\newcommand{\fix}{\mathrm{fix}}
\newcommand{\ghost}{\mathrm{ghost}}

\newcommand{\rYuk}{\mathrm{Yuk}}

\newcommand{\yukUP}[1]{{\mathbf \lambda_{u_{#1}}} }
\newcommand{\yukDO}[1]{{\mathbf \lambda_{d_{#1}}} }

\newcommand{\vev}{v}
\newcommand{\tvev}{\tilde{v}}
\newcommand{\hvev}{\hat{v}}
\newcommand{\VevInsert}{  \vev^k }

\newcommand{\gfive}{\gamma_5}

\newcommand{\VevInsertJ}{  \vev^j }
\newcommand{\tVevInsertKJ}{  \tvev^{k-j} }

\newcommand{\singlearg}[1]{
\ifx&#1&
\else
(#1)   
\fi
}

\newcommand{\doublearg}[2]{
\ifx&#2&
(#1)  
\else
(#1,#2)   
\fi
}

\newcommand{\ampindices}[5]{{#1}_{{#2,#3}}^{#4 }\singlearg{#5}}

\newcommand{\amp}[4]{{\ampindices{\calA}{#1}{#2}{#3}{#4}}}
\newcommand{\ampbar}[4]{{\ampindices{\bar\calA}{#1}{#2}{\hspace{0.6pt}#3}{#4}}}
\newcommand{\ratamp}[4]{{\ampindices{\delta \calR}{#1}{#2}{#3}{#4}}}
\newcommand{\deltaZ}[4]{{\ampindices{\delta Z}{#1}{#2}{#3}{#4}}}
\newcommand{\deltaZtilde}[4]{{\ampindices{\delta \tilde Z}{#1}{#2}{#3}{#4}}}

\newcommand{\barq}{\bar q}

\newcommand{\tildeqidx}[2]{{\tilde q}_{#1}^{\hspace{0.6pt}#2}}

\newcommand{\gpar}{{\mathrm{gp}}}

\def\tad{\mathrm{tad}}

\def\alpham{\alpha}

\newcommand{\TrQ}{\Tr_q}

\newcommand{\Tr}{\mathrm{Tr}}

\newcommand{\srp}{{\scriptscriptstyle \mathrm P}}
\newcommand{\srG}{{\scriptscriptstyle \mathrm G}}
\newcommand{\srm}{{\scriptscriptstyle \mathrm m}}

\newcommand{\rc}{\mathrm c}
\newcommand{\rI}{\mathrm{I}}
\newcommand{\rII}{\mathrm{II}}

\newcommand{\rH}{\mathrm H}
\newcommand{\rHexp}{\rH}
\newcommand{\rR}{\mathrm R}

\newcommand{\rL}{\mathrm L}
\newcommand{\rS}{\mathrm S}
\newcommand{\rT}{\mathrm T}
\newcommand{\ri}{\mathrm i}
\newcommand{\rw}{\mathrm w}
\newcommand{\rd}{\mathrm d}
\newcommand{\rV}{\mathrm V}
\newcommand{\ord}{\mathcal O}

\definecolor{bluemar}{rgb}{0,0,.5}
\definecolor{redmar}{rgb}{.8,0,0}
\definecolor{greenmar}{rgb}{0,.5,0}

\graphicspath{{./figures/}}

\preprint{
\begin{flushright}
PSI-PR-21-14\\
ZU-TH 31/21\\
\end{flushright}
}

\title{\boldmath Two-Loop Rational Terms 
for Spontaneously Broken Theories}

\author[a]{Jean-Nicolas Lang}
\author[a]{Stefano Pozzorini}
\author[a]{Hantian Zhang}
\author[b]{Max F. Zoller}

\affiliation[a]{Physik-Institut, Universit\"at Z\"urich, 
Winterthurerstrasse 190, 
CH-8057 Z\"urich, Switzerland}
\affiliation[b]{Paul Scherrer Institut, Forschungsstrasse 111, CH-5232 Villigen PSI, Switzerland}

\emailAdd{jlang@physik.uzh.ch}
\emailAdd{pozzorin@physik.uzh.ch}
\emailAdd{hantian.zhang@physik.uzh.ch}
\emailAdd{max.zoller@psi.ch}

\abstract{Rational counterterms are a key ingredient for the automation of 
loop calculations through numerical methods.
Building on the recently established properties of rational terms 
of UV origin
at two loops, 
in this paper we present a systematic method for the determination of rational
counterterms within spontaneously broken theories.
In particular we introduce a generalised vev-expansion approach that makes it 
possible to obtain the rational counterterms 
of UV origin
for a spontaneously 
broken theory by means of calculations in the unbroken phase.
The drastic simplifications that result from the underlying symmetry
open the door to the efficient determination of rational counterterms
for the full Standard Model at two loops.
The renormalisation-scheme dependence is analysed in detail, and we
show that rational counterterms need to be 
determined only once and for all in a 
generic renormalisation scheme for the symmetric phase 
and, a posteriori, they can be easily adapted to a wide range of 
physical renormalisation schemes for the spontaneously broken 
phase.
As a first application we determine the full set of 
$\ord(\alphas^2)$ rational counterterms 
of UV origin
for the full Standard Model, \ie
for all 
superficially UV-divergent two-loop vertex functions 
involving combinations of gluons, quarks, electroweak vector bosons and scalar 
bosons.
}

\arxivnumber{}

\begin{document}

\maketitle
\flushbottom

\newpage

\section{Introduction}

The ultraviolet (UV) and  infrared (IR) singularities of scattering
amplitudes in perturbative quantum-field theory are typically 
regularised dimensionally, i.e.~via analytic continuation in the number of space-time
dimensions, 
$D=4-2\eps$~\cite{tHooft:1972tcz}.
Within numerical frameworks, the need of computing loop amplitudes as a
function of a continuous dimensional parameter $D$ can
represent a significant obstacle.
For this reason, a variety of methods have been proposed that aim
at restricting---in part or entirely---the
calculations of loop amplitudes
to an integer number of space-time dimensions~\cite{%
,Bern:1991aq%
,Kilgore:2012tb%
,Siegel:1979wq%
,Signer:2005iu%
,Xiao:2006vr%
,Binoth:2006hk%
,Ossola:2008xq%
,Badger:2008cm%
,Giele:2008ve%
,Abreu:2017hqn%
,Pittau:2012zd%
,Page:2015zca%
,Fazio:2014xea%
,Cherchiglia:2010yd%
,Soper:1999xk%
,Catani:2008xa%
,Becker:2010ng%
,Anastasiou:2018rib%
,Capatti:2019edf%
}.

At one loop, the most efficient and flexible 
automated tools on the market~\cite{Buccioni:2019sur,Denner:2017wsf,
vanHameren:2009dr,Hirschi:2011pa} rely on an implementation of 
dimensional regularisation based on rational 
counterterms~\cite{Ossola:2008xq}.
The key idea is that loop amplitudes can be computed through  
numerical algorithms that build the numerator of loop integrands in 
$\numdim=4$ dimensions, while the 
remaining contributions stemming from the interplay of 
the $(\numdim-4)$-dimensional parts of the loop numerators with UV poles
are reconstructed a posteriori 
by means of process-independent rational
counterterms~\cite{Draggiotis:2009yb,Garzelli:2009is,Pittau:2011qp}.

This method was a key ingredient for the automation 
of one-loop calculations,
and the foundations for its extension to two loops have been
established in~\cite{Pozzorini:2020hkx,Lang:2020nnl}, where it was shown
that renormalised two-loop amplitudes can be constructed through a modified
version of the $\bfR$-operation for loop amplitudes with $\numdim=4$
dimensional integrand numerators.
In this approach, the usual one-loop and two-loop 
counterterms for the subtraction of UV poles 
are supplemented by corresponding rational counterterms,
which reconstruct the contributions that arise from the interplay 
of UV poles with the $(D-4)$-dimensional parts of loop
numerators.
The subtraction of 
quadratic one-loop subdivergences
in $\numdim=4$ dimensions
requires additional UV counterterms
proportional to $\tilde q^2/\eps$, 
where $\tilde q$ is the $(D-4)$-dimensional part of the loop momentum.
All required counterterms are local and process independent. 
Within a given theoretical model, each potentially UV-divergent 1PI vertex function gives rise to a 
rational counterterm, which needs to be determined only once and for all.
A general method for the determination of two-loop rational counterterms 
has been presented in~\cite{Pozzorini:2020hkx,Lang:2020nnl}.

To date, the study of two-loop rational terms was restricted to 
contributions stemming from UV divergences, assuming that 
IR divergences are irrelevant.
This expectation is supported by the fact that 
one-loop amplitudes are free from 
rational terms of IR origin~\cite{Bredenstein:2008zb}.
In general, this is not the case beyond one loop. However, 
we expect that the standard procedure for the 
subtraction of IR divergences should lead to a 
cancellation of rational terms of IR origin.
This conjecture will be the subject of future investigations,
while in this paper we will consider only rational terms 
of UV origin. For convenience, the latter will be referred to 
simply as rational terms.

One-loop rational counterterms $\delta \calR_1$
do not depend on the choice of UV
renormalisation scheme, while two-loop rational counterterms $\delta
\calR_2$ are scheme
dependent~\cite{Lang:2020nnl}.  This dependence involves a trivial
contribution, which corresponds to the naive renormalisation of 
$\delta \calR_1$ counterterms, 
plus a non-trivial scheme dependence, which is due to
the fact that the multiplicative renormalisation of UV subdivergences does
not commute with the projection of loop numerators to $\numdim=4$
dimensions.
As shown in~\cite{Lang:2020nnl}, both contributions are 
process independent, and the scheme-dependent part of $\delta
\calR_2$ counterterms can be expressed as
a linear combination of generic one-loop renormalisation constants,
which can be adapted a posteriori 
to any desired scheme.
The methods presented in~\cite{Pozzorini:2020hkx,Lang:2020nnl}
are briefly reviewed in \refse{sec:oneloop} of the present paper.

So far, two-loop rational counterterms are known explicitly 
only for U(1) and SU($N$) gauge theories~\cite{Pozzorini:2020hkx,Lang:2020nnl}.
In principle, corresponding results for the full Standard Model (SM) can be 
derived using the same techniques.
However, due to symmetry breaking, the required calculations
can become prohibitively involved in the electroweak
(EW) sector of the SM.
In particular, the fact that the various states of SU(2)$\times$U(1)
multiplets acquire different masses, can mix with one another, and receive
different field- and mass-renormalisation constants, represent  
major sources of additional complexity as compared to unbroken theories.

In order to circumvent these difficulties, in this paper we present a new method 
for the efficient derivation of two-loop rational counterterms 
within spontaneously broken (SB) theories. 
This approach is based on the systematic expansion of 
loop amplitudes
in the vacuum expectation value (vev).
As is well known, the vev expansion of loop amplitudes
in a SB theory
can be expressed in terms of corresponding amplitudes 
in the unbroken phase with 
additional external Higgs lines that carry zero momentum.
These so-called vev insertions provide the correct 
expansion of loop amplitudes in $\numdim=D$ dimensions, while
in $\numdim=4$ dimensions, as we will show,
they need to be supplemented 
by auxiliary vev-insertion counterterms
proportional to $\tilq^2$. This is due to the fact that 
the vev expansion of fermionic loop propagators does not commute 
with the projection to $\numdim=4$ dimensions. 
Based on this approach, in \refse{se:sb} we present a vev-expansion formula
that relates the rational counterterms for a SB theory 
to corresponding counterterms in the underlying symmetric theory
by means of standard vev insertions and auxiliary vev insertions.
This makes it possible to carry out the bulk of the 
derivations of rational counterterms
in the symmetric phase, while symmetry-breaking effects are 
reconstructed through a few vev insertions.

The effect of renormalisation-scheme transformations
on the vev expansion of rational counterterms is investigated in
\refse{se:schdep}.
In particular we demonstrate that our  
vev-expansion formula for rational counterterms
can be applied to a wide class of non-trivial renormalisation schemes for 
SB theories, such as schemes of on-shell type.
In this context we analyse the connection between the 
renormalisation of the unbroken and SB phases of the theory, 
and we discuss subtleties related to the renormalisation 
of the Higgs sector, 
with emphasis on the renormalisation of tadpoles and the vev.
In a first step we consider SB theories that respect rigid invariance, i.e.~theories
where the entire vev dependence of the Lagrangian is generated from the
symmetric phase via shifts of the Higgs field, $H\to H+\vev$. 
Subsequently we extend our analysis to SB gauge theories that 
violate gauge invariance through the gauge-fixing procedure.
This happens, for instance, in the case of the widely used
't~Hooft gauge fixing. As we will show, 
the vev-dependent parts of the gauge-fixing terms that violate rigid invariance 
can be accounted for by supplementing our formula for the vev expansion of 
rational counterterms with a new kind of auxiliary vev insertions.
This extends the applicability of our vev-expansion method to 
a wide range of SB gauge theories, including the full SM with EW and mixed QCD--EW corrections
at two loops.

As a first application, in~\refse{se:results} we derive all 
rational counterterms of $\ord(\als^2)$ for vertices
that involve quarks or gluons in combination with at least one EW vector
boson or scalar boson in the SM. 
Together with the counterterms presented in~\cite{Pozzorini:2020hkx,Lang:2020nnl}
this provides the complete set of rational counterterms for 
two-loop QCD calculations in the full SM.

\section{Rational terms at one and two loops} \label{sec:oneloop}

In this section we introduce our notation and we summarise the 
properties of rational terms that are relevant for this paper. For more
details we refer to~\cite{Pozzorini:2020hkx,Lang:2020nnl}.

\subsection{Notation and conventions}
\label{se:oneloopratterms}

For the regularisation of UV divergences, following the 
't~Hooft--Veltman scheme~\cite{tHooft:1972tcz} we
keep external states\footnote{By external states we mean the external
wave functions and momenta as well as all tree-level 
parts that connect the various 1PI loop
subdiagrams of a (multi-)loop Feynman diagram.
For more details on the treatment of external states we refer
to Sect.~2 of~\cite{Pozzorini:2020hkx}.
}
in four dimensions, while 
loop momenta, metric tensors and Dirac
matrices inside the loops are extended to 
$\dendim=4-2\eps$
dimensions.
For the decomposition of these objects into 
four-dimensional parts and
$(D-4)$-dimensional remnants we use the notation
\bea
\label{eq:ddimnotG}
\denbar q^\mu &=& q^\mu+\tilde q^{\tilde\mu}\,,
\qquad
\denbar \gamma^\mu = \gamma^\mu+\tilde \gamma^{\tilde\mu}\,,
\qquad
\denbar g^{\denbar \mu\denbar \nu} = g^{\mu\nu}+\tilde g^{\tilde\mu\tilde
\nu}\,,
\eea
where the bar and the tilde are used to mark, respectively,
the $D$-dimensional and $(D-4)$-dimensional parts, 
while objects without a bar or tilde are four-dimensional.
For $\gamma_5$ we use the KKS
scheme~\cite{Korner:1989is,Kreimer:1993bh,Korner:1991sx}, as detailed 
in~\refse{se:gamma5}.

The amplitude of a
one-loop diagram $\Gamma$ in $D$ dimensions has the form
\bea 
\label{eq:rtoneloopA}
\ampbar{1}{\Gamma}{}{} &=& 
\int\!\rd\barq_1\, \f{{\barN}(\bar{q}_1)}{\Dbar{0}(\barq_1)\cdots
\Dbar{N-1}(\barq_1)}\,,
\eea
with the integration measure
\bea
\rd\barq & = & \mu_0^{2\eps} 
\f{\rd^{^D}\! \bar
q}{(2\pi)^{^D}}\,, 
\label{eq:intmeasure}
\eea
where $\mu_0$ is the scale of dimensional regularisation.
The denominators in~\refeq{eq:rtoneloopA} read
\bea
\label{eq:rtoneloopB}
\Dbar{j}(\barq_1)&=& (\barq_1+p_j)^2-m_j^2\,,
\eea
and $p_j$ are combinations of four-dimensional external momenta.
The corresponding renormalised amplitude reads
\bea
\label{eq:oneloopdivE}
\bfR\, \ampbar{1}{\Gamma}{}{}
&=&
\ampbar{1}{\Gamma}{}{}
+\deltaZ{1}{\Gamma}{}{}\,,
\eea
where $\deltaZ{1}{\Gamma}{}{}$ denotes the UV counterterm.

\begin{figure}[t]
\begin{center}
\includegraphics[width=0.3\textwidth]{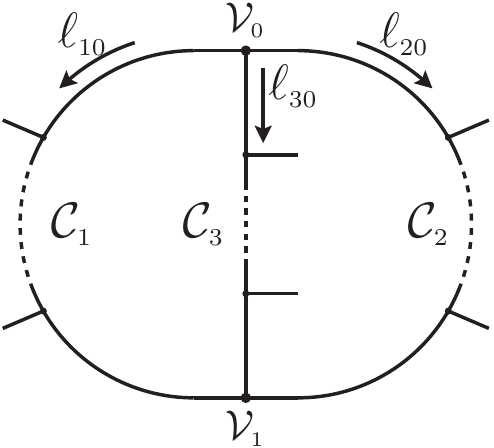}
\end{center}
\caption{A generic irreducible two-loop diagram consists of 
three chains $\calC_1$,  $\calC_2$,  $\calC_3$. Each chain 
$\calC_i$ involves the various propagators depending on the loop momenta 
$\ell_{ia}= q_{i}+p_{ia}$, and $q_1+q_2+q_3=0$. The two connecting vertices, 
$\calV_0$, $\calV_1$, can be 
triple or quartic vertices.
} 
\label{fig:twoloop_irred}
\end{figure}

At two loops, the discussion of rational terms can be restricted to 
two-loop diagrams of irreducible kind, \ie diagrams that cannot be 
factorised into two one-loop diagrams~\cite{Pozzorini:2020hkx}.
An irreducible two-loop diagram $\Gamma$ 
consists of three chains, $\calC_{1},\calC_{2},\calC_{3}$, 
that are connected to each
other by two vertices, $\calV_0,\calV_1$ (see~\reffi{fig:twoloop_irred}).
Its amplitude in $D$ dimensions has the form
\bea
\label{eq:twoloopnotA}
\ampbar{2}{\Gamma}{}{} 
&=&
\int\rd\barq_1
\int \rd\barq_2\,
\frac{
\bar\calN(\barq_1,\barq_2,\barq_3)}
{\calD{1}\,\calD{2}\,\mathcal{D}^{(3)}(\barq_3)}\bigg|_{\barq_3\,=\,-\barq_1-\barq_2}\,,
\eea 
where 
\bea
\label{eq:twoloopnotB}
\calD{i}&=&
D^{(i)}_0(\barq_i)\cdots
D^{(i)}_{N_i-1}(\barq_i)
\eea
is the denominator factor associated with the chain $\calC_i$, and
\bea
\label{eq:twoloopnotB2}
D^{(i)}_a(\barq_i) \,=\, 
\bar \ell_{ia}^{\,2}-m_{ia}^2\,,
\qquad\mbox{with}\qquad
\bar \ell_{ia}\,=\,\barq_i+p_{ia}\,.
\eea
The form of the loop numerator is
\bea
\label{eq:twoloopnumA}
\bar\calN(\barq_1,\barq_2,\barq_3)
&=&
\bar\Gamma^{\bar\alpha_1\bar\alpha_2\bar\alpha_3}(\barq_1,\barq_2,\barq_3)\,
\bar\calN^{(1)}_{\bar\alpha_1}(\barq_1)\,
\bar\calN^{(2)}_{\bar\alpha_2}(\barq_2)\,
\bar\calN^{(3)}_{\bar\alpha_3}(\barq_3)\,,
\eea 
where $\bar \Gamma^{\bar\alpha_1\bar \alpha_2\bar \alpha_3}$
embodies the two vertices $\calV_0$ and $\calV_1$, which are 
connected 
to the numerator factors $\bar\calN^{(i)}_{\bar\alpha_i}(\barq_i)$ 
of the three chains $\calC_i$
through multi-indices
$\bar \alpha_i\equiv (\bar \alpha_{i1},\bar \alpha_{i2})$.

At two loops, UV divergences can be subtracted through the 
$\bfR$-operation~\cite{Bogoliubov:1957gp,Hepp:1966eg,Zimmermann:1969jj,Caswell:1981ek}.
The renormalised  amplitude for 
a two-loop diagram $\Gamma$ in $D$ dimensions  
has the form
\bea 
\label{eq:twoloopren} 
{\textbf{R}}\, \ampbar{2}{\Gamma}{}{}   
&=&  \ampbar{2}{\Gamma}{}{} + 
\sum_{\gamma_i} \deltaZ{1}{\gamma_i}{}{} \cdot \ampbar{1}{\Gamma/\gamma_i}{}{}
+ \deltaZ{2}{\Gamma}{}{}\,,
\eea
where the sum on the rhs runs over the one-loop subdiagrams $\gamma_i$, 
with $i=1,2,3$. The subdiagram $\gamma_i$
results from $\Gamma$ by 
truncating the chain $\calC_i$, and the 
associated subdivergence is cancelled by the subtraction term 
$\deltaZ{1}{\gamma_i}{}{}\cdot \ampbar{1}{\Gamma/\gamma_i}{}{}$,
which is obtained from the original two-loop diagram
$\Gamma$ by replacing $\gamma_i$ 
through its one-loop counterterm $\deltaZ{1}{\gamma_i}{}{}$,
while keeping unchanged its complement $\Gamma/\gamma_i$.
The latter corresponds to the chain 
$\calC_i$, and the explicit form of such subtraction terms is
\bea
\label{eq:subdiagnotB}
\sum_{\gamma_i}
\deltaZ{1}{\gamma_i}{}{}\cdot
\ampbar{1}{\Gamma/\gamma_i}{}{}
&=&
\sum_{\gamma_i}
\int\rd\barq_i\,
\deltaZ{1}{\gamma_i}{\bar\alpha_i}{\barq_i}\,
\frac{\bar\calN^{(i)}_{\bar\alpha_i}(\barq_i)}{\calD{i}}
\,.
\eea
The local two-loop divergence that is left after subtraction of
all subdivergences is removed by the 
counterterm $\deltaZ{2}{\Gamma}{}{}$ in~\refeq{eq:twoloopren}.

The renormalisation operator $\bfR$ should be understood as a linear
operator. Thus, when $\Gamma$ 
is a set of diagrams the formulas \refeq{eq:oneloopdivE} and
\refeq{eq:twoloopren} should be understood as 
the sum of the contributions of individual diagrams.
When $\Gamma$ is a full vertex function, the
linearity of $\bfR$ implies that 
 the rhs of~\refeq{eq:subdiagnotB} can also be understood as 
the sum of all possible vertex functions $\gamma_i$
that can be inserted into 
the one-loop vertex function $\calA_{1,\Gamma}$. In this case,
$\delta Z_{1,\gamma_i}$ corresponds to the 
full one-loop counterterm
for the vertex function $\gamma_i$,
and each term $\deltaZ{1}{\gamma_i}{}{} \cdot \ampbar{1}{\Gamma/\gamma_i}{}{}$
embodies all possible insertions of 
$\delta Z_{1,\gamma_i}$ into the various one-loop diagrams
that contribute to $\calA_{1,\Gamma}$.

\subsection{Rational terms at one and two loops}
\label{se:irredtwoloop}

The calculation of scattering amplitudes can be automated through numerical
algorithms that build the numerators of loop integrands in four dimensions. 
In this approach, the missing contributions stemming from the interplay of
$1/(D-4)$ poles with the $(D-4)$ dimensional parts of loop numerators can be
reconstructed by means of rational counterterms.
With this motivation in mind, we define rational terms through 
a splitting of loop numerators into four- and $(D-4)$-dimensional parts.

At one loop, the numerator of the amplitude~\refeq{eq:rtoneloopA} 
in $D$-dimensions is split into
\bea
\label{eq:rtoneloopD}
\bar \calN(\barq_1)&=& \calN(q_1) + \tilde \calN(\barq_1)\,,
\eea
where 
$\calN(q_1)$
is the four-dimensional part,
obtained by projecting the metric tensor, Dirac matrices
and the loop momentum to four dimensions.
The remnant $\tilde \calN(\barq_1)$ is of $\ord(\eps, \tilde q_1)$
and will be referred to as the $(\dendim-4)$-dimensional part of the numerator.
To keep track of the dimensionality of loop numerators we use the parameter
$\numdim$. The amplitude $\calA_{1,\Gamma}$, defined
in~\refeq{eq:rtoneloopA}, is referred to as 
amplitude in $\numdim=D$ dimensions, while its
counterpart in $\numdim=4$ dimensions corresponds to 
\bea 
\label{eq:rtoneloopH}
\calA_{1,\Gamma} &=& 
\int\!\rd\barq_1\, \f{\calN(q_1)}{\Dbar{0}(\barq_1)\cdots
\Dbar{N-1}(\barq_1)}\,.
\eea
Note that here the numerator is projected to four dimensions, while
retaining the full $\dendim$-dependence of the loop momentum in the
denominator. 
Renormalised one-loop amplitudes in $\numdim=D$ and
$\numdim=4$ dimensions are related to each other by
\bea 
\label{eq:masterformula1}
\textbf{R}\,\ampbar{1}{\Gamma}{}{} &=& 
\amp{1}{\Gamma}{}{} + \deltaZ{1}{\Gamma}{}{} + \ratamp{1}{\Gamma}{}{}
\,,
\eea
where $\deltaZ{1}{\Gamma}{}{}$ are usual UV counterterms,
and $\delta \calR_{1,\Gamma}$ are 
rational counterterms~\cite{Ossola:2008xq,Draggiotis:2009yb,Garzelli:2009is,Pittau:2011qp}.
Since $\delta \calR_{1,\Gamma}$ counterterms
originate only from the interplay of $\ntilde(\barq_1)$
with poles of UV kind~\cite{Bredenstein:2008zb}, they
can be derived once and for all
by computing
\bea 
\label{eq:r1form}
\ratamp{1}{\Gamma}{}{}
&=&  
\ampbar{1}{\Gamma}{}{}  
-\amp{1}{\Gamma}{}{}\,,
\eea
at the level of UV divergent 1PI vertex 
functions.

At two loops, in analogy with~\refeq{eq:rtoneloopD}
we split the $D$-dimensional loop numerator~\refeq{eq:twoloopnumA} 
into four-dimensional and $(D-4)$ dimensional parts,
\bea
\label{eq:rttwoloopDD}
\bar \calN(\barq_1, \barq_2, \barq_3)&=& \calN(q_1, q_2, q_3) + \tilde \calN(\barq_1, \bar
q_2, \bar q_3)\,.
\eea
Similarly as in the one-loop case, 
the two-loop amplitude $\amp{2}{\Gamma}{}{}$, defined
in~\refeq{eq:twoloopnotA}, is referred to as 
an
amplitude in $\numdim=D$ dimensions, and its 
counterpart in $\numdim=4$ dimensions is 
\bea
\label{eq:twoloopnotZ}
\amp{2}{\Gamma}{}{} 
&=&
\int\rd\barq_1
\int \rd\barq_2\,
\frac{
\calN(q_1,q_2,q_3)}
{\calD{1}\,\calD{2}\,\mathcal{D}^{(3)}(\barq_3)}\bigg|_{\barq_3\,=\,-\barq_1-\barq_2}\,,
\eea 
where the numerator is projected to four dimensions, while keeping
the denominator in $D$ dimensions.
As demonstrated in~\cite{Pozzorini:2020hkx}, 
renormalised two-loop amplitudes in $\numdim=D$ and $\numdim=4$
dimensions are related to each other by the following 
generalisation of the 
$\bfR$-operation,
\bea 
\label{eq:masterformula2} 
{\textbf{R}}\, \ampbar{2}{\Gamma}{}{}   
&=&  \amp{2}{\Gamma}{}{} + 
\sum  \limits_{\gamma_i} \lb \deltaZ{1}{\gamma_i}{}{} +\deltaZtilde{1}{\gamma_i}{}{} + \ratamp{1}{\gamma_i}{}{} \rb \cdot \amp{1}{\Gamma/\gamma_i}{}{}
\,+\, 
\deltaZ{2}{\Gamma}{}{} + \ratamp{2}{\Gamma}{}{}
\,,
\eea
where the standard UV counterterms, $\deltaZ{1}{\gamma_i}{}{}$ 
and $\deltaZ{2}{\Gamma}{}{}$, 
are supplemented by associated rational counterterms,
$\ratamp{1}{\gamma_i}{}{}$ and 
$\ratamp{2}{\Gamma}{}{}$, which 
reconstruct the contributions stemming from the interplay of 
$\tilde \calN(\barq_1, \barq_2, \bar q_3)$
with  
one-loop subdivergences and local two-loop divergences.
The extra counterterms $\deltaZtilde{1}{\gamma_i}{}{}$
are required for the full cancellation of 
the UV subdivergences of the one-loop subdiagrams $\gamma_i$ 
in $\numdim=4$ dimensions.
In renormalisable theories such $\deltaZtilde{1}{\gamma_i}{}{}$ counterterms
are needed only for
quadratically divergent 
self-energy
subdiagrams, and their general form
is~\cite{Pozzorini:2020hkx}
\bea
\label{eq:4dimsubdiagE8}
\deltaZtilde{1}{\gamma_i}{\alpham}{\tilde q_i}
&=&
v^\alpham \frac{\tildeqidx{i}{2}}{\eps}\,, 
\eea
where $q_i$ is the external loop momentum that flows through the 
complement $\Gamma/\gamma_i$ of the subdiagram $\gamma_i$

The identity~\refeq{eq:masterformula2}
is valid for any renormalisable theory
assuming that $\ampbar{2}{\Gamma}{}{}$ is free
from IR divergences,  or 
that the latter are subtracted in a way 
that also rational terms of IR origin cancel.
This implies that the $\ratamp{2}{\Gamma}{}{}$ counterterms in~\refeq{eq:masterformula2} 
arise only from divergences of UV kind. More precisely, 
they originate only from local UV divergences.
Thus they can be
derived once and for all
by inverting the
master formula~\refeq{eq:masterformula2}, \ie by computing
\bea 
\label{eq:r2form}
\ratamp{2}{\Gamma}{}{}
&=&
\ampbar{2}{\Gamma}{}{}  
-\amp{2}{\Gamma}{}{} + 
\sum  \limits_{\gamma} \deltaZ{1}{\gamma}{}{}
\cdot \ampbar{1}{\Gamma/\gamma}{}{} 
-
\sum  \limits_{\gamma} \lb \deltaZ{1}{\gamma}{}{}
+\deltaZtilde{1}{\gamma}{}{} + \ratamp{1}{\gamma}{}{} \rb \cdot
\amp{1}{\Gamma/\gamma}{}{}\,,\qquad
\eea
for all superficially divergent 1PI vertex functions $\Gamma$ 
in the theory at hand.

Similarly as for UV counterterms,
the rational counterterms $\ratamp{k}{\Gamma}{}{}$ 
at one and two loops 
have the form of homogeneous 
polynomials of degree $X(\Gamma)$
in the external momenta  
and internal masses, %
 where
$X(\Gamma)$ is the superficial degree of divergence of 
$\Gamma$.

Finally, we point out that the $(D-4)$-dimensional parts of the
loop numerators \refeq{eq:rtoneloopD} and \refeq{eq:rttwoloopDD}, as well as the
associated rational terms, depend on the employed variant of dimensional
regularisation. In this respect we remind the reader that the analysis 
of~\cite{Pozzorini:2020hkx,Lang:2020nnl} and the new results
presented in this paper are based on the 
't~Hooft--Veltman scheme, where all algebraic
objects inside the loops live in $D$
dimensions. In other words, the rational terms presented in this paper
embody the effects resulting from all $(D-4)$-dimensional parts 
of loop momenta, $\gamma$-matrices and metric tensors that appear inside 
loop numerators.

\subsection{Derivation of rational counterterms}
\label{se:r2derivation}

The derivation of $\delta \calR_{2,\Gamma}$ counterterms 
can be simplified by means of expansions 
that capture the relevant UV divergences in the form of massive tadpole
integrals~\cite{Pozzorini:2020hkx,Lang:2020nnl}.
In the following, as a basis for~\refse{se:sb}, we
summarise the Taylor expansion approach
presented in App.~A.3--A.4 of~\cite{Lang:2020nnl}.
The idea is that the local two-loop divergence 
of degree $X=X(\Gamma)$, which is responsible
for $\delta \calR_{2,\Gamma}$, 
can be isolated through a Taylor expansion 
in the external momenta $\{p_{ia}\}$ and internal masses 
$\{m_{ia}\}$, 
which can be applied at the integrand level on the rhs
of~\refeq{eq:r2form}.
Technically, 
one can rescale all external momenta and internal masses by a global parameter
$\lambda$, 
\bea
\label{eq:globrescaling}
p_{ia} \,\to\, \lambda\, p_{ia}\,,
\qquad
m_{ia} \,\to\, \lambda\, m_{ia}
\qquad\forall\;\;
i,a\,,
\eea
and the relevant terms of order $\lambda^X$ 
can be selected 
through the expansion operator\footnote{The simplified notation for the 
expansion operators on the 
rhs should be understood as
\bea
\label{eq:Texpop}
\left(x\frac{\partial}{\partial x}\right)^k  \Bigg|_{x=0}\,F(x)&:=&
x^k \partial_x^k F(0)\,.
\eea
}
\bea
\label{eq:lambdaexp}
\bfT_{X} 
&=&
\frac{1}{X!}
\left(\frac{\rd}{\rd
\lambda}\right)^{\hspace{-1mm}X} \Bigg|_{\lambda=0}
\,=\,
\frac{1}{X!}
\left(\sum_{a,i}
p_{ia}^\mu \frac{\partial}{\partial \ell_{ia}^\mu}
+
\sum_{a,i}m_{ia} \frac{\partial}{\partial m_{ia}}
\right)^{\hspace{-1mm}X} \Bigg|_{p_{ia}\;=\; m_{ia}\;=\;0}\,,
\eea
where we have used the fact that 
loop integrands depend on $p_{ia}$ only through 
$\ell_{ia} = \bar q_i
+p_{ia}$. 
Applying the $\bfT_X$ expansion to 
generic $k$-loop integrals yields massless 
tadpole integrals of the form
\bea
\label{eq:tadexpintform}
\bfT_{X}\, 
\bar\calA_{k,\Gamma}
\,=\,
\sum_{\vec P}
\int
\prod_{i}\rd\barq_i \frac{\calT_{\vec P}(\{\bar q_k, p_{ka}, m_{ka}\})}
{\prod_j(\barq_j^2)^{P_i}}\,,
\eea
where $\vec P=(P_1,\dots)$ describes the denominator 
powers of the various loop chains, and
$\calT_{\vec P}(\{\bar q_k, p_{ka}, m_{ka}\})$ are
polynomials of homogeneous degree $X$ in 
$\{p_{ia}, m_{ia}\}$. 
The Taylor expansion~\refeq{eq:tadexpintform}
generates only scaleless tadpole integrals,
since all momenta and masses are set to zero in the denominators.
This can be avoided by 
introducing an auxiliary mass $M$ through the identity
\bea
\frac{1}{\barq_i^2} &=&
\frac{1}{\left(\barq_i^2-M^2\right)}
\left(1+ \frac{M^2}{\barq_i^2-M^2}\right)^{-1}\,,
\eea
and expanding the second term on the rhs
in $M^2/(\bar q_i^2-M^2)$, including 
all terms that are sufficiently UV divergent to 
contribute to~\refeq{eq:r2form}.
To capture all 
relevant UV singularities,
each loop chain $\calC_i$ of the two-loop diagram $\Gamma$ needs to be expanded 
up to relative order
\bea
\label{eq:XboundB}
\quad X_i(\Gamma) \,= \,
\text{max}\left\{ X(\Gamma), 
X(\gamma_j),X(\gamma_k)\right\}\,,
\eea
in $M/\barq_i$,
where $X(\Gamma)$, $X(\gamma_j)$ and $X(\gamma_k)$ are, respectively, the
superficial degrees of divergence of the full two-loop diagram $\Gamma$ and
of the one-loop subdiagrams with $j,k\neq i$.
In renormalisable theories $X_i(\Gamma)\in [0,2]$, 
and the relevant expansions for the terms $(1/\barq_i^2)^{P_i}$ 
in~\refeq{eq:tadexpintform} are
\bea
\label{eq:pmexp}
\bfM^{(i)}_{[0,X_i]}\,\,
\frac{1}{(\barq_i^2)^{P_i}}\,
&=&\,
\frac{1}{(\barq_i^2- M^2)^{P_i}}\times
\begin{cases}
1& 
\mbox{for}\quad X_i\le 1\,,
\\[3mm]
1- P_i\,\frac{M^2}{\bar q_i^2-M^2}
& \mbox{for}\quad X_i = 2\,.
\end{cases}
\eea
Note that the $\bfT_{X}$ and $\bfM_{[0,X_i]}^{(i)}$ expansions commute.

In summary, for the derivation of rational counterterms,
$k$-loop amplitudes $\bar\calA_{k,\Gamma}$ 
(and their counterparts in $\numdim=4$ dimensions)
can be replaced by 
\bea
\label{eq:genoptexp}
\bar\calA_{k,\Gamma_\tad} &=&
\bigg(\prod_{i}
\bfM_{[0,X_i]}^{(i)}\bigg)\, 
\bfT_{X}\, 
\bar\calA_{k,\Gamma}\,,
\eea
where the product includes all relevant loop chains, \ie one chain at one loop
and three chains at two loops. More explicity, 
applying~\refeq{eq:genoptexp} to~\refeq{eq:r2form} yields
\bea
\ratamp{2}{\Gamma}{}{}
&=&
\bfT_X
\int\rd\barq_1
\int \rd\barq_2
\left[
\bar\Gamma^{\bar\alpha_1\bar\alpha_2\bar\alpha_3}(\barq_1,\barq_2,\barq_3)\,
\prod_{i=1}^3 
\left({\bf M}^{(i)}_{[0,X_i]}\,
\frac{\bar\calN^{(i)}_{\bar\alpha_i}(\barq_i)}{\calD{i}}
\right)
-\Gamma^{\alpha_1\alpha_2\alpha_3}(q_1,q_2,q_3)\,
\right.
\nonumber\\[3mm]
&&\left.{}\times
\prod_{i=1}^3 
\left({\bf M}^{(i)}_{[0,X_i]}\,
\frac{\calN^{(i)}_{\alpha_i}(q_i)}{\calD{i}}
\right)
\right]_{q_3=-q_1-q_2}
+\;\sum_{i=1}^3
\bfT_X
\int\rd\barq_i\,
\Bigg[
\deltaZ{1}{\gamma_i}{\bar\alpha_i}{\barq_i}\,
{\bf M}^{(i)}_{[0,X_i]}
\,
\frac{\bar\calN^{(i)}_{\bar\alpha_i}(\barq_i)}{\calD{i}}
\nonumber\\[3mm]
&&
{}-
\left(\deltaZ{1}{\gamma_i}{\alpha_i}{q_i}
+\deltaZtilde{1}{\gamma_i}{\alpha_i}{\tilde q_i}+\ratamp{1}{\gamma_i}{\alpha_i}{q_i}\right)
{\bf M}^{(i)}_{[0,X_i]}\,
\frac{\calN^{(i)}_{\alpha_i}(q_i)}{\calD{i}}
\Bigg]
\,,
\label{eq:R2calcexpl}
\eea 
where $X=X(\Gamma)$ and $X_i=X_i(\Gamma)$.
Here the $\bfT_{X}$ expansion applies to all mass and momentum-dependent 
terms, including one-loop counterterms. The latter
are non-zero only when the corresponding subdiagrams $\gamma_i$
are divergent, in which case $X_i=X$~\cite{Lang:2020nnl}. 
The  dependence on the auxiliary mass $M$
cancels in~\refeq{eq:R2calcexpl}.
Moreover,
all higher-order terms generated by the $M^2/\barq_i^2$ expansions~\refeq{eq:pmexp}
are cancelled by corresponding 
one-loop counterterm contributions.
Thus, as discussed in App A.4 of~\cite{Lang:2020nnl}, 
the above expansion can be further simplified by 
setting $X_i=0$ throughout and 
compensating the 
effect of the missing subdivergences 
through ad-hoc modifications of the
one-loop counterterms 
$\deltaZ{1}{\gamma_i}{}{}$,
$\deltaZtilde{1}{}{}{}$, and
$\ratamp{1}{\gamma_i}{}{}$.

\section{Symmetry breaking and rational terms}
\label{se:sb}

The general properties of rational terms established 
in~\cite{Pozzorini:2020hkx,Lang:2020nnl} are valid for
any renormalisable theory, 
but so far the explicit expressions of 
$\ratamp{2}{\Gamma}{}{}$ counterterms have been worked out 
only for U(1) and SU(N) gauge theories.
For such theories, the task of determining $\ratamp{2}{\Gamma}{}{}$
via~\refeq{eq:r2form} 
is greatly simplified by the underlying gauge symmetry,
which allows one to treat all fields in terms of 
generic fermion and gauge-boson multiplets.
For instance, the rational terms for the 
quartic gauge-boson vertex $A_\mu^a A_\nu^b A_\rho^c A_\sigma^d$ in a
symmetric SU(N) theory can be derived 
through a single calculation, 
where the dependence from the SU(N) indices 
$a,b,c,d$ can be easily factorised.

In the presence of symmetry breaking this is no longer possible
due to the fact that the different states of each multiplet 
can acquire different masses and 
can mix with one another.\footnote{%
For instance, in the EW sector of the SM
the total number of two-loop Feynman diagrams that contribute to the 
quartic gauge-boson vertex $V_\mu^a V_\nu^b V_\rho^c V_\sigma^d$ 
grows by more than a factor fourty when the external and internal 
vector bosons $V=A,Z,W^\pm$ are handled as 
different mass eigenstates in the broken phase.
}
Moreover, also mass and field-renormalisation constants 
can vary within a multiplet. 
For these reasons, the direct determination of 
$\ratamp{2}{\Gamma}{}{}$ terms for the entire SM,
including contributions of order
$\alpha_S^2$, $\alpha \alpha_S$ and
$\alpha^2$, is a quite arduous task.
However, as discussed in the following, this problem can be 
greatly simplified by relating $\ratamp{2}{\Gamma}{}{}$ 
counterterms in a SB  theory
to corresponding rational counterterms 
in the underlying symmetric theory
via vev expansions.

\subsection{Mass and vev expansions}
\label{se:mvevexp}

As discussed in~\refse{se:r2derivation}, calculations of 
$\ratamp{2}{\Gamma}{}{}$  counterterms can be simplified by means of 
Taylor expansion in the external momenta and internal masses.
Such expansions, 
see~\refeq{eq:globrescaling}--\refeq{eq:tadexpintform},
involve only contributions 
from order zero up to $X$ in the internal masses $\{m_{ia}\}$, 
where $X$ is the superficial degree of divergence of
$\Gamma$ and corresponds to its mass dimension.
For this reason, 
all ingredients that enter the calculation of rational terms 
can be replaced by their expansion 
in $\{m_{ia}\}$ up to order $X$.
Thus, defining the 
expansion operator\footnote{Using again the simplified notation defined in \eqref{eq:Texpop}.}
\bea
\label{eq:massexp}
\bfm_{[0,X]}&=&
\sum_{k=0}^X
\frac{1}{k!}
\left(\sum_{a,i}m_{ia} \frac{\partial}{\partial m_{ia}}
\right)^{\hspace{-1mm}k} \Bigg|_{m_{ia}\;=\;0}\,{},
\eea
we can turn the formulas~\refeq{eq:r1form} and~\refeq{eq:r2form} 
into 
\bea 
\label{eq:expr1form}
\ratamp{1}{\Gamma}{}{}
&=&  
\bfm_{[0,X]}\,\Big[
\ampbar{1}{\Gamma}{}{}  
-\amp{1}{\Gamma}{}{}
\Big]\,,
\\[2mm]
\label{eq:expr2form}
\ratamp{2}{\Gamma}{}{}
&=&  
\bfm_{[0,X]}\,\Big[
\ampbar{2}{\Gamma}{}{}  
-\amp{2}{\Gamma}{}{} + 
\sum  \limits_{\gamma} \deltaZ{1}{\gamma}{}{}
\cdot \ampbar{1}{\Gamma/\gamma}{}{} 
-
\sum  \limits_{\gamma} \lb \deltaZ{1}{\gamma}{}{}
+\deltaZtilde{1}{\gamma}{}{} + \ratamp{1}{\gamma}{}{} \rb \cdot
\amp{1}{\Gamma/\gamma}{}{}
\Big]\,.\nonumber\\
\eea
In the following, we consider such mass expansions in 
the context of a generic SB  theory
with a scalar multiplet $\Phi_i(x)$ that acquires a 
vev
\bea
\label{eq:vevdefA}
\vev_i &=& \langle 0|\Phi_i(x)|0\rangle\,.
\eea
The scalar excitations around the vev are denoted
\bea
\label{eq:vevdefA2}
\Phi_i(x) = \vev_i + \widetilde\Phi_i(x)\,.
\eea
For simplicity we assume that the scalar sector has a single vev 
and a single Higgs boson, and for the scalar oscillations
in the vev direction we use the notation
\bea
\label{eq:vevdefB}
\phi (x) = \vev +H(x),\qquad
\vev &=& \langle 0|\phi(x)|0\rangle\,.
\eea
In this Section,
we assume that the SB theory is described by a
Lagrangian $\calL(H)$, which is connected to 
the Lagrangian $\calL_{\ubk}(H)$ of an underlying symmetric theory
through a shift of the Higgs field, $H\to H+\vev$ , \ie
\bea
\label{eq:brokentheory}
\mathcal{L}_{\bk}(H) &=&
\mathcal{L}_{\ubk}(H+\vev)\,.
\eea 
This implies that the symmetry of $\calL_{\ubk}$ 
is inherited by  the SB Lagrangian in the form of 
rigid invariance,\footnote{Violations of rigid invariance 
induced by gauge fixing are discussed in~\refse{se:thooftgfix}.} 
\ie invariance wrt 
combined gauge transformations of $\vev_i + \widetilde\Phi_i(x)$.
For $\mathcal{L}_{\ubk}$ we do not assume any specific 
symmetry group, 
and we
only assume that in the SB theory
all particle masses are generated by the 
vev of the scalar field $\phi$.
 Thus all masses 
are proportional to $\vev$, and
the mass expansion~\refeq{eq:massexp} 
can be identified with a vev expansion,
\bea
\label{eq:vevexp}
\bfm_{[0,X]}\,=\,
\sum_{k=0}^X
\frac{\vev^k}{k!}\,
\frac{\partial^k}{\partial \vev^k}
\bigg|_{\vev\;=\;0} 
\,.
\eea
Since UV divergences in the SM are at 
most quadratic, expansions  
truncated at second order will be sufficient. 

Concerning the propagators of massive particles
we will first assume that all mass eigenstates
have well defined tree-level propagators, \ie that the mixing 
between different fields can be removed by means of appropriate 
unitary transformations.
In this respect we note that in the SM 
the cancellation of mixing between gauge bosons and Goldstone bosons 
requires also appropriate gauge choices, and the 
corresponding gauge-fixing terms 
can be in conflict with the
assumption of rigid invariance, \ie with the property~\refeq{eq:brokentheory}
of the SB Lagrangian.
A systematic approach to deal with such violations of 
rigid invariance is discussed in~\refse{se:thooftgfix}.

\subsection{Vev-expansion in $\numdim=D$ dimensions}
\label{se:ddvevexp}

As is well known, the vev expansion~\refeq{eq:vevexp} can be implemented at
the level of the Feynman rules.  This is achieved by 
turning the usual
perturbative expansion in the coupling constants $g_i$ 
into a double expansion
in $g_i$ and $\vev$. 
In this approach, 
vev-dependent terms of $\mathcal{L}(H)$ 
are treated as interaction
vertices, which can be generated from corresponding 
vertices of the symmetric theory by replacing 
external Higgs lines with vev insertions.
 
To introduce our notation, in the following we consider the mass expansion 
for the two-point vertex function\footnote{$\Gamma_{aa}$ should not be confused with the 
generic loop diagram (or vertex function) $\Gamma$ that enters 
identities like~\refeq{eq:expr1form}--\refeq{eq:expr2form}.
Note also that in~\cite{Lang:2020nnl} 
we used the symbol $\Gamma_{a}$ instead of
$\Gamma_{aa}$, while the latter is more appropriate
for the discussion at hand.
}
$\bar\Gamma_{aa}(\barq, m_a)$ 
of a generic field $\varphi_a$ with  mass $m_a$, 
and for the related propagator $\bar G_{aa}^{\bk}(\bar q, m_a)$,
defined through
\bea
\label{eq:propdefDdim}
\label{eq:DdimpropC}
\bar G_{aa}^{\bk}(\barq,m_a) \,
\bar \Gamma_{aa}^{\bk}(\barq,m_a)=-1\,.
\eea
The field $\varphi_a$ may be a fermion, a vector boson, or a scalar field.
For later convenience, the momentum $\barq$ that flows 
through the two-point function at hand
is assumed to be a $D$-dimensional loop momentum. 
Possible Lorentz/Dirac indices associated with $\varphi_a$ 
are also assumed to be in $D$ dimensions
but are kept implicit.
At tree level, the propagator has the form
\bea
\label{eq:Ddimprop}
\bar G_{aa}^{\bk}(\bar q, m_a) = \f{\bar g_{aa}^{\bk}(\barq,m_a)}{\bar q^2 -m_a^2}\,,
\eea
and its numerator fulfils 
\bea
\label{eq:DdimpropB}
\bar g_{aa}^{\bk}(\bar q,m_a) \, \bar \Gamma_{aa}^{\bk}(\bar q,m_a) & = &  -
(\bar q^2 - m_a^2)\,.
\eea
The mass dependence of $\bar\Gamma_{aa}(\barq, m_a)$
in the SB  theory
arises from the triple $H \varphi_a^2$ 
and/or quartic 
$H^2 \varphi_a^2$ vertices of the symmetric theory,
\bea
\label{eq:aaHvertices}
\bar\Gamma^{\ubk}_{aa \rHexp }(\bar q)
\;\;=\;\;
\vcenter{\hbox{\raisebox{0pt}{\includegraphics[width=0.11\textwidth]{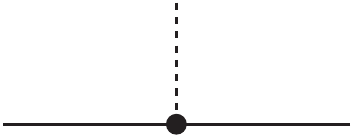}} }} 
\qquad\qquad
\bar\Gamma^{\ubk}_{aa \rHexp \rHexp }(\bar q)
\;\;=\;\;
\vcenter{\hbox{\raisebox{0pt}{\includegraphics[width=0.11\textwidth]{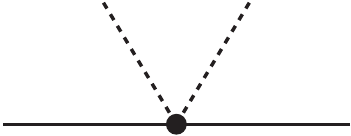}}}}
\,,
\eea
where solid and dashed lines correspond, respectively, to the $\varphi_a$ and 
Higgs fields.
Note that in the SM the triple vertex 
$\bar \Gamma_{aa\rHexp }^{\ubk}$ is vanishing for scalars and gauge bosons,
while in the case of fermions the quartic vertex $\bar \Gamma_{aa \rHexp \rHexp }^{\ubk}$ 
vanishes.
As a result of symmetry breaking, 
the vertices~\refeq{eq:aaHvertices} give rise to mass terms 
of the form $\vev \varphi_a^2$ and $\vev^2 \varphi_a^2$, which 
can be described through the vev-insertion vertices
\bea
\label{eq:aavevvertices}
\bar\Gamma^{\ubk}_{aa\vev}(\bar q)
\;&=&\;
\vcenter{\hbox{\raisebox{0pt}{\includegraphics[width=0.11\textwidth]{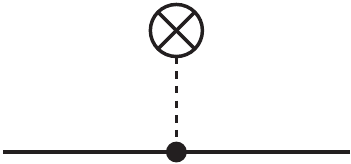}} }} 
\,,
\qquad\qquad
\bar\Gamma^{\ubk}_{aa\vev\vev}(\bar q)
\;=\;
\vcenter{\hbox{\raisebox{0pt}{\includegraphics[width=0.11\textwidth]{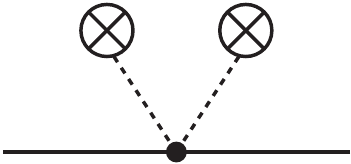}} }}
\,,
\eea
where vev insertions are indicated by crossed blobs. Each 
vev insertion corresponds to a factor $\vev$ together with the
assignment of zero external momentum for the corresponding 
Higgs line.
Moreover, diagrams with $k$ vev insertions receive a $1/k!$ factor.
More explicitly, 
$\bar\Gamma^{\ubk}_{aa\vev}(\bar q)
=
\vev\,\bar\Gamma^{\ubk}_{aa \rHexp }(\bar q)
\big|_{p_{\rHexp }=0}$
and
$\bar\Gamma^{\ubk}_{aa\vev\vev}(\bar q)
=
\frac{\vev^2}{2}\bar\Gamma^{\ubk}_{aa \rHexp \rHexp }(\bar q)
\big|_{p_{\rHexp }=0}
$.
With these vev-insertion vertices at hand, the expansion of the 
$\varphi_a$ two-point function up to order $\vev^2$ can be expressed as
\bea
\label{eq:2ptvevexpE}
\bfm_{[0,2]}\, \bigg[\,
\vcenter{\hbox{\raisebox{-15pt}{\includegraphics[width=0.09\textwidth]{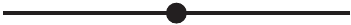}} }}\bigg]^{\bk}_{\numdim=D}  \!\!\!\!\! 
&=&
\; \bigg[
\vcenter{\hbox{\raisebox{-22pt}{\includegraphics[width=0.1\textwidth]{fermion_amp}} }} 
+
\vcenter{\hbox{\raisebox{0pt}{\includegraphics[width=0.11\textwidth]{fermionSVev_amp}} }} 
+
\vcenter{\hbox{\raisebox{0pt}{\includegraphics[width=0.11\textwidth]{bosonDVev_amp}} }}
\bigg]^{\ubk}_{\numdim=D}\,,
\eea
or, equivalently,
\bea
\label{eq:2ptvevexpD1}
\bfm_{[0,2]}\, 
\bar\Gamma_{aa}(\bar q, m_a) &=&
\bfV_{[0,2]}\, 
\bar\Gamma_{aa}(\bar q, m_a)\,, 
\eea
with
\bea
\label{eq:2ptvevexpD}
\bfV_{[0,2]}\, 
\bar\Gamma_{aa}(\bar q, m_a) &=&
\bar\Gamma^{\ubk}_{aa}(\bar q)
+
\bar\Gamma^{\ubk}_{aa\vev}(\bar q)
+
\bar\Gamma^{\ubk}_{aa\vev\vev}(\bar q)\,.
\eea
Note that in~\refeq{eq:aavevvertices}--\refeq{eq:2ptvevexpE} 
only tree contributions are considered,
while~\refeq{eq:2ptvevexpD1}--\refeq{eq:2ptvevexpD} hold to any order in the coupling constant.

With this approach, the vev expansion of 
any $l$-loop amplitude in $\numdim=D$ dimensions 
can be expressed as a series of vev insertions, 
\bea 
\label{eq:ddimvevexp}
\bfm_{[0,X]}\, \ampbar{l}{\Gamma}{\bk}{} 
&=&
\bfV_{[0,X]}\, \ampbar{l}{\Gamma}{\bk}{}\,,
\eea
with 
\bea  
\label{eq:ddimvevexp2}
\bfV_{[0,X]}\, \ampbar{l}{\Gamma}{\bk}{} 
&=& \sum_{k=0}^{X}  
\ampbar{l}{\Gamma \VevInsert }{\ubk}{}
\qquad\mbox{and}\qquad
\ampbar{l}{\Gamma \VevInsert }{\ubk}{}
\,:=\,
\f{v^k}{k!} \, \ampbar{l}{\Gamma  \rHexp ^k }{\ubk}{}  \Big|_{p_{\rHexp }=0} 
\,,
\eea
where amplitudes carrying the superscript YM
are defined through the Feynman rules 
of the symmetric theory
with Lagrangian $\calL_{\ubk}(H)$.
The shorthands $\vev^k$ and $\rHexp ^k$ denote 
$k$ vev  and Higgs insertions, respectively,
and $p_{\rHexp }=0$ applies only to such insertions,
while it does not apply to
possible physical Higgs lines that enter the original vertex
$\Gamma$.
Note that~\refeq{eq:ddimvevexp}--\refeq{eq:ddimvevexp2} apply also to
vertices that do not exist at tree
level in the symmetric theory and are induced by symmetry breaking,
such as the $W^+W^-H$ vertex in the SM.
In the case of vertices $\Gamma$ of order $\vev^n$ at tree level,
the vev insertions $\Gamma \vev^k$ with $k< n$ 
vanish in~\refeq{eq:ddimvevexp2}.
Note also that when vev insertions are applied to 
mass-dependent vertices inside the loops, the global degree of 
singularity is not reduced. However, this happens only 
when---due to the presence of
mass-dependent interactions---the superficial degree of
singularity is smaller than what is expected from 
dimensional analysis. 
In general, in order to 
capture all UV singularities, 
it is sufficient to set the order 
$X$ of the vev expansion 
equal to the 
maximum degree of UV singularity
that is allowed by dimensional analysis, \ie
\bea
X=4-\mathrm{dim}(\Gamma)\,,
\eea
where $\mathrm{dim}(\Gamma)$ is the
mass dimension of the
vertex function  at hand.

As a preparation for
the discussion in~\refse{se:4dvevexp},
let us now consider the vev expansion up to order $\vev^2$
for the propagator defined in~\refeq{eq:propdefDdim}, 
\bea
\label{eq:mexpGaa}
\bfm_{[0,2]}\, \bar G_{aa}^{\bk}(\bar q, m_a) 
&=&
\sum_{k=0}^2
\frac{m^k_a}{k!}\,
\frac{\partial^k}{\partial m_a^k}
\bigg|_{m_a\,=\,0}\,
\bar G_{aa}^{\bk}(\bar q, m_a)\,.
\eea
The vev derivatives of $\bar G_{aa}$ on the rhs
can be expressed in terms of vev derivatives of the two-point 
function $\bar \Gamma_{aa}^{\bk}$ by using~\refeq{eq:propdefDdim}
and the related identities
\bea
\frac{\partial^k}{\partial m_a^k}\left[
\bar G_{aa}^{\bk}(\barq,m_a) \, \bar \Gamma_{aa}^{\bk}(\barq,m_a)\right]
&=&0\,.
\eea
In this way one can show that 
\bea
\label{eq:GvevexpDa}
\bfm_{[0,2]}\, \bar G_{aa}^{\bk}(\bar q, m_a) 
&=&
\bar G_{aa}^{\ubk} (\bar q)
\,+\, 
\bar G_{aa}^{\ubk} (\bar q)\,
\bar \Gamma_{aa\vev}^{\ubk} (\bar q)\,
\bar G_{aa}^{\ubk} (\bar q)
\,+\, 
\left[\bar G_{aa}^{\ubk} (\bar q)\,
\bar \Gamma_{aa\vev}^{\ubk} (\bar q)\right]^2\,
\bar G_{aa}^{\ubk} (\bar q)
\nonumber\\[1mm]&&{}
+\,
\bar G_{aa}^{\ubk} (\bar q)\,
\bar \Gamma_{aa\vev\vev}^{\ubk} (\bar q)\,
\bar G_{aa}^{\ubk} (\bar q)\,,
\eea
where $\bar G_{aa}^{\ubk} (\bar q)$ is the propagator of the symmetric
theory, and $\bar \Gamma_{aa\vev^k}^{\ubk} (\bar q)$ are the vev-insertion
vertices defined in~\refeq{eq:aavevvertices}.
This relation can be expressed diagrammatically as
\bea
\label{eq:GvevexpDb} 
\bfm_{[0,2]}\, \bigg[\, \vcenter{\hbox{\raisebox{-15pt}{\includegraphics[width=0.09\textwidth]{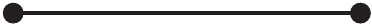}} }}\bigg]^{\bk}_{\numdim=D}  \!\!\!\!\! 
&=&
\; \bigg[ \vcenter{\hbox{\raisebox{-22pt}{\includegraphics[width=0.1\textwidth]{fermion}} }} 
+  \vcenter{\hbox{\raisebox{0pt}{\includegraphics[width=0.11\textwidth]{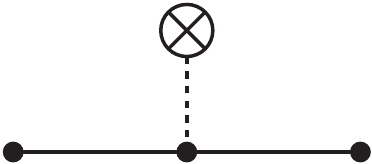}} }} 
+  \vcenter{\hbox{\raisebox{0pt}{\includegraphics[width=0.11\textwidth]{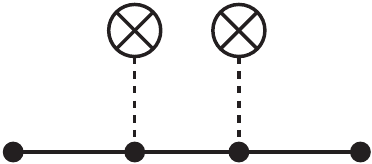}} }}
+  \vcenter{\hbox{\raisebox{0pt}{\includegraphics[width=0.11\textwidth]{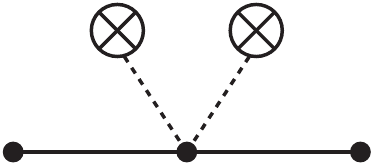}} }}  \bigg]^{\ubk}_{\numdim=D} 
\,, \nonumber \\
\eea
and analogously as for the expansion
of the two-point function $\bar\Gamma_{aa}$ 
it can be directly obtained 
from~\refeq{eq:ddimvevexp}--\refeq{eq:ddimvevexp2}
in the form
\bea
\label{eq:GvevexpDc}
\bfm_{[0,2]}\, 
\bar G_{aa}(\bar q, m_a) &=&
\bfV_{[0,2]}\, 
\bar G_{aa}(\bar q, m_a) 
\,=\,
\bar G^{\ubk}_{aa}(\bar q)
+
\bar G^{\ubk}_{aa\vev}(\bar q)
+
\bar G^{\ubk}_{aa\vev\vev}(\bar q)\,,
\eea
where $\bar G^{\ubk}_{aa\vev\vev}$ embodies the contributions of the
last two diagrams in~\refeq{eq:GvevexpDb}.

\subsection{Vev-expansion in $\numdim=4$ dimensions}
\label{se:4dvevexp}

In order to simplify the calculation of rational 
counterterms~\refeq{eq:expr1form}--\refeq{eq:expr2form}
for SB  theories,
the  vev-insertion approach~\refeq{eq:ddimvevexp}--\refeq{eq:ddimvevexp2}
needs to be extended to amplitudes in $\numdim=4$ dimensions.
To this end, let us first consider the naive
conversions to $\numdim=4$ of the  
lhs and rhs of~\refeq{eq:ddimvevexp}.
The $\numdim=4$ counterpart of the vev expansion 
on the lhs of~\refeq{eq:ddimvevexp} is
\bea
\label{eq:mexp4Da}
\bfm_{[0,X]}\, \amp{l}{\Gamma}{\bk}{} 
&=&  
\bfm_{[0,X]}\, \bfP_4 \,\ampbar{l}{\Gamma}{\bk}{}\,,
\eea
where the operator $\bfP_4$ projects
loop numerators to $\numdim=4$.
As for the vev insertion on the rhs 
of~\refeq{eq:ddimvevexp} we define 
its naive $\numdim=4$ projection as 
\bea
\label{eq:mexp4Dc}
\bfV_{[0,X]}\, \amp{l}{\Gamma}{\bk}{}
&:=&  
\bfP_4\, 
\bfV_{[0,X]}\, \ampbar{l}{\Gamma}{\bk}{} 
\,=\, \sum_{k=0}^{X}  
\amp{l}{\Gamma \VevInsert }{\ubk}{}\,,
\eea
with 
\bea
\label{eq:mexp4Dd}
\amp{l}{\Gamma \VevInsert }{\ubk}{}
\,=\,
\f{v^k}{k!} \, \amp{l}{\Gamma  \rHexp ^k }{\ubk}{}  \Big|_{p_{\rHexp }=0}\,.
\eea
As we will see, at variance with the
$\numdim=D$ case, 
\refeq{eq:mexp4Da} and~\refeq{eq:mexp4Dc} are not identical
in $\numdim=4$.
This is due to the fact that the vev expansion~\refeq{eq:vevexp}
and the projection to $\numdim=4$ do not commute.
Thus, the extension of~\refeq{eq:ddimvevexp} to 
$\numdim=4$ assumes the form
\bea
\label{eq:mexp4Db}
\bfm_{[0,X]}\, \amp{l}{\Gamma}{\bk}{} 
&=&  
\bfV_{[0,X]}\, \amp{l}{\Gamma}{\bk}{}
+
\Delta\bfV_{[0,X]}\, \amp{l}{\Gamma}{\bk}{}\,,
\eea
where the second term on the rhs embodies the non-commuting terms. 
More explicitly, we can write
\bea
\label{eq:mexp4De}
\Delta\bfV_{[0,X]}\, \amp{l}{\Gamma}{\bk}{}
&=&  
\bfm_{[0,X]}\, \amp{l}{\Gamma}{\bk}{} 
-
\bfV_{[0,X]}\, \amp{l}{\Gamma}{\bk}{}
\,=\,
\left(\bfm_{[0,X]}\bfP_4
-
\bfP_4\,\bfV_{[0,X]}\right) \ampbar{l}{\Gamma}{\bk}{}\,,
\eea
and using~\refeq{eq:ddimvevexp} we arrive at
\bea
\label{eq:commutator}
\Delta\bfV_{[0,X]}\, \amp{l}{\Gamma}{\bk}{}
&=&
\big[\bfm_{[0,X]}\,,\bfP_4\big]
\, \ampbar{l}{\Gamma}{\bk}{}\,.
\eea
To gain more insights into the commutator $\Delta\bfV_{[0,X]}$, 
we will consider its effect on the basic building blocks
of loop diagrams, \ie
loop-momentum dependent 
tree vertices and propagators.
Tree vertices inside loops are simple polynomials of the form 
$\sum_{k=0}^n m^k\,\bar \Lambda_k(\bar q)$, where 
the projection to $\numdim=4$ acts only 
on the mass-independent coefficients $\bar \Lambda_k(\bar q)$.
For this reason   the commutator~\refeq{eq:commutator} vanishes, 
which means that, in the case of tree vertices, 
the mass expansion
$\bfm_{[0,X]}$
and the $\bfV_{[0,X]}$ vev insertions
are equivalent to each other both in $\numdim=D$ and $\numdim=4$.
For instance, for the expansion of a two-point vertex function 
up to order $\vev^2$, in $\numdim=4$  we have
\bea
\label{eq:mexp2pt4d}
\bfm_{[0,2]} \, \Gamma_{aa}^{\bk}(q, m_a ) 
&\,\loeq\,& \bfV_{[0,2]} \, \Gamma_{aa}^{\bk}( q, m_a ) 
\,=\,
\Gamma^{\ubk}_{aa}( q)
+
\Gamma^{\ubk}_{aa\vev}( q)
+
\Gamma^{\ubk}_{aa\vev\vev}( q)\,.
\eea

Let us now discuss the loop propagators~\refeq{eq:Ddimprop} and their
projection to $\numdim=4$,
\bea
\label{eq:4dimprop}
G_{aa}^{\bk}(\bar q, m_a) = \f{g_{aa}^{\bk}(q,m_a)}{\bar q^2 -m_a^2}\,.
\eea
In this case, due to the non-polynomial dependence on $m_a$
and the different effect of the $\numdim=4$ projection on the
loop momentum in the numerator and denominator,
the commutator~\refeq{eq:commutator} does not vanish.
Its explicit expression can be
derived from   
\bea
\label{eq:commutid}
\Delta\bfV_{[0,2]} \, G_{aa}^{\bk}(\bar q, m_a ) 
&=&
\bfm_{[0,2]} \, G_{aa}^{\bk}(\bar q, m_a ) 
-
\bfV_{[0,2]} \, G_{aa}^{\bk}(\bar q, m_a )\,,
\eea
\ie by comparing the mass expansion
\bea
\label{eq:massexpG4}
\bfm_{[0,2]}\,  G_{aa}^{\bk}(\bar q, m_a) 
&=&
\sum_{k=0}^2
\frac{m_a^k}{k!}
\frac{\partial^k}{\partial m_a^k}
\bigg|_{m_a\;=\;0}\,
G_{aa}^{\bk}(\bar q, m_a)\,,
\eea
to the vev insertions
\bea
\bfV_{[0,2]}\,  G_{aa}(\bar q, m_a) 
&=&
 G^{\ubk}_{aa}(\bar q)
+
 G^{\ubk}_{aa\vev}(\bar q)
+
 G^{\ubk}_{aa\vev\vev}(\bar q)\,.
\eea
The latter can be written, similarly as on the rhs of~\refeq{eq:GvevexpDa}
and~\refeq{eq:GvevexpDb},
as
\bea
\label{eq:2ptder4d}
\bfV_{[0,2]}\,  G_{aa}(\bar q, m_a) 
&=&
 G_{aa}^{\ubk} (\bar q)
\,+\, 
 G_{aa}^{\ubk} (\bar q)\,
 \Gamma_{aa\vev}^{\ubk} ( q)\,
 G_{aa}^{\ubk} (\bar q)
\,+\, 
\left[ G_{aa}^{\ubk} (\bar q)\,
 \Gamma_{aa\vev}^{\ubk} ( q)\right]^2\,
 G_{aa}^{\ubk} (\bar q)
\nonumber\\[1mm]&&{}
\,+\,
 G_{aa}^{\ubk} (\bar q)\,
 \Gamma_{aa\vev\vev}^{\ubk} ( q)\,
 G_{aa}^{\ubk} (\bar q)\,,
\eea
where $ G_{aa}^{\ubk} (\bar q)$ is the propagator of the symmetric
theory, while, according to~\refeq{eq:mexp2pt4d}, 
the vev insertions on the rhs correspond to 
derivatives of the two-point function,
\bea
\label{eq:mderGamma4}
\Gamma_{aa\vev^k}^{\ubk} (q)
&=&
\frac{m_a^k}{k!}
\frac{\partial^k}{\partial m_a^k}
\bigg|_{m_a\;=\;0}\,
\Gamma_{aa}(q,m_a)\,.
\eea
Following similarly lines as in the $\numdim=D$ case, 
mass expansions and vev insertions 
can be related to each other via identities that 
connect the derivatives of $G_{aa}(q,m_a)$ 
in~\refeq{eq:massexpG4}
to the ones of 
$\Gamma_{aa}(q,m_a)$
in~\refeq{eq:2ptder4d}--\refeq{eq:mderGamma4}.
In $\numdim=D$ such identities 
follow from the trivial relations~\refeq{eq:propdefDdim} and
\refeq{eq:DdimpropB} between propagator and two-point function, 
while in $\numdim=4$ we have
\bea
\label{eq:4dimpropB}
g_{aa}^{\bk}(q,m_a) \, \Gamma_{aa}^{\bk}(q,m_a) & = &  - (q^2 - m_a^2) \,,
\eea
and
\bea
\label{eq:4dimpropC}
 \quad G_{aa}^{\bk}(\bar q,m_a) \,\Gamma_{aa}^{\bk}(q,m_a) &=& -\f{q^2 - m_a^2}{\bar q^2 - m_a^2} \, = \, \f{\tilde q^2}{\bar q^2-m_a^2}-1
\,.
\eea
Here we see that, contrary to the $\numdim=D$ case,
$G_{aa}^{\bk}(\bar q,m_a)$ does not correspond to the
inverse of $\Gamma_{aa}^{\bk}(q,m_a)$. In particular, the 
$\tilde q^2$ terms on the rhs of~\refeq{eq:4dimpropC} 
indicate that the inversion of the two-point function
does not commute with the projection to $\numdim=4$.
As shown in~\refapp{app:DeltaV2}, such non-commuting 
terms lead to a $\tilde q^2$-dependent 
difference between mass expansion and vev insertions
in~\refeq{eq:commutid}, and 
for a generic numerator function
$g_{aa}(q,m_a)$ we find
\bea
\label{eq:deltaV02gen}
\Delta\bfV_{[0,2]} \, G_{aa}^{\bk}(\bar q, m_a ) 
&=&
g_{aa \vev}\, 
\f{\tilde q^2}{\bar q^4} 
\,-\, 
g^2_{aa\vev}\,
\Gamma_{aa}^{\ubk}(q)\, 
\f{\tilde  q^2}{\bar q^6} 
\,+\, 
g_{aa \vev\vev}\,
\f{\tilde q^2}{\bar q^4}\,,
\eea
with
\bea
\label{eq:gmassder}
g_{aa\vev}
\,=\, m_a \, \frac{\partial g_{aa}(q,m_a)}{\partial m_a}  \bigg|_{m_a=0}\,,
\qquad
g_{aa \vev\vev}
\,=\, \frac{m_a^2}{2}\frac{\partial^2 g_{aa}(q,m_a)}{\partial m_a^2}
\bigg|_{m_a=0}\,.
\eea
The above derivations assume propagators of the form 
\refeq{eq:4dimprop}, where the numerator $g_{aa}(q,m_a)$ is a polynomial 
in $q$ and $m_a$. In the SM they are applicable
to all fermion, scalar and ghost propagators.
Explicit expressions for the various types of propagators are 
easily obtained from the mass derivatives~\refeq{eq:gmassder} 
of the corresponding numerators.
In the case of fermion propagators we have 
$g_{ff}^{\bk}(q,m_f) = \ri \, (\slashed q + m_f)$
and 
$\Gamma_{ff}^{\ubk}(q) = \ri \, \slashed q$, which yields
\bea
\label{eq:deltaV02fer}
\Delta \bfV_{[0,2]}\, G_{ff}^{\bk}(\bar q, m_f) & = &
\ri  m_f \, \f{\tilde q^2}{\bar q^4}  
\,+\, \ri m_f^2\, \f{\slashed q \,  \tilde q^2 }{\bar q^6} 
\qquad\mbox{for fermions}\,. 
\eea
The numerators of scalar and ghost propagators are 
free from mass terms. Thus
\bea
\label{eq:eq:deltaV02bos}
\Delta \bfV_{[0,2]}\, G_{aa}^{\bk}(\bar q, m_a)
&=& 0
\qquad\mbox{for scalars and ghosts}\,.
\eea
The same holds also for gauge-boson propagators%
\footnote{For a
discussion of gauge fixing and mixing between vector bosons and 
Goldstone bosons we refer to~\refse{se:thooftgfix}.
} in the Feynman gauge.
For gauge bosons in the $R_\xi$ gauge we have
\bea
\Gamma_{_{VV}}^{\mu \nu} (q,m_V, \xi) &=&
 - \ri \left[  q^2 g^{\mu \nu} + \lb \f{1}{\xi} - 1 \rb q^\mu q^\nu - m_V^2 \, g^{\mu \nu}\right]\,, \\
 G_{_{VV}}^{\mu \nu} (\bar q, m_V, \xi) &=& 
 \f{g_{_{VV}}^{\mu \nu} (\bar q, m_V, \xi) }{\bar{q}^2-m_V^2} \,, \quad 
 g_{_{VV}}^{\mu \nu} (\bar q, m_V, \xi) \;=\;
  - \ri \left[  g^{\mu \nu} + (\xi-1) \f{q^\mu q^\nu}{\bar{q}^2 - \xi m_V^2}
    \right].\qquad
\eea
Here the formula~\refeq{eq:deltaV02gen}
is not applicable for $\xi\neq 1$ 
since $g_{_{VV}}^{\mu \nu}(\bar q, m_V, \xi)$ involves an 
additional ($\bar{q}^2 - \xi m_V^2$) denominator.
An explicit derivation yields,
\bea
\label{eq:Rxitvev}
\Delta \bfV_{[0,2]} \, G_{_{VV}}^{\mu \nu} (\bar q, m_V, \xi) &=& 
\displaystyle \; - \, \ri \, m_V^2 \, (\xi-1)^2 \, \f{q^\mu q^\nu \tilde{q}^2}{\bar{q}^8}
\quad\mbox{for gauge bosons in the $R_\xi$ gauge.}
\nonumber\\
\eea
This confirms that $\Delta \bfV_{[0,2]} \, G_{_{VV}}^{\mu \nu}$
vanishes in the Feynman gauge, as expected from~\refeq{eq:deltaV02gen}.
Note also that 
$\Delta \bfV_{[0,2]}\, G_{aa}=0$ 
for any massless propagator since the mass expansion does not generate any
term beyond zeroth order.

\subsection{Auxiliary vev insertions in $\numdim=4$}
\label{eq:tvevins}

In $\numdim=D$ 
dimensions  the mass expansion~\refeq{eq:ddimvevexp}--\refeq{eq:ddimvevexp2}
can be generated in a systematic way by supplementing the 
Feynman rules of the symmetric theory with the 
vev-insertion vertices~\refeq{eq:aavevvertices}.
As shown in the following, this approach can be easily extended to 
the expansion~\refeq{eq:mexp4Db} in $\numdim=4$. To this end, 
in order to account for the additional
$\Delta \bfV_{[0,X]}$ corrections to fermion-loop propagators,
we introduce 
auxiliary vev-insertion vertices of the form
\bea
\label{eq:vtildevertices}
G_{aa\tvev}^{\ubk}(\bar q) \;=\; 
\vcenter{\hbox{\raisebox{8pt}{\includegraphics[width=0.125\textwidth]{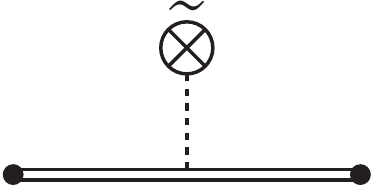}}
}}\;\;,\qquad\quad 
G_{aa\tvev\tvev}^{\ubk} (\bar q) \;=\;
\vcenter{\hbox{\raisebox{8pt}{\includegraphics[width=0.125\textwidth]{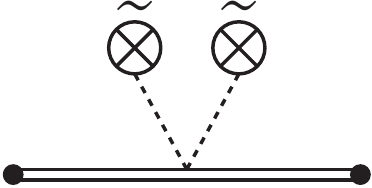}}
}}\;\;.
\eea
The crossed blobs carrying a tilde 
will be
referred to as auxiliary vev insertions, or $\tvev$ insertions. 
Similarly as for standard vev insertions,
each $\tvev$ insertion is associated with a
zero-momentum external state and with a factor $\vev$.
Using $\tvev$ insertions in the Feynman rules~\refeq{eq:vtildevertices}
makes it possible to reduce the power counting in $\vev$
to a trivial counting of the total number of 
$\vev$ and $\tvev$ insertions.
Otherwise, the external $\tvev$ states have no physical 
significance. 
In fact, the diagrams in~\refeq{eq:vtildevertices}
should be regarded as auxiliary propagators that embody the 
$\Delta \bfV_{[0,2]}$ corrections to standard fermion-loop propagators.
In particular, the dots on the two ends of the double lines 
should be connected to interaction vertices in the same way as for standard fermion
propagators.

In this approach, the second-order expansion of loop propagators in $\numdim=4$, 
\bea
\label{eq:Gexpd4}
\bfm_{[0,2]} \, G_{aa}^{\bk}(\bar q, m_a ) 
&=&
\bfV_{[0,2]} \, G_{aa}^{\bk}(\bar q, m_a )
+
\Delta\bfV_{[0,2]} \, G_{aa}^{\bk}(\bar q, m_a ) 
\,,
\eea
can be generated through the extended Feynman rules of the symmetric theory
via combinations of $\vev$ and $\tvev$ insertions
up to order $\vev^2$.
Pure $\vev$ insertions yield the part
\bea
\bfV_{[0,2]} \, G_{aa}^{\bk}(\bar q, m_a )&=&
G^{\ubk}_{aa}(\bar q)
+
G^{\ubk}_{aa\vev}(\bar q)
+
G^{\ubk}_{aa\vev\vev}(\bar q)\,,
\eea
while  $\tvev$ insertions and mixed $\vev\tvev$ insertions
yield the remaining part 
\bea
\label{eq:DVtvevins}
\Delta\bfV_{[0,2]} \, G_{aa}^{\bk}(\bar q, m_a )&=&
G^{\ubk}_{aa\tvev}(\bar q)
+
G^{\ubk}_{aa\vev\tvev}(\bar q)
+
G^{\ubk}_{aa\tvev\tvev}(\bar q)\,.
\eea
Diagrammatically, the latter identity reads
\bea
\label{eq:DVtvevinsdia}
\Delta\bfV_{[0,2]} 
\, \bigg[\, \vcenter{\hbox{\raisebox{-15pt}{\includegraphics[width=0.09\textwidth]{fermion}} }}\bigg]^{\bk}_{\numdim=4}  
&=&
\bigg[
\vcenter{\hbox{\raisebox{2pt}{\includegraphics[width=0.11\textwidth]{fermiondtVI}} }}
\,+\,  \vcenter{\hbox{\raisebox{2pt}{\includegraphics[width=0.11\textwidth]{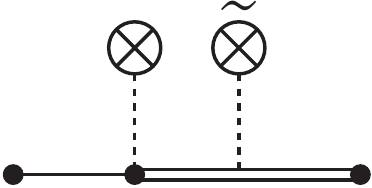}} }} 
\,+\,  
\vcenter{\hbox{\raisebox{2pt}{\includegraphics[width=0.11\textwidth]{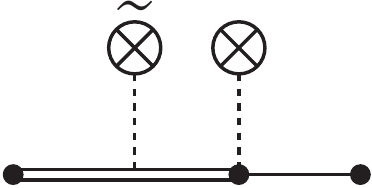}} }} 
\,+\,  \vcenter{\hbox{\raisebox{2pt}{\includegraphics[width=0.11\textwidth]{fermiondtVII}} }} 
\bigg]^{\ubk}_{\numdim=4}\,, \nonumber \\[1mm] 
\eea
and the Feynman rules for $\tvev$ insertions
can be obtained by
matching~\refeq{eq:DVtvevins}--\refeq{eq:DVtvevinsdia}
to the explicit 
results~\refeq{eq:deltaV02gen}--\refeq{eq:deltaV02fer}.
Due to~\refeq{eq:eq:deltaV02bos}, $\tvev$ insertions 
apply only to fermion propagators, and 
matching to~\refeq{eq:deltaV02fer}
at first order in $\vev$ we get
\bea
\label{eq:v1ct}
G^{\ubk}_{ff\tvev}(\bar q)
&\;=\;&
\vcenter{\hbox{\raisebox{2pt}{\includegraphics[width=0.11\textwidth]{fermiondtVI}} }}
\;=\;
\ri  m_f \, \f{\tilde q^2}{\bar q^4}\,.  
\eea
Combining this $\tvev$ insertion with a standard $\vev$ insertions results
into
\bea
\label{eq:ffvtvterm}
G^{\ubk}_{ff\vev\tvev}(\bar q)
&\;=\;&
\vcenter{\hbox{\raisebox{2pt}{\includegraphics[width=0.11\textwidth]{fermiondtVIV}} }} 
\,+\,  
\vcenter{\hbox{\raisebox{2pt}{\includegraphics[width=0.11\textwidth]{fermionVdtVI}} }} 
\;=\;
2 \ri m_f^2\, \f{\slashed q \,  \tilde q^2 }{\bar q^6}\,. 
\eea
Finally, matching all terms of order $\vev^2$ yields\footnote{Note that, for an efficient implementation, 
the contribution~\refeq{eq:ffvtvterm} can be absorbed  
into the counterterm~\refeq{eq:v2cta}, 
while vetoing diagrams with 
a mixed $\vev\tvev$ insertions into the same 
propagator.}
\bea
\label{eq:v2cta}
G^{\ubk}_{aa\tvev\tvev}(\bar q)
&\;=\;&
\vcenter{\hbox{\raisebox{2pt}{\includegraphics[width=0.11\textwidth]{fermiondtVII}} }} 
\;=\;
\ri m_f^2\, \f{\slashed q \,  \tilde q^2 }{\bar q^6}  
-G^{\ubk}_{ff\vev\tvev}(\bar q)
\,=\,
{}- \ri m_f^2\, \f{\slashed q \,  \tilde q^2 }{\bar q^6} 
\,.
\eea

Since the correction~\refeq{eq:commutator} arises only 
from fermion-loop propagators,
at the level of multi-loop amplitudes 
it can be implemented by applying the $\tvev$ insertions
\refeq{eq:DVtvevinsdia} to all 
relevant propagators and restricting 
the number of $\vev$ and $\tvev$ insertions
according the desired order in $\vev$.
Thus the complete mass expansion~\refeq{eq:mexp4Db}
in $\numdim=4$ can be obtained 
from the full set of diagrams 
with $\vev$ and $\tvev$ insertions 
including up to a total number 
$X$ of insertions.
More explicitly, for a generic $l$-loop amplitude
in $\numdim=4$ we have
\bea 
\label{eq:VDVexpD4}
\bfm_{[0,X]}\, \amp{l}{\Gamma}{\bk}{} 
& = &  \sum_{k=0}^{X}    \sum_{j=0}^{k} \amp{l}{\Gamma \VevInsertJ
\tVevInsertKJ}{\ubk}{}\,,   
\eea
where the summand on the rhs represent terms of
order $\vev^k$ with $j$ vev insertions and 
$k-j$ auxiliary $\tvev$ insertions, which are
controlled, respectively, by the Feynman rules
\refeq{eq:aavevvertices}
and
\refeq{eq:vtildevertices}.

\subsection{Vev expansion of UV counterterms}
\label{se:vevexpdZ}

Let us now discuss vev expansions 
for UV counterterms. In this section we assume 
counterterms of $\msbar$ or $\ms$ kind, which 
include only UV poles, while generic 
renormalisation schemes are discussed in~\refse{se:schdep}.

According
to~\refeq{eq:ddimvevexp}--\refeq{eq:ddimvevexp2}, the
mass expansion of 
$l$-loop amplitudes in $\numdim=D$ 
can be expressed as a linear combination of symmetric-theory
amplitudes with vev insertions,
\bea 
\label{eq:vevexpvertD}
\bfm_{[0,X]}
\, \ampbar{l}{\Gamma}{\bk}{}
&=&
\sum_{k=0}^{X}  
\ampbar{l}{\Gamma \VevInsert }{\ubk}{}\,.
\eea
When $X$ is set equal to the superficial degree of divergence
of $\Gamma$, its local divergence is entirely captured by the
above vev expansion.
This implies that the $l$-loop $\msbar$ counterterms can be obtained 
through a corresponding vev expansion,
\bea
\label{eq:deltaZvevexp}
\deltaZ{l}{\Gamma}{\bk}{} &=& 
\sum_{k=0}^{X} \, \deltaZ{l}{\Gamma \VevInsert}{\ubk}{}\,,  
\eea
where
\bea
\label{eq:dZGammavevk}
\deltaZ{l}{\Gamma \VevInsert}{\ubk}{} 
&=&
\f{v^k}{k!} \, \deltaZ{l}{\Gamma \rHexp ^k}{\ubk}{} \, \Big|_{p_{\rHexp }=0} \,.
\eea

In preparation for the discussion of two-loop rational counterterms 
it is instructive to derive~\refeq{eq:deltaZvevexp} 
for $\deltaZ{2}{\Gamma}{}{}$
starting from its definition 
\bea
\label{eq:dZ2vevexpA}
 \deltaZ{2}{\Gamma}{\bk}{} &=&  
- \bfK \, 
\bigg( \ampbar{2}{\Gamma}{\bk}{} 
+ \sum  \limits_{\gamma \in \Omega(\Gamma)}  
\deltaZ{1}{\gamma}{\bk}{}  \cdot 
\ampbar{1}{\Gamma/\gamma}{\bk}{} 
\bigg)\,.
\eea
Here the second term between brackets 
subtracts all one-loop subdivergences, and 
the operator $\bfK$ extracts the 
remaining UV pole.\footnote{For a detailed definition of
the $\bfK$ operator  we refer to~\cite{Pozzorini:2020hkx}.}
As discussed at the end of~\refse{se:oneloopratterms},
the set $\Omega(\Gamma)$ includes all one-loop vertices 
$\gamma$ that can be inserted
into the one-loop vertex function $\ampbar{1}{\Gamma}{\bk}{}$.
In the following, the will refer to $\gamma$ as 
subvertex of the two-loop vertex function $\ampbar{2}{\Gamma}{\bk}{}$.
Before extracting the local divergence, 
the terms between brackets in~\refeq{eq:dZ2vevexpA} 
can be replaced by their 
mass expansions.
For the subtraction terms 
$\deltaZ{1}{\gamma}{\bk}{}  \cdot 
\ampbar{1}{\Gamma/\gamma}{\bk}{}$ this yields 
\bea
\label{eq:dZ2vevexpC1}
\bfm_{[0,X]}
\,\sum  \limits_{\gamma \in \Omega(\Gamma)}  
\deltaZ{1}{\gamma}{\bk}{}  \cdot \ampbar{1}{\Gamma/\gamma}{\bk}{}
&=&
\sum_{k=0}^X 
\sum_{j=0}^k
\sum  \limits_{\gamma \in \Omega(\Gamma)}\,  
\deltaZ{1}{\gamma\vev^j}{\ubk}{}  \cdot
\ampbar{1}{(\Gamma/\gamma)\,\vev^{k-j}}{\ubk}{}\,,
\eea
where terms of order $\vev^k$ 
arise from the interplay of 
$j$ vev insertions into the one-loop vertex $\gamma$ 
and 
$k-j$ 
vev insertions into its complement
$\Gamma/\gamma$. For later convenience, such combinations can be
recast into
\bea
\label{eq:dZ2vevexpC2}
\bfm_{[0,X]}\,
\sum  \limits_{\gamma \in \Omega(\Gamma)}  
\deltaZ{1}{\gamma}{\bk}{}  \cdot \ampbar{1}{\Gamma/\gamma}{\bk}{}
&=&
\sum_{k=0}^X 
\sum  \limits_{\gamma' \in \Omega(\Gamma \VevInsert)}\,  
\deltaZ{1}{\gamma'}{\ubk}{}   
\cdot \ampbar{1}{\Gamma \VevInsert /\gamma'}{\ubk}{}\,,
\eea
where $\Omega(\Gamma \vev^k)$ on the rhs
corresponds to the full set of 
subvertices $\gamma'= \gamma\,\vev^j$ 
with $\gamma\in \Omega(\Gamma)$ 
and $j\in[0,k]$ vev insertions.
With this representation one finds
\bea
\label{eq:vevsubdiv}
\deltaZ{2}{\Gamma}{\bk}{} &=&  
- \bfK \, 
\bfm_{[0,X]}
\, 
\bigg( \ampbar{2}{\Gamma}{\bk}{} 
+ \sum  \limits_{\gamma \in \Omega(\Gamma)}  
\deltaZ{1}{\gamma}{\bk}{}  \cdot 
\ampbar{1}{\Gamma/\gamma}{\bk}{} 
\bigg)
\nonumber\\[1mm]
&=&- 
\sum_{k=0}^X\,
\bfK \bigg(
\ampbar{2}{\Gamma \VevInsert}{\ubk}{}
-
\sum  \limits_{\gamma' \in \Omega(\Gamma \VevInsert)}\,  
\deltaZ{1}{\gamma'}{\ubk}{}   
\cdot \ampbar{1}{\Gamma \VevInsert /\gamma'}{\ubk}{}\bigg)\,
\,=\,
\sum_{k=0}^X
\deltaZ{2}{\Gamma \VevInsert}{\ubk}{}\,,
\eea
consistently with~\refeq{eq:deltaZvevexp}.

Let us now consider the 
mass expansion of the 
UV counterterms 
$\deltaZtilde{1}{\gamma}{}{}$, which contribute to~\refeq{eq:expr2form}.
As discussed in Sect.~4.2 of~\cite{Pozzorini:2020hkx},
such counterterms
are required 
for the subtraction of 
the UV divergences 
of the one-loop subdiagrams of 
two-loop diagrams in $\numdim=4$.
More precisely, 
\bea
\label{eq:dZtildevevexpa}
\deltaZ{1}{\gamma}{\bk}{} +
\deltaZtilde{1}{\gamma}{\bk}{}
&=&
-\bfK\, \amp{1}{\gamma}{\bk}{}\,.
\eea
Applying~\refeq{eq:mexp4Db} we have
\bea
\label{eq:dZtildevevexpa}
\deltaZ{1}{\gamma}{\bk}{} +
\deltaZtilde{1}{\gamma}{\bk}{}
&=&
-\bfK\,\bfm_{[0,X]}\, \amp{1}{\gamma}{\bk}{}
\,=\,
-\bfK\, \Big(
\bfV_{[0,X]}\,\amp{1}{\gamma}{\bk}{}
+
\Delta\bfV_{[0,X]}\,\amp{1}{\gamma}{\bk}{}
\Big)\,,
\eea
and the $\Delta\bfV_{[0,X]}$ contribution results into vanishing
$\tvev$ insertions, 
\bea
\label{eq:dZtildevevexpb}
-\bfK\,\Delta\bfV_{[0,X]}\,\amp{1}{\gamma}{\bk}{}
&=&
\sum_{k=1}^X\sum_{j=1}^k
\bfK\, 
\amp{1}{\gamma \vev^{k-j}\tvev^j}{\ubk}{}
\,=\,0\,.
\eea
This is due to the fact that the 
$\tvev$ insertions~\refeq{eq:v1ct} and \refeq{eq:v2cta}
involve $\tilde q^2$ factors that 
lead only to finite rational contributions of 
order $(D-4)/\eps$ at one-loop level.
For the $\bfV_{[0,X]}$ part of~\refeq{eq:dZtildevevexpa} we have
\bea
\label{eq:dZtildevevexpc}
-\bfK\,\bfV_{[0,X]}\,\amp{1}{\gamma}{\bk}{}
\,=\, -
\bfK\,\sum_{k=0}^X \amp{1}{\gamma \vev^{k}}{\ubk}{}  
\,=\,
\sum_{k=0}^X
\left(
\deltaZ{1}{\gamma\vev^k}{\ubk}{} +
\deltaZtilde{1}{\gamma\vev^k}{\ubk}{}
\right)
\,. 
\eea
Here the terms $\deltaZ{1}{\gamma\vev^k}{\bk}{}$ correspond
to the expansion~\refeq{eq:deltaZvevexp} 
of $\deltaZ{1}{\gamma}{\bk}{}$, which implies that 
\bea
\label{eq:dZtildevevexpd}
\deltaZtilde{1}{\gamma}{\bk}{}
&=&
\sum_{k=0}^X
\deltaZtilde{1}{\gamma\vev^k}{\ubk}{}\,.
\eea
Given that each $\vev$ insertion reduces the degree of 
divergence by one, and  $\delta \tilde Z_1 \neq 0$
only for quadratically divergent (sub)diagrams,
in renormalisable theories we simply have
\bea
\label{eq:dZtildevevexpe}
\deltaZtilde{1}{\gamma}{\bk}{}
&=&
\deltaZtilde{1}{\gamma}{\ubk}{}\,.
\eea

\subsection{Vev expansion of rational counterterms in the $\msbar$ scheme}
\label{se:vevexpdRms}

In the following, exploiting the 
mass expansions~\refeq{eq:ddimvevexp} and \refeq{eq:mexp4Db}
we 
show that the rational counterterms 
of a SB  theory can be  
related to corresponding counterterms of the
symmetric theory through general formulas of the
form 
\bea
\label{eq:dRvexpgen}
\delta \calR^{\bk}_{l,\Gamma}
&=&
\sum_{k=0}^{X}   \, \sum_{j=0}^{k} \, 
\ratamp{l}{\Gamma \vev^{k-j}\tvev^j}{\ubk}{}\,,
\eea
where the summands on the rhs involve similar combinations of 
$\vev$ and $\tvev$ insertions 
as in~\refeq{eq:VDVexpD4}.

At one loop, applying~\refeq{eq:ddimvevexp} and~\refeq{eq:mexp4Db}
to~\refeq{eq:expr1form} we have
\bea
\label{eq:dR1SB2YM}
 \ratamp{1}{\Gamma}{\bk}{} 
& = & \bfm_{[0,X]}\lb \ampbar{1}{\Gamma}{\bk}{} - \amp{1}{\Gamma}{\bk}{}  \rb 
\,=\,
\bfV_{[0,X]}\,\lb \ampbar{1}{\Gamma}{\bk}{} - \amp{1}{\Gamma}{\bk}{}  \rb 
-
\Delta\bfV_{[0,X]}\,\amp{1}{\Gamma}{\bk}{}\,.
\eea
The contribution of the vev-insertion operator $\bfV_{[0,X]}$ yields
\bea
\bfV_{[0,X]}\,\lb \ampbar{1}{\Gamma}{\bk}{} - \amp{1}{\Gamma}{\bk}{}  \rb 
&=&
\sum_{k=0}^{X}  \, \bigg(  \ampbar{1}{\Gamma  \vev^k}{\ubk}{}    
- \amp{1}{\Gamma  \vev^{k}}{\ubk}{} \bigg)
\; = \;
\sum_{k=0}^X \ratamp{1}{\Gamma \VevInsert}{\ubk}{}\,,
\eea
where the rhs consists of rational counterterms 
for vertices $\Gamma$ with extra (static) Higgs 
insertions in the symmetric theory, 
\bea
\label{eq:r1vevinsd}
\ratamp{1}{\Gamma \vev^k}{\ubk}{}
&=&
\frac{\vev^k}{k!}\,\ratamp{1}{\Gamma \rHexp ^k}{\ubk}{}\bigg|_{p_\rHexp =0}\,.
\eea
The additional $\Delta \bfV_{[0,X]}$ contribution 
to~\refeq{eq:dR1SB2YM} corresponds  to the 
terms with $j>0$ in~\refeq{eq:VDVexpD4}, 
\bea
\Delta\bfV_{[0,X]}\,
\amp{1}{\Gamma}{\bk}{}
\; = \;
\sum_{k=1}^{X} \sum_{j=1}^{k} \amp{1}{\Gamma  \vev^{k-j}\tvev^j}{\ubk}{}\,,
\eea
\ie to one-loop amplitudes in $\numdim=4$ 
with combinations of $\vev$ and $\tvev$ insertions.
For a systematic bookkeeping of such terms we define
\bea
\label{eq:r2vevinsone}
 \ratamp{1}{\Gamma \vev^{k-j}\tvev^j}{\ubk}{}  
& := &  
- \amp{1}{\Gamma \vev^{k-j}\tvev^j}{\ubk}{} \,.
\eea
In this way the complete one-loop counterterm can be written as
\bea
\label{eq:r1vevinsb}
\ratamp{1}{\Gamma}{\bk}{}
&=&
\sum_{k=0}^{X}
\sum_{j=0}^{k} \, \ratamp{1}{\Gamma \vev^{k-j}\tvev^j}{\ubk}{}\,.
\eea

At two loops, applying~\refeq{eq:ddimvevexp} and~\refeq{eq:mexp4Db} 
to~\refeq{eq:expr2form} results into
\bea
\label{eq:r2vevinsa}
\ratamp{2}{\Gamma}{}{}
&=&  
\bfV_{[0,X]}\,\Big[
\ampbar{2}{\Gamma}{}{}  
-
\amp{2}{\Gamma}{}{} + 
\sum  \limits_{\gamma} \deltaZ{1}{\gamma}{}{}
\cdot \ampbar{1}{\Gamma/\gamma}{}{} 
\,-\,
\sum  \limits_{\gamma} \lb \deltaZ{1}{\gamma}{}{}
\,+\,
\deltaZtilde{1}{\gamma}{}{} 
+ \ratamp{1}{\gamma}{}{} \rb \cdot
\amp{1}{\Gamma/\gamma}{}{}
\Big]
\nonumber\\[2mm]
&&{}-\Delta\bfV_{[0,X]}\,\Big[
\amp{2}{\Gamma}{}{}
+
\sum  \limits_{\gamma} \lb \deltaZ{1}{\gamma}{}{}
+\deltaZtilde{1}{\gamma}{}{} + \ratamp{1}{\gamma}{}{} \rb \cdot
\amp{1}{\Gamma/\gamma}{}{}
\Big]\,. \label{eq:dR2vevexp}
\eea
The $\bfV_{[0,X]}$ contribution embodies the 
pure $\vev$-insertion parts of the expansions of the various 
amplitudes and the associated subtraction terms. Combining the 
latter in a similar way as
in~\refeq{eq:dZ2vevexpC1}--\refeq{eq:dZ2vevexpC2}, 
for the first line of~\refeq{eq:dR2vevexp} we obtain 
\bea
\label{eq:r2vevinsb}
\bfV_{[0,X]}\,\Big[\dots\Big]
&=&
\sum_{k=0}^X
 \bigg[  
\ampbar{2}{\Gamma \VevInsert}{\ubk}{}
\,-\,
\amp{2}{\Gamma \VevInsert}{\ubk}{} 
\,+\, 
\sum  \limits_{\gamma' \in \Omega(\Gamma \VevInsert)}  
\deltaZ{1}{\gamma'}{\ubk}{} \cdot  
\ampbar{1}{\Gamma \VevInsert /\gamma'}{\ubk}{}
\nonumber\\ &&{}
-  \sum  \limits_{\gamma' \in \Omega(\Gamma \VevInsert)} 
 \lb    \deltaZ{1}{\gamma'}{\ubk}{} +\deltaZtilde{1}{\gamma'}{\ubk}{} + \ratamp{1}{\gamma'}{\ubk}{}   \rb 
  \cdot
 \amp{1}{\Gamma \VevInsert /\gamma'}{\ubk}{}  \bigg]
\,=\, \sum_{k=0}^X\ratamp{2}{\Gamma \vev^k}{\ubk}{}
\,, 
\nonumber\\
\eea
where the summands correspond 
to the rational counterterms for the vertex $\Gamma$ with 
$k$ vev insertions in the symmetric theory, 
\bea
\label{eq:r2vevinsc}
\ratamp{2}{\Gamma \vev^k}{\ubk}{}
&=&
\frac{\vev^k}{k!}\,\ratamp{2}{\Gamma \rHexp ^k}{\ubk}{}\bigg|_{p_\rHexp =0}\,.
\eea
The remaining $\Delta \bfV_{[0,X]}$ part of~\refeq{eq:r2vevinsa}
embodies all $\tvev$-insertion contributions to the expansion of
the $\numdim=4$ dimensional two-loop amplitude $\amp{2}{\Gamma}{}{}$ and the associated 
subtraction terms,
\bea
\label{eq:r2vevinsd}
-\Delta\bfV_{[0,X]}\,\Big[\dots\Big]
&=&
-\,\sum_{k=1}^X\sum_{j=1}^k
\Big[
\amp{2}{\Gamma \vev^{k-j}\tvev^j}{\ubk}{}  
\nonumber\\[2mm]&&{}+
\sum  \limits_{\gamma' \in \Omega(\Gamma \vev^{k-j}\tvev^j)} 
\lb    
\deltaZ{1}{\gamma'}{\ubk}{} 
+\deltaZtilde{1}{\gamma'}{\ubk}{} 
+\ratamp{1}{\gamma'}{\ubk}{}
\rb  
\cdot
\amp{1}{\Gamma  \vev^{k-j}\tvev^j / \gamma'}{\ubk}{} 
\Big]
\,.
\eea
Here the first term between square brackets corresponds to 
the two-loop amplitude in $\numdim=4$ with
$\vev^{k-j}\tvev^j$ insertions, while in the second term
such insertions are distributed between the 
one-loop subvertices $\gamma'\in\Omega(\Gamma \vev^{k-j}\tvev^j)$ and their one-loop complements.
Since the expansion of UV counterterms~\refeq{eq:deltaZvevexp} is
free from $\tvev$ insertions, 
\bea
\label{eq:r2vevinse}
\deltaZ{1}{\gamma\vev^n\tvev^m}{\ubk}{} &=& 0
\qquad\mbox{for}\quad m>0\,,
\eea
while for renormalisable
theories~\refeq{eq:dZtildevevexpd}--\refeq{eq:dZtildevevexpe} imply 
\bea
\label{eq:r2vevinse2}
\deltaZtilde{1}{\gamma\vev^n\tvev^m}{\ubk}{}
&=&0
\qquad\mbox{for}\quad \mbox{$n>0$\; or\; $m>0$}\,.
\eea
Taking this into account, we can recast the $\vev^{k-j}\,\tvev^j$
insertions on the rhs of~\refeq{eq:r2vevinsd} into
\bea
\label{eq:r2vevinsf}
\ratamp{2}{\Gamma \vev^{k-j}\tvev^j}{\ubk}{}
&:=&
{}-\amp{2}{\Gamma \vev^{k-j}\tvev^j}{\ubk}{}  
\,-\,
\sum  \limits_{\gamma' \in \Omega(\Gamma)}   
\deltaZtilde{1}{\gamma'}{\ubk}{} 
\cdot
\amp{1}{(\Gamma/ \gamma')\,  \vev^{k-j} \tvev^j }{\ubk}{} 
\nonumber \\[2mm]
&&{} 
\;-\;
\sum  \limits_{\gamma' \in \Omega(\Gamma \vev^{k-j})}   
\deltaZ{1}{\gamma'}{\ubk}{} 
\cdot
\amp{1}{(\Gamma  \vev^{k-j}/ \gamma')\, \tvev^j }{\ubk}{} 
\;-\;
\!\!\!\!\!\!\!\! \sum  \limits_{\gamma' \in \Omega(\Gamma \vev^{k-j}\tvev^j)}   \!\!\!\!\!\!\!\!
\ratamp{1}{\gamma'}{\ubk}{}
 \cdot
  \amp{1}{\Gamma  \vev^{k-j}\tvev^j / \gamma'}{\ubk}{} 
\,,\nonumber\\
\eea
and the full expansion of 
two-loop rational counterterms becomes
\bea
\label{eq:r2vevinsg}
\ratamp{2}{\Gamma}{\bk}{}
&=&
\sum_{k=0}^{X}
\sum_{j=0}^{k} \, \ratamp{2}{\Gamma \vev^{k-j}\tvev^j}{\ubk}{}\,.
\eea
Here all contributions with $j=0$ correspond to 
rational counterterms~\refeq{eq:r2vevinsc} of the symmetric theory
with static Higgs insertions,
while all other effects of symmetry breaking 
are accounted for by $\tvev$ insertions, which affect only 
the propagators of massive fermions.
In practice, for logarithmically divergent vertices 
we have
$\ratamp{l}{\Gamma}{\bk}{}=\ratamp{l}{\Gamma}{\ubk}{}$,
\ie rational terms are independent of symmetry breaking.
Linearly divergent vertices require 
extra contributions with a single $\vev$ or $\tvev$ insertion, 
and in the presence of quadratic divergences
also insertions of type $\vev^2$,  $\vev\tvev$ and $\tvev^2$
are needed.

Note that the vev-expansion 
identities~\refeq{eq:deltaZvevexp}, \refeq{eq:r1vevinsb} and \refeq{eq:r2vevinsg}  
are only applicable to the derivation of UV and rational counterterms,
while the other building blocks of the renormalised 
amplitudes \refeq{eq:masterformula1} and \refeq{eq:masterformula2}
require exact calculations in the SB phase.

\section{Renormalisation-scheme dependence}
\label{se:schdep}

In the previous section, the relation~\refeq{eq:dRvexpgen} 
between rational counterterms in the SB and symmetric phase 
has been proven assuming rigid invariance and
for renormalisation schemes of $\msbar$ or $\ms$ kind.
However, the applicability of~\refeq{eq:dRvexpgen} is not restricted to 
such special cases.
As demonstrated in \refses{se:vevexpdRY}{se:abc}
the relation~\refeq{eq:dRvexpgen}
is valid in a very wide 
class of renormalisation schemes, 
and can be easily adapted 
to account for 
mixing effects.
Moreover, in~\refse{se:thooftgfix} we present an extension 
of~\refeq{eq:dRvexpgen} that is applicable 
also in the presence of gauge-fixing terms that violate 
rigid invariance, such as in the case of the widely used
't~Hooft gauge.

\subsection{Vev expansion of rational counterterms in a generic
scheme}
\label{se:vevexpdRY}

As pointed out in~\refse {se:irredtwoloop}, the renormalisation
identities~\refeq{eq:masterformula1} and \refeq{eq:masterformula2}
are valid in any renormalisation scheme.
Moreover, the calculations of $\delta\calR_{l,\Gamma}$ counterterms 
through~\refeq{eq:expr1form}--\refeq{eq:expr2form} 
can be carried out once and for all in a generic renormalisation
scheme, where the finite parts of UV renormalisation constants 
are treated as free parameters~\cite{Lang:2020nnl}.

Concerning the relation~\refeq{eq:dRvexpgen} between $\delta\calR_{l,\Gamma}$ counterterms 
in the SB and symmetric phases, the
only property of the renormalisation scheme that 
was assumed in the proof of~\refse{se:vevexpdRms} 
is that the UV counterterms should fulfil the mass-expansion
identity~\refeq{eq:deltaZvevexp}. 
This requirement is relevant for the two-loop 
rational counterterms~\refeq{eq:expr2form}, whose scheme
dependence arises only from the 
one-loop UV counterterms $\delta Z_{1,\gamma}$. The one-loop 
rational counterterms~\refeq{eq:expr1form}
are instead scheme-independent.\footnote{More
precisely, one-loop rational counterterms include trivial 
factors of order $\eps$ associated with the change of renormalisation 
scale~\cite{Pozzorini:2020hkx,Lang:2020nnl}, which
are relevant when $\delta\calR_{1,\gamma}$ 
factors are applied in the two-loop formulas~\refeq{eq:masterformula2}
and~\refeq{eq:expr2form}. However, this form of scheme dependence 
is consistent with the proof of~\refeq{eq:dRvexpgen}.} 
Thus the validity of~$\refeq{eq:dRvexpgen}$ at one loop is trivially
guaranteed for 
any renormalisation scheme. 
For these reasons, the presented proof of the relation~\refeq{eq:dRvexpgen} 
up to two loops is valid in any scheme where the one-loop counterterms in the 
SB theory are 
related to the ones of the symmetric phase via
\bea
\label{eq:dZ1vevexp}
\deltaZ{1}{\Gamma}{\bk}{} &=& 
\sum_{k=0}^{X} \, \deltaZ{1}{\Gamma \vev^k}{\ubk}{}\,.
\eea
In addition, in the derivation of~\refeq{eq:dRvexpgen} we have assumed
that $l$-loop amplitudes in $\numdim=D$ and $\numdim=4$ 
obey the mass-expansion identities
\bea 
\label{eq:massexpforamps}
\bfm_{[0,X]}\, \ampbar{l}{\Gamma}{\bk}{} 
& = &  \sum_{k=0}^{X}   \ampbar{l}{\Gamma \vev^k}{\ubk}{}
\,,\\
\label{eq:massexpforampsb}
\bfm_{[0,X]}\, \amp{l}{\Gamma}{\bk}{} 
&=&
\sum_{k=0}^{X}    \sum_{j=0}^{k} \amp{l}{\Gamma \vev^{k-j}
\tvev^j}{\ubk}{}\,,
\eea
and that the mass expansion of subdivergences
fulfils~\refeq{eq:dZ2vevexpC1}--\refeq{eq:dZ2vevexpC2}.

As discussed in the following, the relation~\refeq{eq:dZ1vevexp}
is fulfilled in
any scheme 
where the renormalisation of the SB phase is equivalent to 
a renormalisation of the 
symmetric phase,\footnote{For the SM such a renormalisation scheme
was first proposed in \cite{Bohm:1986rj} and worked out to one-loop order.
For a general discussion of the renormalisation of the SM at two loops 
in the SB phase see e.g.~\cite{Actis:2006ra,Actis:2006rb}.
} 
where the independent renormalisation constants 
can assume arbitrary finite parts, provided that the underlying symmetry is
preserved, while the vev should be
renormalised in the same way as the Higgs
field.
Since we are only interested in the extension of~\refeq{eq:dZ1vevexp} 
to generic renormalisation schemes, for simplicity 
we will restrict ourselves to one loop.

Let us start with the definition of a renormalisation scheme in the
symmetric phase.
For the renormalisation of fields and parameters we use a 
similar notation as in~\cite{Lang:2020nnl}.
The symmetric phase can be described by a certain set of independent
parameters
\bea
\label{eq:paramubk}
\param^{\ubk}_b
\in \{\dots, g^{\ubk}_k, \dots, \xi_j\,, \dots\}
\cup\{\mu\}\,,
\eea
which consist of dimensionless couplings $g^{\ubk}_k$,
gauge-fixing parameters $\xi_j$,
and a mass scale $\mu$ that enters the scalar potential. 
At one loop 
their renormalisation reads
\bea
\label{eq:PARRCsubk} 
\param^{\ubk}_{b,0} \,=\,
\param^{\ubk}_{b}
+
\delta\param^{\ubk}_{b}\,,\qquad
\delta \param^{\ubk}_{b}\,=\,
\delta \calZ^{\ubk}_{1,\param_b}\,
\param^{\ubk}_{b}
\,.
\eea
As usual, symbols with and without a zero correspond, respectively, 
to bare quantities and their renormalised counterparts,
and $\delta \theta_b^{\ubk}=
\delta \calZ^{\ubk}_{1,\param_b}\theta_b^{\ubk}$ 
can depend on 
multiple couplings or gauge parameters $\theta_{b'}^{\ubk}$.
To describe the effect of parameter renormalisation at one loop we 
introduce the differential operator
\bea
\label{eq:D1thetaubk}
D^{\ubk}_{1,\theta}
&=& 
\sum_b\delta \param^{\ubk}_b
\;\frac{\partial}{\partial\param^{\ubk}_b}
\,=\,
\sum_{b} \delta\calZ^{\ubk}_{1,\param_b}\;\lndev{\param^{\ubk}_b}\,.
\eea
Similarly as in~\cite{Lang:2020nnl}, we assume that 
the
gauge-fixing parameters are renormalised in such a way that 
the gauge-fixing part of the Lagrangian remains effectively unrenormalised,
both in the symmetric and in the SB phase.

For the fields of the symmetric theory we use the symbols
$\widetilde\varphi_i$, which represent generic multiplets 
of scalar, fermion, gauge-boson or ghost fields. 
The tilde is introduced in order to distinguish 
gauge eigenstates of the
symmetric theory from mass eigenstates 
in the SB phase.
The components $\fieldcomp{(\widetilde\varphi_i)}{\beta}$
of a multiplet represent individual fields. For a systematic 
description of mixing effects, all fields that can possibly mix together are 
assigned to the same (generalised) multiplet. For instance, in the SM the 
SU(2)$\times$U(1) gauge bosons are combined in a single multiplet
$\widetilde\varphi_i=
\tilde V$ with components
$\fieldcomp{(\tilde V^\mu)}{\beta}
=
W_1^{\mu}, W_2^\mu, W_3^\mu, B^\mu$.
The renormalisation of a generic multiplet $\widetilde \varphi_i$ 
at one loop 
reads
\bea
\label{eq:FRCsubk}
\widetilde \varphi_{i,0} &=& \Big( 1+\frac{1}{2}
\delta\calZ^{\ubk}_{1,\widetilde\varphi_i}
\Big)\,\widetilde \varphi_{i}\,,
\eea
{where $\delta\calZ^{\ubk}_{1,\widetilde\varphi_i}$ is a diagonal
matrix.

For $l$-loop vertex functions in the symmetric phase we use the notation
\bea
\label{eq:ampnotationubk}
\ampbar{l}{\widetilde \Gamma}{\ubk}{}
&\equiv&
\ampbar{l}{
\widetilde \varphi_1\cdots
\widetilde \varphi_n
}{\ubk}{}\,,
\eea
where 
$\widetilde\varphi_{1}, \dots, \widetilde\varphi_{n}$
are the fields associated with the (incoming) external lines 
of the vertex $\widetilde\Gamma$.
The one-loop UV counterterm
for the generic vertex $\widetilde \Gamma$ 
can be generated via renormalisation of the
corresponding tree-level amplitude as
\bea
\label{eq:D1ubk}
\deltaZ{1}{\widetilde \Gamma}{\ubk}{}
&=& 
\ampbar{0}{\widetilde \Gamma}{\ubk}{}
\,\bigg(\sum_{i=1}^n
\frac{1}{2}
\delta\calZ^{\ubk}_{1,\widetilde \varphi_i}
\bigg)
+
D^{\ubk}_{1,\theta}\,
\ampbar{0}{\widetilde \Gamma}{\ubk}{}\,,
\eea
where 
the usual field-renormalisation 
factors are written on the rhs 
for later convenience, 
and $D^{\ubk}_{1,\theta}\,$ is defined in~\refeq{eq:D1thetaubk}.

Let us now connect the renormalisation of the symmetric theory with 
the renormalisation of its SB  counterpart. The SB
phase is
described by a set of independent parameters
consisting of dimensionless couplings $g_k$, 
gauge-fixing parameters $\xi_j$,
and physical masses $m_l$,
\bea
\label{eq:parambk}
\param^{}_a \in \{\dots, g_k\,,\dots, \xi_j\,, \dots
\}\cup\{\dots, m_l,\dots\}\,.
\eea
At one loop the corresponding renormalisation identities read
\bea
\label{eq:PARRCsbk}
\param^{}_{a,0} \,=\,
\param^{}_{a}+\delta\param^{}_{a},\qquad
\delta\param^{}_{a}\,=\,
\delta \calZ^{}_{1,\param_a}%
\,\param^{}_{a}\,,
\eea
and the renormalisation of all parameters can be encoded in the operator
\bea
\label{eq:D1thetabk}
D^{\bk}_{1,\theta}
&=& 
\sum_a\delta \param^{\bk}_a
\;\frac{\partial}{\partial\param^{\bk}_a}
\,=\,
\sum_{a} \delta\calZ^{\bk}_{1,\param_a}\;\lndev{\param^{\bk}_a}\,.
\eea

The two sets of parameters \refeq{eq:paramubk}, \refeq{eq:parambk}
and their renormalisation are connected 
through
 tree-level relations of the following form, 
which are assumed to hold both for renormalised and bare parameters,
\bea
\label{eq:treerelations}
\param_a\,=\, f_a(\{\param_b^\ubk\}, v)\,,
\qquad
\param_{a,0} = f_a(\{\param_{b,0}^\ubk\}, \vev_0)\,.
\eea
In these identities the bare and renormalised vev parameters,
\bea
\label{eq:vevren}
\vev_{0} \,=\,
\vev +\delta\vev,\qquad
\delta\vev\,=\,
\delta \calZ^{}_{1,\vev}
\,\vev\,,
\eea
play a special role.
The renormalised vev, $v=\phi-H$, 
is fixed by the requirement that the minimum of the
tree-level potential is located at $H=0$.
Thus $\vev$ is connected to the renormalised parameters of the symmetric
theory,
\bea
\label{eq:vevparamrel}
\vev &=& f_v(\{\param_{b}^\ubk\})\,.
\eea 
However, the position of the minimum of the potential is not protected by any
symmetry. 
Therefore the vev can be renormalised in 
a way that $\vev_0 \neq f_v(\{\param_{b,0}^\ubk\})$.
As discussed below, 
in order to guarantee the validity of the vev-expansion
formula~\refeq{eq:deltaZvevexp}, the vev needs to be renormalised in the same way as the
Higgs field 
in the symmetric phase,
\ie by setting
\bea
\label{eq:vevphiren}
\delta \calZ_{1,\vev} = 
\frac{1}{2}\delta \calZ^{\ubk}_{1,\phi}\,.
\eea
The relations~\refeq{eq:treerelations} and \refeq{eq:vevren} 
connect the one-loop renormalisation of the SB 
and unbroken theories  
through 
\bea
\delta\param_{a}
&=&
\sum_{b} 
\delta\param^{\ubk}_b\,
\frac{\partial
\param_{a}}{\partial\param^{\ubk}_b}
+
\delta\vev\,
\frac{\partial
\param_{a}}{\partial\vev}\,,
\eea
or, equivalently,
\bea
\delta\calZ^{\bk}_{1,\param_a}
\,=\,
\param^{-1}_a
\Bigg[\sum_{b} \delta\calZ^{\ubk}_{1,\param_b}\;
\param^{\ubk}_b\;
\frac{\partial \param_a}{\partial\param^{\ubk}_b}
\,+\,
\delta\calZ^{}_{1,\vev}
\,\vev\,
\frac{\partial \param_a}{\partial\vev}
\Bigg]\,,
\eea
which implies
\bea
\label{eq:D1paridentity}
D_{1,\theta}^{\bk}
&=&
D_{1,\theta}^{\ubk}+
\delta\calZ^{}_{1,\vev}
\,\vev\,
\frac{\partial}{\partial\vev}\,.
\eea

Let us now turn to the renormalisation of fields 
in the SB phase. 
In this case,  for multiplets of 
fields we use the symbols $\varphi_i$.
Their components 
$\fieldcomp{(\varphi_i)}{\alpha}$
correspond to individual mass-eigenstate fields.
In general, they are 
related to the 
gauge eigenstates 
$\fieldcomp{(\widetilde\varphi_i)}{\beta}$
through mixing transformations
as detailed below.
As a consequence, the 
generic renormalisation of mass-eigenstate fields,
\bea
\label{eq:FRCsbk}
\varphi_{i,0} &=& \Big( 1+\frac{1}{2} \delta\calZ^{}_{1,\varphi_i} \Big)\,\varphi_{i}\,,
\qquad
\eea
involves a renormalisation matrix
$\delta\calZ^{}_{1,\varphi_i}$.
More explicitly,
\bea
\label{eq:FRCsbk}
\fieldcomp{(\varphi_{i,0})}{\alpha}
&=& 
\Big( 
\fieldcomp{\delta}{\,\alpha\alpha'}+
\frac{1}{2} 
\fieldcomp{\big(\delta\calZ^{}_{1,\varphi_i}\big)}{\alpha \alpha'}
\Big)\,
\fieldcomp{(\varphi_{i})}{\alpha'}\,,
\eea
where $\delta\calZ^{}_{1,\varphi_i}$ can involve off-diagonal elements.

In the SB phase, an amputated vertex function with 
external fields $\varphi_1,\dots,\varphi_n$ is written as
\bea
\label{eq:ampnotationbk}
\ampbar{l}{\Gamma}{\bk}{}
&\equiv&
\ampbar{l}{
\varphi_1\cdots
\varphi_n
}{\bk}{}\,,
\eea
where the indices that characterise the individual components 
$\fieldcomp{(\varphi_i)}{\alpha}$
of $\varphi_{i}$ are implicitly understood.
Similarly as in~\refeq{eq:D1ubk}, for one-loop UV counterterms 
we have
\bea
\label{eq:D1bk}
\deltaZ{1}{\Gamma}{\bk}{}
&=& 
\ampbar{0}{\Gamma}{\bk}{}
\,\bigg(\sum_{i=1}^n
\frac{1}{2}
\delta\calZ^{\bk}_{1,\varphi_i}
\bigg)
+
D^{\bk}_{1,\theta}\,
\ampbar{0}{\Gamma}{\bk}{}\,,
\eea
where the product of tree amplitude and field-renormalisation matrices 
should be understood as 
\bea
\ampbar{0}{\varphi_{1}\dots\varphi_{i}\dots\varphi_{n}}{\bk}{}\;
\delta\calZ^{\bk}_{1,\varphi_i}&\equiv&
\fieldcomp{\Big(\ampbar{0}{\varphi_{1}\dots\varphi_{i}\dots\varphi_{n}}{\bk}{}\Big)}{\alpha_1\dots
\alpha'_i\,\dots\alpha_n}
\;\fieldcomp{\big(\delta\calZ^{\bk}_{1,\varphi_i}\big)}{{\alpha'_i}
{\alpha_i}}\,.
\eea

Let us now consider the relation between the counterterms~\refeq{eq:D1bk}
of the SB theory and their counterparts~\refeq{eq:D1ubk}
in the symmetric phase. To this end we need the 
mixing transformations that connect the corresponding fields,
\bea
\label{eq:mixing1}
\varphi_i &=& 
\calU_{i}\,
\widetilde \varphi_i\qquad\mbox{or}\qquad
\fieldcomp{(\varphi_{i})}{\alpha} \,=\,
\fieldcomp{\big(\calU_{i}\big)}{\alpha\beta}
\fieldcomp{(\widetilde\varphi_{i})}{\beta}\,.
\eea
The mixing angles that enter $\calU_{i}$ 
are fixed such as to diagonalise the related 
tree-level mass matrix. In practice  
they depend on 
the symmetry-breaking pattern
and the dimensionless couplings of the symmetric theory, \ie
\bea
\label{eq:mixing2}
\calU_{i}&\equiv&
\calU_{i}(\{\theta^{\ubk}_b\})\,.
\eea
In order to ensure a direct correspondence between
the renormalisation of the SB and symmetric phases
we require that~\refeq{eq:mixing1}--\refeq{eq:mixing2}
hold also for bare quantities,\footnote{In general this 
requirement does not allow for an on-shell renormalisation of 
all mass-eigenstate fields. However this limitation can be 
circumvented by means of additional LSZ factors as explained in 
\refse{se:abc}.} 
\ie
\bea
\label{eq:mixing3}
\varphi_{i,0} &=& 
\calU_{i,0}\,
\widetilde \varphi_{i,0}
\qquad\mbox{with}\qquad
\calU_{i,0}\,=\,
\calU_{i}(\{\theta^{\ubk}_{0,b}\})\,=\,
\calU_{i}(\{\theta^{\ubk}_{b}+\delta\theta^{\ubk}_{b}\})\,.
\eea
For the one-loop renormalisation of the mixing matrix this implies
\bea
\label{eq:mixing4}
\calU_{i,0}\,=\,\calU_{i}+\delta\, \calU_{i}
\qquad\mbox{with}\qquad
\delta\,\calU_{i}
\,=\,
\sum_{b} 
\delta\param^{\ubk}_b\,
\frac{\partial\,\calU_{i}}{\partial\param^{\ubk}_b}
\,=\,
D^{\ubk}_{1,\theta}\,\calU_{i}\,.
\eea
Combining~\refeq{eq:mixing1} and~\refeq{eq:mixing3} it is easy to show that
the field renormalisation constants in the SB and symmetric phases
are connected by
\bea
\label{eq:mixing5}
\frac{1}{2}\delta\calZ_{1,\varphi_i}^{\bk}
&=&\Big(\delta\,\calU_{i}+
\frac{1}{2}\,\calU_{i}\,
\delta\calZ_{1,\widetilde\varphi_i}^{\ubk}
\Big)\,\calU^{-1}_{i}\,.
\eea
For later convenience, using
\bea
\delta\,\calU_{i}\,\calU^{-1}_{i}
\,=\,
-\calU_{i}\,\delta\,\calU^{-1}_{i}
\,=\,
-\calU_{i}\,D^{\ubk}_{1,\theta}\,\calU^{-1}_{i}\,,
\eea
the above relation can be turned into
\bea
\label{eq:mixing6}
\frac{1}{2}\,\calU^{-1}_{i}\,
\delta\calZ_{1,\varphi_i}^{\bk}
&=&\Big[
\frac{1}{2}\,%
\delta\calZ_{1,\widetilde\varphi_i}^{\ubk}
-\Big(D^{\ubk}_{1,\theta}\,\calU^{-1}_{i}\Big)
\,\calU_i
\Big]\,\calU^{-1}_{i}
\,.
\eea

We are now ready to derive the relation~\refeq{eq:dZ1vevexp}
between the UV counterterms~\refeq{eq:D1bk} in the SB phase
and their counterparts~\refeq{eq:D1ubk} in the symmetric phase. To this end we exploit the fact that in the SB
theory $n$-point tree vertices, \ie 1PI amputated tree amplitudes
that connect $n$ lines,  
obey the simple mixing transformation
\bea
\label{eq:genren1}
\ampbar{0}{\Gamma}{}{}&\equiv&
\ampbar{0}{\varphi_1\dots \varphi_n}{}{}
\,=\,
\ampbar{0}{\widetilde\varphi_1 \dots \widetilde\varphi_n}{}{}
\,\prod_{i=1}^n\,\calU^{-1}_i
\,\equiv\,
\ampbar{0}{\widetilde\Gamma}{}{}\,\prod_{i=1}^n\,\calU^{-1}_i\,.
\eea
Here the vertex $\ampbar{0}{\widetilde\Gamma}{}{}\equiv
\ampbar{0}{\widetilde\varphi_1 \dots \widetilde\varphi_n}{}{}
$ 
connects 
multiplets of gauge-eigenstate  
fields $\widetilde\varphi_i$ 
according to the interaction Lagrangian of the 
SB theory. 
It is related to the tree vertices of the symmetric phase by means of the 
vev expansion
\bea
\label{eq:genren2}
\ampbar{0}{\widetilde\Gamma}{}{}&=&
\sum_{k=0}^X \ampbar{0}{\widetilde\Gamma\vev^k}{\ubk}{}\,,
\eea
where $X$ is the mass dimension of the vertex at hand.
Applying these identities on the rhs of~\refeq{eq:D1bk} yields
\bea
\label{eq:genren3}
\deltaZ{1}{\Gamma}{\bk}{}
&=& 
\sum_{k=0}^X 
\bigg[
\ampbar{0}{\widetilde\Gamma\vev^k}{\ubk}{}\,
\prod_{i'=1}^n\,\calU^{-1}_{i'}\,
\,\bigg(\sum_{i=1}^n
\frac{1}{2}
\delta\calZ^{\bk}_{1,\varphi_i}
\bigg)
+
\Big(D^{\ubk}_{1,\theta}+
\delta\calZ^{}_{1,\vev}
\,\vev\,
\frac{\partial}{\partial\vev}
\Big)\,
\Big(\ampbar{0}{\widetilde\Gamma\vev^k}{\ubk}{}\,
\prod_{i'=1}^n\,\calU^{-1}_{i'}\,
\Big)\bigg]
\nonumber\\
&=&
\sum_{k=0}^X 
\bigg[
\ampbar{0}{\widetilde\Gamma\vev^k}{\ubk}{}\,
\Big\{
k\,\delta \calZ_{1,\vev}+
\sum_{i=1}^n
\Big[\frac{1}{2}\,%
\delta\calZ_{1,\widetilde\varphi_i}^{\ubk}
-\Big(D^{\ubk}_{1,\theta}\,\calU^{-1}_{i}\Big)
\,\calU_i
\Big]\Big\}
\prod_{i'=1}^n\,\calU^{-1}_{i'}\,
\nonumber\\
&&{}+
D^{\ubk}_{1,\theta}\,
\Big(\ampbar{0}{\widetilde\Gamma\vev^k}{\ubk}{}\,
\prod_{i'=1}^n\,\calU^{-1}_{i'}\,
\Big)\bigg]\,,
\eea
where in the second step we have used~\refeq{eq:mixing6}
and $\vev\partial_\vev \ampbar{0}{\widetilde\Gamma\vev^k}{\ubk}{} = 
k\,\ampbar{0}{\widetilde\Gamma\vev^k}{\ubk}{}$.
Finally, the term $D^{\ubk}_{1,\theta}\,\calU^{-1}_{i}$ 
cancels against an opposite contribution from the last line, and 
we obtain
\bea
\label{eq:genren4}
\deltaZ{1}{\Gamma}{\bk}{}
&=& 
\sum_{k=0}^X 
\bigg[
\ampbar{0}{\widetilde\Gamma\vev^k}{\ubk}{}\,
\Big(
k\, \delta \calZ_\vev
\,+\sum_{i=1}^n
\frac{1}{2}\,%
\delta\calZ_{1,\widetilde\varphi_i}^{\ubk}
\Big)
+
\Big(
D^{\ubk}_{1,\theta}\,
\ampbar{0}{\widetilde\Gamma\vev^k}{\ubk}{}
\Big)\bigg]
\prod_{i'=1}^n\,\calU^{-1}_{i'}\,
\,.
\eea
Based on~\refeq{eq:dZGammavevk}, 
\refeq{eq:D1ubk} and~\refeq{eq:vevphiren},
the expressions between squared brackets 
can be identified with the vev-insertion counterterms 
$\deltaZ{1}{\widetilde\Gamma\vev^k}{\ubk}{}$ 
in the symmetric theory, \ie
\bea
\label{eq:genren5}
\deltaZ{1}{\Gamma}{\bk}{}
&=& 
\bigg(\sum_{k=0}^X 
\,
\deltaZ{1}{\widetilde\Gamma\vev^k}{\ubk}{}
\bigg)\prod_{i=1}^n\,\calU^{-1}_i\,
\,.
\eea
Note in particular that the vev-renormalisation prescription~\refeq{eq:vevphiren}
guarantees that the term $\delta \calZ_\vev$ in~\refeq{eq:genren4} 
supplies the required field-renormalisation factors 
$\frac{1}{2} \delta \calZ_\phi^{\ubk}$
for each of the external Higgs lines associated with vev insertions
in $\deltaZ{1}{\widetilde\Gamma\vev^k}{\ubk}{}$.
This ensures the
correct cancellation of the UV 
divergences of the related vev-insertion amplitudes
$\ampbar{1}{\widetilde\Gamma\vev^k}{\ubk}{}$. 
The above equation is equivalent to~\refeq{eq:dZ1vevexp} with
\bea
\label{eq:genren5b}
\qquad
\deltaZ{l}{\Gamma\vev^k}{\ubk}{}\,=\,
\deltaZ{l}{\widetilde\Gamma\vev^k}{\ubk}{}\,\prod_{i=1}^n\,\calU^{-1}_i\,,
\eea
where the mixing matrices act only on the $n$ external lines
corresponding to the vertex $\Gamma$ and not on the additional 
vev insertions.
Within the symmetric theory, where all propagators are massless, 
the overall effect of the mixing transformation~\refeq{eq:mixing1} 
on internal vertices and propagators cancels.
Thus loop amplitudes transform as
\bea
\label{eq:genren5c}
\ampbar{l}{\Gamma}{\ubk}{}&=&
\ampbar{l}{\widetilde\Gamma}{\ubk}{}\,\prod_{i=1}^n\,\calU^{-1}_i\,.\qquad
\eea
This holds both in $\numdim=D$ and $\numdim=4$ dimensions, 
as well as in the presence of vev insertions. Moreover, in the presence of
mixing the
mass expansions~\refeq{eq:massexpforamps}--\refeq{eq:massexpforampsb} 
remain valid with
\bea
\label{eq:genren5d}
\ampbar{l}{\Gamma\vev^k}{\ubk}{}&=&
\ampbar{l}{\widetilde\Gamma\vev^k}{\ubk}{}\,\prod_{i=1}^n\,\calU^{-1}_i\,,\qquad
\amp{l}{\Gamma\vev^k\tvev^j}{\ubk}{}\,=\,
\amp{l}{\widetilde\Gamma\vev^k\tvev^j}{\ubk}{}\,\prod_{i=1}^n\,\calU^{-1}_i\,.
\eea
Also here the mixing matrices apply only to the fields associated with $\Gamma$.
Concerning $\tvev$ insertions, since their definition in~\refse{eq:tvevins}
is based on the propagators of mass eigenstates in the SB phase,
the related Feynman rules need to be adapted to 
gauge eigenstates in order to compute
$\amp{l}{\widetilde\Gamma\vev^k\tvev^j}{\ubk}{}$.

The above analysis demonstrates that the 
identities~\refeq{eq:dZ1vevexp}--\refeq{eq:massexpforampsb}
are satisfied. 
This holds also for~\refeq{eq:dZ2vevexpC1}--\refeq{eq:dZ2vevexpC2}.
Therefore all prerequisites 
for the proof in~\refse{se:vevexpdRms} are fulfilled, and 
rational counterterms 
can be obtained through the vev-expansion
formula~\refeq{eq:dRvexpgen}.
To this end, the 
required building blocks $\ratamp{l}{\Gamma\vev^k\tvev^j}{\ubk}{}$
can be constructed using the explicit formulas
in~\refse{se:vevexpdRms} together
with~\refeq{eq:genren5c}--~\refeq{eq:genren5d}.
Alternatively, the various terms of the vev expansion can be 
derived in the gauge-eigenstate basis and converted to the 
mass-eigenstate basis via
\bea
\label{eq:genren5e}
\ratamp{l}{\Gamma\vev^k\tvev^j}{\ubk}{}\,=\,
\ratamp{l}{\widetilde\Gamma\vev^k\tvev^j}{\ubk}{}\,\prod_{i=1}^n\,\calU^{-1}_i\,.
\eea

\subsection{Mixing effects, tadpoles and vev renormalisation}
\label{se:abc}

As proven in the previous section,
rational counterterms 
can be obtained through the vev expansion~\refeq{eq:dRvexpgen}
in any renormalisation scheme that fulfils the following conditions:

\begin{itemize}
\item[(a)]  The field-renormalisation constants in the SB theory are 
connected to the ones of the symmetric theory via mixing as described 
in~\refeq{eq:mixing1}--\refeq{eq:mixing5}.

\item[(b)] 
The renormalisation procedure preserves all tree-level relations between parameters in the SB and 
symmetric theory with the 
only
exception of the relation between the
vev and the parameters of the unbroken Higgs sector.
See \refeq{eq:treerelations}--\refeq{eq:vevparamrel}.

\item[(c)] The vev is renormalised in the same way as the Higgs field. See \refeq{eq:vevren}--\refeq{eq:vevphiren}.

\end{itemize}
In practical calculations of scattering amplitudes, the renormalisation 
scheme is specified through renormalisation conditions in the SB phase.
The corresponding renormalisation constants of the symmetric phase,
which enter the vev expansion of 
rational counterterms~\refeq{eq:dRvexpgen},
need to be adapted to the ones of the SB phase
consistently with (a)--(c).
However, the above conditions are too
restrictive to satisfy the needs of non-trivial renormalisation schemes
for SB theories.
Nonetheless, 
such limitations can be easily circumvented as discussed in the following.

In the presence of mixing, such as the mixing 
between photons and $Z$ bosons in the SM,
field-renormalisation constants 
$\calZ_{\varphi_i}$ that satisfy (a)
do not allow one to achieve an on-shell field
renormalisation such that all
mixing effects between on-shell states cancel.
In general, this requires an
additional finite renormalisation, 
\bea
\label{eq:LSZren1}
\varphi_i &=&
\left(\calZ^{(\Delta\mathrm{OS})}_{\varphi_i}\right)^{\frac{1}{2}}
\varphi_i^{(\mathrm{OS})}\qquad\mbox{with}\qquad
\calZ^{(\Delta\mathrm{OS})}_{\varphi_i}\,=\,
\calZ^{-1}_{\varphi_i}
\calZ^{(\mathrm{OS})}_{\varphi_i}\,,
\eea
where 
$\calZ^{(\Delta\mathrm{OS})}_{\varphi_i}$
are finite non-diagonal renormalisation matrices. 
Such a finite field renormalisation can be easily implemented 
at the end of the calculation of the renormalised amplitude
by means of appropriate LSZ factors,
\bea
\label{eq:LSZren2}
\bfR\, \bar \calA_\Gamma^{(\mathrm{OS})} &=&
\bfR\, \bar \calA_\Gamma
\prod_{i=1}^n 
\left(%
\calZ^{(\Delta \mathrm{OS})}_{\varphi_i}\right)^{\frac{1}{2}}\,,
\eea
where 
$\bfR\, \bar\calA_{\Gamma} =
\bar\calA_{0,\Gamma}+
\bfR\, \bar\calA_{1,\Gamma}
+\bfR\, \bar\calA_{2,\Gamma}
$
is the renormalised amplitude in a scheme 
fulfilling (a)--(c).
In practice, 
$\bfR\, \bar\calA_{\Gamma}$ 
can first be 
computed in $\numdim=4$ dimensions using~\refeq{eq:masterformula1} and~\refeq{eq:masterformula2}
with the field-renormalisation constants 
$\calZ_{\varphi_i}$ 
and the related 
UV and rational counterterms. 
The finite
renormalisation $\calZ^{(\Delta \mathrm{OS})}_{\varphi_i}$ can be applied a
posteriori, exploiting the fact that the behaviour of 
$\bfR\, \bar\calA_{\Gamma}$ 
under final renormalisation transformations 
is the same as in $\numdim=D$
dimensions.

Let us now consider the requirements (b)--(c) 
and their 
interplay with the vev and tadpole renormalisation 
in non-trivial schemes, e.g.~in schemes where 
the mass parameters are renormalised 
through on-shell conditions in the SB phase.
To this end, 
we focus on the tadpole and Higgs-mass terms in the SB 
scalar sector,
\bea
\calL(H) &=& - t_0 H_0 +\frac{1}{2} m^2_{H,0} H_0^2 +\dots
\nonumber\\
&=& - 
(t+\delta t) 
\Big(1+\frac{1}{2}\delta \calZ_{1,H}\Big)
H  
+\frac{1}{2}m^2_{H}H^2 
\big(1+2\delta \calZ_{1,M_H}\big) \big(1+\delta \calZ_{1,H}\big)
+\dots\,,\qquad
\eea
where $\delta \calZ_{1,H} = \delta \calZ_{1,\phi}^{\ubk}$.
As discussed in~\refse{se:vevexpdRY}, 
we can restrict ourselves to one-loop level.
For simplicity we assume that, in the symmetric phase, 
the scalar potential depends on a
dimensionless coupling $\lambda$ and a dimensionful scale $\mu$, which
are related to the vev in the same way as in the SM.
This implies that
the bare and renormalised tadpole parameters 
are related to the parameters of the unbroken scalar sector via 
\bea 
t_0 &=& \vev_0\Big(\mu_0^2-\frac{\lambda_0}{4}\vev_0^2\Big)\,,
\qquad
t \,=\, \vev\Big(\mu^2-\frac{\lambda}{4}\vev^2\Big)\,,
\eea
while for the Higgs mass we have
\bea 
m_{H,0}^2 &=& 
2\mu_0^2-3 \frac{t_0}{\vev_0}\,,
\qquad
m_{H}^2 
\,=\,
2\mu^2-3 \frac{t}{\vev}\,.
\eea
The requirement that the minimum of the renormalised
tree-level potential is located at $H=0$ implies the vanishing 
of the renormalised
tadpole, 
\bea
\label{eq:tadvanish}
t &=& 0\,,%
\eea
and thus 
\bea
\label{eq:smvev}
\vev^2\,=\,4\mu^2/\lambda\,, 
\eea
which corresponds to~\refeq{eq:vevparamrel}.
Since the minimum of the potential is shifted by quantum corrections, there
is no reason to require that also the bare tadpole should vanish.
The associated counterterm, $\delta t = t_0-t$, is given by 
\bea
\label{eq:tadCT}
\delta t &=& 
\vev
\Big[
\mu^2\big(\delta \calZ_{1,\vev} +2\delta \calZ_{1,\mu} \big)
-
\frac{\lambda}{4}
\big(\delta \calZ_{1,\lambda} +3\delta \calZ_{1,\vev} \big)
\Big]
\nonumber\\[1mm]
&=& 
\vev\mu^2 
\big(2\delta \calZ_{1,\mu}-2\delta \calZ_{1,\vev}  -\delta \calZ_{1,\lambda}\big)\,,
\qquad
\eea
and $\delta t= 0$ only if $\vev_0^2\,=\,4\mu_0^2/\lambda_0$,
which corresponds to 
$\delta \calZ_{1,\vev} = \delta \calZ_{1,\mu}  -\delta
\calZ_{1,\lambda}/2$. 
As shown above, 
the choice
$\delta \calZ_{1,\vev} = \frac{1}{2}\delta \calZ_{1,\phi}$
in combination with the property~\refeq{eq:brokentheory}
ensures that each UV counterterm $\delta Z_{l,\Gamma}$
fulfils~\refeq{eq:deltaZvevexp}, 
which guarantees the UV finiteness
of all renormalised 1PI vertex functions in the SB theory.
This holds also for one-point vertex functions. Thus $\delta t$ as defined
in~\refeq{eq:tadCT} ensures the cancellation of all 
UV poles stemming from tadpole diagrams.
In addition, one can impose that renormalised 
tadpoles cancel exactly while respecting the conditions (b)--(c).}
To this end, the finite parts of the
independent renormalisation constants 
that generate the tadpole counterterm, \ie
$\delta \calZ_{1,\vev}$, 
$\delta \calZ_{1,\mu}$ and $\delta \calZ_{1,\lambda}$,
can be fixed according to the following
three renormalisation conditions in the SB phase:

\begin{enumerate}
\item Renormalised tadpoles are required to vanish exactly, 
\ie $\delta t = -\delta T$ where $\delta T$ stands for the unrenormalised 
Higgs one-point function.

\item The finite part of $\delta
\calZ_{1,\vev}$ 
is fixed according to a physical renormalisation scheme in the SB phase
(e.g.~through an on-shell 
renormalisation condition for a certain gauge-boson mass), and the
scalar-field renormalisation is adapted to $\delta \calZ_{1,\vev}$ as
required 
by~\refeq{eq:vevphiren}, 
\ie $\delta \calZ_{1,H}=\delta \calZ^{\ubk}_{1,\phi}
=2\delta \calZ_{1,\vev}$.
This unphysical choice for the finite part of  $\delta \calZ_{1,H}$ 
can be corrected a posteriori 
by applying a finite LSZ factor
$\sqrt{\calZ^{(\Delta \mathrm{OS})}_{H}}$ 
for each external Higgs particle as
in~\refeq{eq:LSZren1}--\refeq{eq:LSZren2}.

\item The above choices fix $\delta \calZ_{1,\vev}$ and the 
combination $2\delta \calZ_{1,\mu}-\delta \calZ_{1,\lambda}$,
which enters~\refeq{eq:tadCT}. 
The remaining free parameter in the scalar sector can be 
fixed through a renormalisation condition for the Higgs mass, 
whose counterterm reads
\bea
\label{eq:mHCT}
\delta m_H^2 = 
2 \calZ_{1,m_H} m_H^2 &=& 
4\mu^2 \delta\calZ_{1,\mu} 
-
3\frac{\delta t}{\vev}\,.
\eea
Alternatively a renormalisation condition for the scalar self-coupling can
be imposed.
 
\end{enumerate}

This approach makes it possible to derive  
the rational counterterms for a SB  theory 
by means of~\refeq{eq:dRvexpgen}, \ie 
through calculations in the symmetric phase, for the case of 
fully realistic renormalisation schemes.
In the derivations of rational counterterms, 
all independent renormalisation constants ($\delta \calZ_{1,\vev}$, 
$\delta \calZ_{1,\mu}$, $\delta \calZ_{1,\lambda}$, \dots)
can be handled as free parameters as discussed in~\cite{Lang:2020nnl}.
In this way the resulting $\delta \calR^{\bk}_{l,\Gamma}$
counterterms can be easily adapted to the desired renormalisation
scheme in the SB phase. Finally, the on-shell renormalisation of fields can
be imposed at the level of renormalised amplitudes through finite LSZ factors.

\subsection{Gauge fixing}
\label{se:thooftgfix}

The vev-expansion approach introduced in~\refses{se:mvevexp}{se:abc}
is based on rigid invariance, \ie on the assumption that the
SB Lagrangian is symmetric wrt gauge transformation of the full 
Higgs multiplet~\refeq{eq:vevdefA2}, or, equivalently,
that the vev dependence of the SB Lagrangian is entirely generated 
from the symmetric Lagrangian 
through shifts of the Higgs field, \mbox{$H(x)\to H(x)+\vev$}. 
While this property is trivially fulfilled by the classical SB Lagrangian,
the gauge-fixing procedure can break
rigid invariance.

A well-established approach that  
preserves rigid invariance at the level of the quantised Lagrangian 
is the background-field method (BFM)
\cite{DeWitt:1964yg,DeWitt:1967ub} (see also \cite{Abbott:1981ke}).
In the BFM the broken phase can be generated
through a shift of the Higgs 
background field, \mbox{$\widehat \Phi_i(x) \to \widehat \Phi_i(x)+\vev_i$},
while background-field Ward identities
guarantee that, consistently with~\refeq{eq:vevphiren}, 
the vev-renormalisation constant and the 
field-renormalisation constants associated with the 
various components of the
Higgs doublet are related by~\cite{Denner:1994xt}
\begin{align}
\calZ_{\vev}\,=\,  \left(\calZ_{\widehat \Phi_i}\right)^{\frac{1}{2}}\,.
  \label{eq:vevrenBFM}
\end{align}
Although not mandatory, one can impose background-field gauge invariance
on the renormalised vertex function 
that fixes the finite parts of all fields
entering the gauge fixing-procedure.
This ensures, as assumed in \refse{se:vevexpdRY}, a consistent renormalisation of fields in 
the SB and symmetric phases, while
differences in the finite parts of the field-renormalisation constants can be
implemented a posteriori in the form of
LSZ factors~\refeq{eq:LSZren1}--\refeq{eq:LSZren2}.
In summary, the BFM ensures that the requirements for the
validity of the vev-expansion approach of~\refses{se:mvevexp}{se:abc}
are fulfilled.
Moreover it supports the usual 't~Hooft--Feynman gauge in both the symmetric and SB phase,
making practical calculations simple and efficient.

\begin{figure}[t]
\begin{center}
\includegraphics[width=28mm]{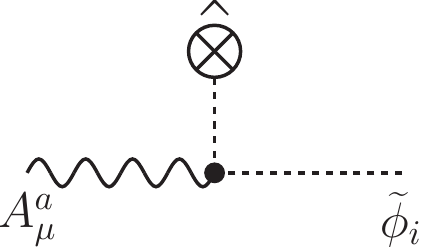}\qquad
\includegraphics[width=28mm]{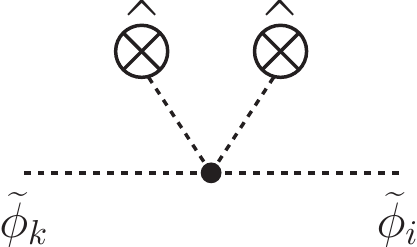}\qquad
\includegraphics[width=28mm]{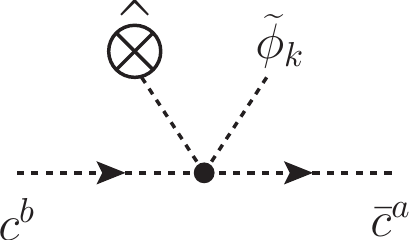}\\[3mm]
\includegraphics[width=28mm]{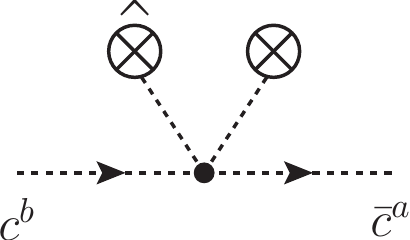}

\end{center}
\caption{The three $\hvev$-insertion vertices on the 
first line correspond to the $\hvev$-dependent 
two- and three-point vertices associated with 
the gauge-fixing term $\Delta\calL_\fix (\hvev, H)$, defined in~\refeq{eq:gfixlagI}.
The double $\vev\hvev$-insertion vertex on the second line 
contributes to the $\vev$ expansion 
of the SB theory~\refeq{eq:gfixlagH}.
Like for the case of the $\tvev$ insertions introduced 
in~\refse{eq:tvevins}, the (pseudo) external lines associated with
$\hvev$ insertions do not involve any propagator and do not
correspond to any field.
Such external lines should be regarded as a technical
tool to keep track of the total 
power in $\hvev$ 
induced by vertices stemming from 
$\Delta\calL_\fix (\hvev, H)$. 
}
\label{fig:gfixvertices}
\end{figure}

Let us now consider the standard quantisation approach, where all fields are
quantised,
and let us assume that the vev dependence of the gauge-fixing Lagrangian  
violates rigid invariance. 
In this case the vev renormalises differently wrt 
the Higgs
field~\cite{Sperling:2013eva}, and some aspects of our vev-expansion 
procedure need to be amended.
As a concrete example we will consider 
the 't~Hooft gauge fixing, but the following approach is
applicable to a wide range of gauge-fixing procedures.
For a generic gauge theory 
with gauge coupling 
$g$ and generators $T^a$, 
the 't~Hooft gauge-fixing Lagrangian reads\footnote{For 
simplicity here we use the form of the gauge-fixing term
corresponding to a multiplet of real scalar fields.
For complex fields one should replace 
\begin{align*}
\widetilde \Phi_i T^a_{ij} \hvev_j
\,\to \,
\frac{1}{2}\left(\tilde\Phi_i^\dagger T_{ij}^a \hat v_j-
\hat v_i T^a_{ij} \tilde \Phi_j
\right)\,.
\end{align*}
}
\bea
\label{eq:gfixlagA}
\calL_{\fix} 
&=&
-\frac{1}{2\xi}\sum_a\Big(\partial^\mu A^a_\mu
+\ri g  \widetilde \Phi_i T^a_{ij} \hvev_j
\Big)^2\,,
\eea
where $\xi$ is the gauge-fixing parameter.
As defined in~\refeq{eq:vevdefA}--\refeq{eq:vevdefB},  
$\widetilde\Phi_i = \Phi_i - \vev_i$, 
are the dynamic components of the scalar multiplet, 
and
$\hvev_i = \xi'\vev_i$, where
$\xi'$ is a gauge-fixing parameter. 
Similarly we define
\bea
\label{eq:gfixlagC}
\hvev &=& \xi'\vev\,.
\eea
The combination $\widetilde\Phi_i T^a_{ij}$ in~\refeq{eq:gfixlagA} 
corresponds to the 
would-be Goldstone bosons associated with the 
gauge bosons $A^a_\mu$ of the SB symmetry group.
The $\widetilde \Phi_i\, \partial^\mu A^a_\mu $ bilinear 
terms resulting 
from~\refeq{eq:gfixlagA} 
make it possible to cancel the mixing between 
gauge bosons and Goldstone bosons,
thereby ensuring 
well-defined propagators for 
these two kinds of fields.
This is achieved by identifying the two renormalised gauge parameters, 
\bea
\label{eq:xiparamchoice}
\xi'&=&\xi\,,
\eea
while,
in general,
the associated renormalisation constants are different.
We will assume that $\calZ_{\xi}$ and $\calZ_{\xi'}$ 
are chosen in
such a way that the gauge-fixing term~\refeq{eq:gfixlagA}
remains effectively unrenormalised. 
This is achieved by setting
\bea
\label{eq:xinonrenid}
\calZ_{\xi}\,=\,\calZ^{}_{A}\,,
\eea
and
\bea
\label{eq:hvevren}
\calZ_{\hvev}
&=& 
\calZ_{\xi'}
\calZ_{\vev}
\,=\,
\frac{1}{\calZ_{g}}
\left(\frac{\calZ^{}_{A}}{\calZ^{}_{\phi}}\right)^{\frac{1}{2}}\,.
\eea
In the 't~Hooft gauge the ghost Lagrangian reads
\bea
\label{eq:gfixlagB}
\calL_{\ghost} 
&=&
-\bar c^a\Big( \partial^\mu D^{ab}_\mu 
- g^2  \hvev_i T^a_{ij}T^b_{jk} (\vev_k+\widetilde\Phi_k)
\Big) c^b\,,
\eea
and depends, through $\vev_k$ 
and $\hvev_i$,
on  the usual vev parameter $\vev$
as well as on $\hvev$.
The former can be generated through a shift $H\to H+\vev$ of the Higgs
component of the scalar multiplet in~\refeq{eq:gfixlagB}, while 
this is not possible 
for the parts of~\refeq{eq:gfixlagA} and~\refeq{eq:gfixlagB} that depend on 
$\hvev$.
Therefore, in the 't~Hooft gauge, the relation~\refeq{eq:brokentheory} between the symmetric and
SB  Lagrangian needs to be replaced by 
\bea
\label{eq:gfixlagH}
\calL(H) &=&
\calL^{\widehat\ubk}(\hvev,\vev+H)\,,
\eea
where $\calL^{\widehat\ubk}$ is a modified Yang--Mills
Lagrangian of the form 
\bea
\label{eq:gfixlagF}
\calL^{\widehat\ubk}(\hvev, H)
&=&
\calL^{\ubk}(H)
+\Delta\calL_\fix (\hvev, H)\,.
\eea
Here $\calL^{\ubk}(H)$ is the usual symmetric Lagrangian,
where the 
gauge-fixing and ghost terms correspond to the Lorentz 
gauge, while the dependence on $\hvev$ is embodied in 
the term $\Delta\calL_\fix (\hvev, H)$.
Its explicit expression in the `t~Hooft gauge is
\bea
\label{eq:gfixlagI}
\Delta\calL_\fix (\hvev, H) 
&=&
-\frac{1}{2\xi}\bigg[\Big(\partial^\mu A^a_\mu
+\ri g  \widetilde \Phi_i T^a_{ij} \hvev_j
\Big)^2
-
\Big(\partial^\mu A^a_\mu\Big)^2\bigg]
+g^2  %
\hvev_i T^a_{ij}T^b_{jk} \widetilde\Phi_k
\,\bar c^a c^b\,.\quad
\eea
The modified symmetric theory described by the 
Yang--Mills Lagrangian~\refeq{eq:gfixlagF}
will be referred to as the $\hYM$
theory.
According to~\refeq{eq:gfixlagH}, 
the SB theory can be related to the 
$\hYM$ theory via standard vev insertions, while
the $\hYM$  
theory is related to the 
symmetric theory through the 
$\hvev$-dependent terms~\refeq{eq:gfixlagI}.
The
Feynman rules of the $\hYM$  
theory correspond to the 
ones of the symmetric theory 
supplemented with 
the three types of $\hvev$-insertion vertices 
depicted in the first line of~\reffi{fig:gfixvertices}.

Let us now discuss the implications of $\hvev$ insertions on 
the vev-expansion formula~\refeq{eq:dRvexpgen}, which connects rational counterterms 
in the SB and symmetric phases.
As we will show in the following, this formula
can be extended to SB theories of type~\refeq{eq:gfixlagH}
by replacing vev expansions through double
expansions in $\vev$ and $\hvev$.
In order to demonstrate this, we will generalise the 
mass-expansion identities~\refeq{eq:dZ1vevexp}--\refeq{eq:massexpforampsb}
in such a way that all requirements for the 
derivation of~\refeq{eq:dRvexpgen} in~\refse{se:vevexpdRms}
remain valid.

The expansion~\refeq{eq:massexpforamps} 
for 1PI vertex functions in $\numdim=D$
can be generalised as 
\bea 
\label{eq:gfixampA}
\bfm_{[0,X]}\, \ampbar{l}{\Gamma}{\bk}{}
&=&
\left(\sum_{k=0}^{X}
\ampbar{l}{\widetilde\Gamma{\vev^k}}{\widehat\ubk}{}
\right)
\prod_{i=1}^n\,\calU^{-1}_{i}\,,
\eea
where we have included the effect of mixing as in~\refse{se:vevexpdRY}, and 
the summands between brackets denote
terms of total order $k$ in the vev.
They correspond to the 
result of a double expansion\footnote{Note 
that in quantities carrying
the superscript $\widehat{\mathrm{YM}}$  the subscript $\vev^k$
denotes the total vev-order, \ie the total power 
in $\vev$ and $\hvev$, while 
for quantities with superscript 
$\mathrm{YM}$ 
the individual orders in
$\vev$ and $\hvev$ 
are
indicated.
} 
in $\vev$ and $\hvev$, \ie
\bea
\label{eq:gfixampB}
\ampbar{l}{\widetilde\Gamma \vev^{k}}{\widehat\ubk}{}
&=&
\sum_{j=0}^{k}  
\ampbar{l}{\widetilde\Gamma \vev^{k-j}\hvev^{j}}{\ubk}{}
\,=\,
\sum_{j=0}^{k}  
\frac{\vev^{k-j}}{(k-j)!}\,
\ampbar{l}{\widetilde\Gamma \rHexp ^{k-j}\hvev^{j}}{\ubk}{}\Big|_{p_\rHexp =0}\,.
\eea
Here the subscripts $\vev^{k-j}$ and 
$\hvev^j$ indicate the number of vev insertions
of each kind,
where the number $j$ of $\hvev$ insertions 
should be understood as the total power in the 
parameter $\hvev$ that 
results from the insertion of $\hvev$-dependent vertices.
The expansion~\refeq{eq:massexpforampsb} in 
$\numdim=4$ can be generalised in a similar way as
\bea
\label{eq:gfixampA2}
\bfm_{[0,X]}\, \amp{l}{\Gamma}{\bk}{} 
&=&
\left(\sum_{k=0}^{X}    \sum_{j=0}^{k} 
\amp{l}{\Gamma \vev^{k-j}
\tvev^j}{\widehat\ubk}{}\right)
\prod_{i=1}^n\,\calU^{-1}_{i}\,.
\eea
Here the terms with fixed $k$ 
correspond to the various contributions of 
total order $k$ in $\vev$, $\hvev$ and $\tvev$. The summands between 
brackets read
\bea
\label{eq:gfixampB2}
\amp{l}{\widetilde\Gamma \vev^{m}\tvev^j}{\widehat\ubk}{}
&=&
\sum_{n=0}^{m}  
\amp{l}{\widetilde\Gamma \vev^{m-n}\hvev^{n}\tvev^j}{\ubk}{}
\,=\,
\sum_{m=0}^{n}  
\frac{\vev^{m-n}}{(m-n)!}\,
\amp{l}{\widetilde\Gamma \rHexp ^{m-n}\hvev^{n}\tvev^j}{\ubk}{}\Big|_{p_\rHexp =0}\,.
\eea
As discussed in~\refse{eq:tvevins}, the 
auxiliary $\tvev$ insertions generate the 
parts of the mass expansion that are not accounted for by 
$\vev$ and $\hvev$ insertions in $\numdim=4$.
Finally, the mass expansion~\refeq{eq:dZ1vevexp}  
for UV counterterms can be generalised as
\bea
\label{eq:gfixampD}
\deltaZ{l}{\Gamma}{\bk}{} &=& 
\left(%
\sum_{k=0}^{X} \, \deltaZ{l}{\widetilde\Gamma \vev^k}{\widehat\ubk}{} 
\right)
\prod_{i=1}^n\,\calU^{-1}_{i}\,,
\eea
where $\deltaZ{l}{\Gamma}{\bk}{}$ is the local counterterm for~\refeq{eq:gfixampA}, 
and the summands on the rhs can be defined 
as the contributions of order $k$ in its mass expansion,
\ie the terms of total order $k$ in $\vev$ and $\hvev$.
This implies that
$\deltaZ{l}{\widetilde\Gamma\vev^k}{\widehat\ubk}{}$ are the local UV counterterms
for $\ampbar{l}{\widetilde\Gamma \vev^{k}}{\widehat\ubk}{}$. 
In this way, similarly as
for~\refeq{eq:dZ1vevexp}--\refeq{eq:massexpforampsb},  
the identities~\refeq{eq:gfixampA}--\refeq{eq:gfixampD} 
are nothing but mass expansions of the corresponding quantities.
Moreover, in analogy with~\refeq{eq:dZ2vevexpC1}--\refeq{eq:dZ2vevexpC2},
for the mass expansion of 
subdivergences we have
\bea
\label{eq:dZ2vevexpC2b}
\bfm_{[0,X]}\,
\sum  \limits_{\gamma \in \Omega(\Gamma)}  
\deltaZ{1}{\gamma}{\bk}{}  \cdot \ampbar{1}{\Gamma/\gamma}{\bk}{}
&=&
\sum_{k=0}^X 
\sum_{j=0}^k
\sum  \limits_{\gamma \in \Omega(\Gamma)}\,  
\deltaZ{1}{\gamma\vev^j}{\widehat\ubk}{}  \cdot
\ampbar{1}{(\Gamma/\gamma)\,\vev^{k-j}}{\widehat\ubk}{}
\nonumber\\
&=&
\sum_{k=0}^X 
\sum  \limits_{\gamma' \in \Omega(\Gamma \VevInsert)}\,  
\deltaZ{1}{\gamma'}{\widehat\ubk}{}   
\cdot \ampbar{1}{\Gamma \VevInsert /\gamma'}{\widehat\ubk}{}
\,.
\eea

The above properties satisfy all assumptions in the
derivation of~\refeq{eq:dRvexpgen}.
Therefore, rational counterterms can be derived
within the $\hYM$ theory 
by means of a generalised expansion of the form 
\bea
\label{eq:gfixampFa}
\delta \calR^{\bk}_{l,\Gamma}
&=&
\left(\sum_{k=0}^{X}   
\ratamp{l}{\widetilde\Gamma \VevInsert}{\widehat\ubk}{}\,
\right)
\prod_{i=1}^n\,\calU^{-1}_{i}\,,
\eea
where $\ratamp{l}{\widetilde\Gamma \VevInsert}{\widehat\ubk}{}$ 
for $l=1,2$
can be obtained from the related formulas in~\refse{se:vevexpdRms} 
through the substitutions
\bea
\label{eq:hvevsubst}
\ampbar{l}{\Gamma \vev^{m}}{\ubk}{}
\to 
\ampbar{l}{\widetilde\Gamma \vev^{m}}{\widehat\ubk}{}\,,
\qquad
\amp{l}{\Gamma \vev^{m}\tvev^j}{\ubk}{}
\to 
\amp{l}{\widetilde\Gamma \vev^{m}\tvev^j}{\widehat\ubk}{}\,,
\qquad
\deltaZ{1}{\Gamma \vev^k}{\ubk}{} 
\to
\deltaZ{1}{\widetilde\Gamma \vev^k}{\widehat\ubk}{}\,,
\eea
while $\delta \tilde Z^{\widehat\ubk}_{1,\gamma}= 
\delta \tilde Z^{\ubk}_{1,\gamma}$
since $\delta \tilde Z$ counterterms do not receive 
any vev insertion in renormalisable theories. See \refeq{eq:r2vevinse2}. 

The one-loop counterterms
$\deltaZ{1}{\widetilde\Gamma\vev^k}{\widehat\ubk}{}$, which are required for
the derivation of two-loop rational counterterms, can be generated through
multiplicative renormalisation within the $\hYM$ theory.
In the $\hvev=0$ case, 
according to~\refeq{eq:genren4}--\refeq{eq:genren5}
we have
\bea
\label{eq:genren4mod}
\deltaZ{1}{\widetilde\Gamma\vev^k}{\ubk}{}
&=& 
\ampbar{0}{\widetilde\Gamma\vev^k}{\ubk}{}\,
\Big(
k\, \delta \calZ_\vev
\,+\sum_{i=1}^n
\frac{1}{2}\,%
\delta\calZ_{1,\widetilde\varphi_i}^{\ubk}
\Big)
+
D^{\ubk}_{1,\theta}\,
\ampbar{0}{\widetilde\Gamma\vev^k}{\ubk}{}
\,,
\eea
and this relation can be extended to the $\hvev\neq 0$ case
following the same derivations as in~\refse{se:vevexpdRY}
with only few modifications as discussed in the following.

Within the SB broken theory~\refeq{eq:gfixlagH}, 
the additional $\hvev$-dependent UV divergences stemming from the `t~Hooft gauge
fixing can be cancelled through a 
shift of the vev renormalisation
constant~\cite{Sperling:2013eva}, \ie by turning~\refeq{eq:vevphiren}
into
\bea
\label{eq:gfixampC}
\delta \calZ_{1,\vev} = 
\frac{1}{2}\delta \calZ^{\ubk}_{1,\phi}
+\delta \calZ^\fix_{1,\vev}\,,
\eea
while all vev-independent renormalisation constants 
can be kept fixed as in the $\hvev=0$ case.
The shift $\delta \calZ^\fix_{1,\vev}$ 
should be propagated to all vev-dependent renormalisation constants, 
such as mass counterterms and the tadpole counterterm~\refeq{eq:tadCT}. 
For convenience the UV divergent and finite parts of
$\delta \calZ^\fix_{1,\vev}$ 
can be chosen in such a way that 
renormalised tadpoles cancel exactly as 
discussed in~\refse{se:abc} for the $\hvev=0$ case.

The $\hvev$-dependent parts of 
the
UV counterterms of the SB 
and $\hYM$ theories can be related to each other 
using the approach of~\refse{se:vevexpdRY}. To this end,
apart from~\refeq{eq:gfixampC} only two 
aspects need to be adapted to the $\hvev\neq 0$ case:
in the tree-level mass expansion~\refeq{eq:genren2} one should replace
\bea
\label{eq:treeamprepl}
\ampbar{0}{\widetilde\Gamma\vev^k}{\ubk}{}
&\to&
\ampbar{0}{\widetilde\Gamma\vev^k}{\widehat\ubk}{}\,,
\eea
and the list of independent parameters of the original 
symmetric theory 
has to be supplemented by 
$\hvev=\xi' \vev$.
Therefore the renormalisation operator~\refeq{eq:D1thetaubk}
has to be extended as
\bea
D^{\widehat\ubk}_{1,\theta}
&=&
D^{\ubk}_{1,\theta}+
\delta\calZ^{}_{1,\hvev}
\,\hvev\,
\frac{\partial}{\partial\hvev}\,,
\eea
with $\delta\calZ^{}_{1,\hvev}=\delta\calZ^{}_{1,\xi'}+\delta\calZ^{}_{1,\vev}$.
For the parameter-renormalisation operator~\refeq{eq:D1paridentity}  of the
SB theory, which depends also on $\vev$, 
we have 
\bea
D_{1,\theta}
&=&
D^{\widehat\ubk}_{1,\theta}+
\delta\calZ^{}_{1,\vev}
\,\vev\,
\frac{\partial}{\partial\vev}
\,=\,
D^{\ubk}_{1,\theta}+
\delta\calZ^{}_{1,\hvev}
\,\hvev\,
\frac{\partial}{\partial\hvev}
+
\delta\calZ^{}_{1,\vev}
\,\vev\,
\frac{\partial}{\partial\vev}\,.
\eea
Note that the renormalisation of $\xi$ is included in
$D^{\ubk}_{1,\theta}$, while 
$\delta\calZ^{}_{1,\hvev}
\,\hvev\,
\frac{\partial}{\partial\hvev}$
is responsible for the renormalisation of $\xi'$. 
A suitable renormalisation of $\hat v$ is given
by~\refeq{eq:hvevren},
which guarantees, together with~\refeq{eq:xinonrenid}, 
that the gauge-fixing term does not renormalise, 
and leads to finite vertex functions for physical and
unphysical fields. 
Note that the choice~\refeq{eq:xinonrenid}--\refeq{eq:hvevren}
fixes the renormalisation of the mass
terms in the ghost sector, \ie it correctly cancels all the remaining poles 
after ghost-field renormalisation.

Applying the above modifications to~\refeq{eq:genren4}--\refeq{eq:genren5}
it turns out that the one-loop counterterms of the $\hYM$ theory 
can be obtained from the renormalisation identity~\refeq{eq:genren4mod}
with the replacements~\refeq{eq:gfixampC} and~\refeq{eq:treeamprepl}.
More explicitly, using a similar $\hvev$-expansion as in~\refeq{eq:gfixampB}
we have 
\bea
\deltaZ{1}{\widetilde\Gamma\vev^k}{\widehat\ubk}{}
&=&
\sum_{j=0}^{k}  
\deltaZ{1}{\widetilde\Gamma\vev^{k-j}\hvev^j}{\ubk}{}\,,
\eea
with
\bea
\label{eq:gfixampG}
\deltaZ{1}{\widetilde\Gamma\vev^{k-j}\hvev^j}{\ubk}{}
&=&
\ampbar{0}{\widetilde\Gamma\vev^{k-j}\hvev^j}{\ubk}{}\,
\bigg(
k\,\delta \calZ_{1,\vev}
\,+\sum_{i=1}^n
\frac{1}{2}\,%
\delta\calZ_{1,\widetilde\varphi_i}^{\ubk}
\bigg)
+
\Big(
D^{\ubk}_{1,\theta}
+j\delta \calZ_{\xi'}
\Big)
\,
\ampbar{0}{\widetilde\Gamma\vev^{k-j}\hvev^j}{\ubk}{}
\,,
\quad\qquad
\eea
where $\delta \calZ_{1,\vev}$ has to be chosen according to~\refeq{eq:gfixampC}.
Note that the tree-level amplitudes on the rhs are proportional
to 
$\vev^{k-j}\hvev^j= (\xi')^j\vev^k$.
The renormalisation of this factor 
generates the term $k\,\delta \calZ_{1,\vev}$ on the rhs of~\refeq{eq:gfixampG}
plus an extra term 
$j\,\delta \calZ_{\xi'}$, which 
originates from
$\delta\calZ^{}_{1,\hvev}=\delta\calZ^{}_{1,\xi'}+\delta\calZ^{}_{1,\vev}$.

In summary, in the presence of gauge-fixing terms
with a $\hvev$-dependence of type~\refeq{eq:gfixlagH}, 
rational terms can be determined through the
generalised vev expansions~\refeq{eq:gfixampFa}--\refeq{eq:hvevsubst}, 
where standard vev insertions are supplemented by $\hvev$ insertions.

\section{Rational counterterms of $\ord(\als^2)$ for the Standard Model}
\label{se:results}

The full set of two-loop rational counterterms for QCD has been presented 
in~\cite{Lang:2020nnl}, and in this section we derive all remaining 
$\delta\calR_2$ counterterms of 
$\ord(\als^2)$ for the full SM.
These correspond to two-, three- or four-point 
counterterms involving quarks ($Q$) and/or gluons ($G$) 
in combination with one or more EW gauge bosons ($V$)
and Higgs or would-be Goldstone bosons ($S$).\footnote{%
Note that the rational counterterms for the vertices involving
ghosts (c) combined with EW gauge bosons ($V$) and scalars ($S$), such as
$c\bar cV$ and $c\bar cS$, are not needed since they 
are induced by two-loop and higher-loop diagrams, which, for amplitudes
with no external ghosts, become only relevant in three-loop QCD
calculations and beyond.
}
In practice 
there are six classes of 
non-vanishing rational counterterms of this kind:
$GGVV$,
$GGGV$,
$Q QV$,
$GGSS$,
$GGS$ and
$Q QS$.

All calculations have been implemented twice and independently 
in two different frameworks.
On the one hand we have used {\sc Geficom}~\cite{GEFICOM},
which is based on {\sc Qgraf}~\cite{Nogueira:1991ex}, {\sc Q2E} and
{\sc Exp}~\cite{Seidensticker:1999bb,Harlander:1997zb},
{\sc Form}~\cite{Vermaseren:2000nd,Tentyukov:2007mu},
{\sc Matad}~\cite{Steinhauser:2000ry} and {\sc Color}~\cite{vanRitbergen:1998pn}.
On the other hand we have employed an in-house framework
implemented in {\sc Python} that uses {\sc Qgraf}~\cite{Nogueira:1991ex}, 
{\sc Form}~\cite{Vermaseren:2000nd,Tentyukov:2007mu} and {\sc
python-Form}\footnote{\url{https://github.com/tueda/python-form}}.
A more detailed description of the methodology implemented in these two frameworks 
can be found in Sects.~5.1--5.2 of~\cite{Lang:2020nnl}.

In addition to the cross checks between the two independent
calculations, we have verified that our results for one-loop rational
counterterms are in agreement with the literature~\cite{Draggiotis:2009yb},
and  that the two-loop results fulfil 
various self-consistency properties described in Sect.~5.1 of~\cite{Lang:2020nnl}.

As a further validation, we have compared two different ways of
handling particle masses in the loops:
direct calculations with massive quarks and, alternatively,
the new vev-expansion technique
introduced in this paper.
The only 
massive states that can circulate in the loops at $\ord(\als^2)$ 
are quarks, 
and the only $\delta\calR$ counterterms that depend on 
quark masses are the ones associated with the 
$Q\bar Q$ and $GG$ two-point vertices~\cite{Lang:2020nnl}
and with the $GGS$ three-point vertex.
For such $\delta\calR$ counterterms, 
in~\refse{se:R2results} we demonstrate that 
the results of explicit calculations with massive quarks
are in agreement with the outcome of vev expansions.
This should be regarded as a simple illustration of 
the usage of vev expansions, keeping in mind that the main goal of 
the vev-expansion approach is the calculation of $\delta\calR$ counterterms 
for two-loop EW 
and mixed QCD--EW corrections.

\subsection{Lagrangian}
The presented results are based on the renormalised Lagrangian 
\bea
\label{eq:qcdewlag}
&&\mathcal L \,=\,
-\frac{1}{4}
\,
G^a_{\mu\nu}G^{a,\mu\nu} 
-
\frac{\calZ_{\gpar}}{2 \, \xi} 
\Big(\partial^\mu G^a_\mu\Big)^2
- \, \calZ_{c} \,
\bar{c}^a \partial_\mu D^\mu_{ab}\, c^b
\nonumber\\
&&{}
+\sum_{i=1}^{N_\gen}
\left[\,
  \QiLbar \Big(\ri \gamma_\mu D^\mu \Big)\calZ_{Q_i} \QiL 
+
\uiRbar \Big(\ri \gamma_\mu D^\mu \Big)\calZ_{u_i} \uiR 
+
\diRbar \Big(\ri \gamma_\mu D^\mu \Big)\calZ_{d_i}\diR 
\right]
\nonumber\\
  &&{}-\,\sum_{i=1}^{N_\gen}\left\{
  \left[
    \yukUP{i} \calZ_{\yukUP{i}} \left(\QiLbar\calZ_{Q_i}^{1/2}\Phi^\rc\right)
    \left(\calZ_{u_i}^{1/2}
    \uiR \right)
+
\yukDO{i} \calZ_{\yukDO{i}}  \left(\QiLbar\calZ_{Q_i}^{1/2} \Phi\right) \left(\calZ_{d_i}^{1/2} \diR
\right)\right]
\,+\,\mathrm{h.c.}  \right\}\,,
\nonumber\\
\eea
with the field-strength tensor and the covariant derivatives
\bea
\label{eq:covdev}
G^a_{\mu\nu}&=&
\calZ_G^{1/2}
\left[
\partial_\mu G^a_\nu -
\partial_\nu G^a_\mu
+ 
\left(\calZ_{\als}
\calZ_G
\right)^{1/2}
g\, f^{abc}G_\mu^b G_\nu^c
\right]
\,,
\nonumber\\
D^{\mu}_{ab} &=& \partial^\mu \delta_{ab} - \left(\calZ_{\als} \calZ_G\right)^{1/2} \gs \,
f^{abc}\, G^{c\mu}\,\nonumber\\
D_\mu &=& \partial_\mu 
- \ri \left(\calZ_{\als} \calZ_G\right)^{1/2}\gs t^a\,
G^{a}_{\mu} - \ri g_2
T^b W^{b}_{\mu} + \ri g_1 \frac{Y}{2} B_\mu,
\eea
where $G^a_\mu$, $W^b_\mu$ and $B_\mu$ are, respectively, the SU(3), SU(2)
and U(1) gauge fields, and $c^a$ stands for the SU(3) ghosts. We
include $N_\gen$ generations of left-chiral 
quark doublets $Q_{i,\rL}=(u_{i,\rL}, d_{i,\rL})^\rT$
and right-chiral quark singlets $u_{i,\rR}$, $d_{i,\rR}$.
This corresponds to an even number of active quark flavours,
$\nq=2N_\gen$. 
For the Higgs doublet and its charge conjugate we use the parametrisation
\bea
  \Phi = 
  \begin{pmatrix}
    \phi^+\\
    \frac{1}{\sqrt{2}} 
    \left(
      \vev + H + \ri \chi
    \right)
  \end{pmatrix}\,,\;\quad
  \Phi^\rc = \ri \sigma^2 \Phi^\dagger
  =
  \begin{pmatrix}
    \frac{1}{\sqrt{2}} 
    \left(
      \vev + H - \ri \chi
    \right)\\
    \phi^-
  \end{pmatrix}\,.
\eea
In~\refeq{eq:qcdewlag} we assume a diagonal
CKM matrix, and quark masses are related to Yukawa couplings via
\bea
\label{eq:quarkmasses}
m_q &=& \frac{\lambda_q\vev}{\sqrt{2}}\qquad\mbox{for}\quad
q=u_i, d_i\,.
\eea
Since we restrict ourselves to QCD corrections, 
the Higgs sector of the SM Lagrangian is not included 
in~\refeq{eq:qcdewlag}.
The same holds  
for the kinetic and gauge-fixing terms for the SU(2)$\times$U(1) 
gauge fields. 
For the gluon fields we adopt the
Feynman gauge, which corresponds to $\xi=1$.

The various renormalisation constants in~\refeq{eq:qcdewlag} 
are expanded in the strong coupling $\als=\gs^2/(4\pi)$ 
up to second order as 
\bea
\label{eq:rcs1}
\calZ_{\rcarg} & = &  1+\sum_{k=1}^2
\lb\frac{\als\, t^\eps}{4\pi} \rb^k 
\delta \hat\calZ_{k,\rcarg}
\qquad\mbox{for}\qquad
\rcarg= \als,\,G,\, c,\, q,\, \lambda_{q},\, \gpar\,,
\eea
where $q=u_i$ or $d_i$. 
The scale factor $t= \msfact \mu_0^2/\mu_\rR^2$
embodies the dependence on the
regularisation scale $\mu_0$,
the renormalisation scale $\mu_\rR$, and a possible 
rescaling factor $\msfact$.\footnote{For example, in the $\msbar$ scheme
$S^\eps=(4\pi)^\eps\Gamma(1+\eps)$ and 
$\delta \hat \calZ_{k,\rcarg}$ involves only pure $1/\eps$ poles. 
See~\cite{Lang:2020nnl} for more details.  
} 

At lowest order in the EW and Yukawa interactions, 
the renormalisation of quark fields is independent of their chirality, 
and quark-doublet fields are renormalised by the diagonal matrix
$\calZ_{Q_i}=\mathrm{diag}(\calZ_{u_i},\calZ_{d_i})$. 
The gauge-fixing term involves a finite renormalisation
constant $\calZ_{\gpar}=\calZ_G/\calZ_\xi$, which is kept free in our
calculations, but in practical applications one can set $\calZ_{\gpar}=1$.

Explicit expressions for the UV poles of the various
renormalisation constants 
in the $\msbar$ scheme\footnote{Note that the symbols for various fields and
parameters used in~\cite{Lang:2020nnl} have been renamed
as follows in order to avoid conflicts: 
$A\to G$, $u\to c$, $f\to q_i=u_i, d_i$, $\lambda\to \xi$, and $\alpha\to\als$.
Note also that the Yukawa renormalisation constants
$\calZ_{\lambda_q}$ are identical to the mass renormalisation constants 
$\calZ_{m_q}$ in~\cite{Lang:2020nnl}.} 
are listed in Appendix~B
of~\cite{Lang:2020nnl}. 
The remaining scheme-dependent parts of all renormalisation constants
are handled as free parameters in our 
calculations,\footnote{See Sect.~5.1 of~\cite{Lang:2020nnl} for technical
details.} 
and all results for the two-loop rational counterterms $\delta\calR_{2,\Gamma}$ are expressed as 
linear combinations of one-loop renormalisation constants 
$\delta \hat\calZ_{1,\rcarg}$ in a generic scheme.

The SU(3) generators in the 
fundamental representation obey 
\bea
\big[ t^a, t^b\big] &=& \ri f^{abc} t^c, \qquad
\Tr\left(t^a t^b\right) \,=\, \TF   \delta^{ab}\,.
\eea
Our results depend on the normalisation factor $\TF$
and on the Casimir eigenvalues  
\bea
\CF &=& \TF\frac{\Nc^2-1}{\Nc}\,,
\qquad
\CA \,=\, \Nc\,,
\eea
where $\Nc=3$ is the number of colours.

The SU(2) and U(1) generators in the fundamental representation are
\bea
\label{eq:RLfundewgen}
T^b&=& \frac{\sigma^b}{2}\,\omega_\rL\,,
\qquad
Y\,=\, Y_{\rR}\,\omega_\rR + Y_{\rL}\,\omega_\rL\,, 
\eea
where $\sigma^b$ are the Pauli matrices, and 
\bea
\omega_\rR 
\,=\, \omega_+ 
\,=\, \frac{1}{2}(1+\gamma_5)\,,
\qquad
\omega_\rL
\,=\, \omega_- 
\,=\, \frac{1}{2}(1-\gamma_5)
\eea
project fermions on right- and left-chiral states.
The $T^3$ and $Y$ generators are related to the electric charge $Q$ through  
\begin{align}
  Q = T^3 + \frac{Y}{2}\,,
\label{eq:gmnrelation}
\end{align}
and the hypercharge eigenvalues for quarks are
$Y_{\rL,q} = 1/3$ and
$Y_{\rR,u} = 4/3$,
$Y_{\rR,d} = -2/3$.

\subsection{Bookkeeping of SU(2)$\times$U(1) interactions}
\label{se:EWbookkeeping}
For an efficient bookkeeping of the  
EW gauge fields and associated generators it is convenient to 
use an extended multiplet
\bea
\label{eq:ewsymmmultiplet}
\tilde V_\mu \,=\, \left(W^{1}_\mu,W^{2}_\mu,W^{3}_\mu,B_\mu\right)^\rT\,,
\eea
and to write the EW part of the covariant derivative as
\bea
\label{eq:EWcovdev1}
g_2
T^b W^{b}_{\mu} - g_1 \frac{Y}{2} B_\mu
\,=\, e\, \sum_{\tilde V^a}
\tilde I^{\tilde V^a}\,\tilde V^a_\mu\qquad\mbox{with}\quad
\tilde V^a =W^{1},W^{2},W^{3},B\,.
\eea
Here and in the following the tilde is used to denote 
SU(2)$\times$U(1) eigenstate fields and related quantities.
The electromagnetic coupling is given by
\bea
e \,=\, \cw g_1 \,=\, \sw g_2\,,
\eea
where $\cw=\cos{\theta_\rw}$ and $\sw=\sin{\theta_\rw}$ are the cosine and
sine of the weak mixing angle. 
The generators on the rhs of~\refeq{eq:EWcovdev1} correspond to 
\bea
\tilde I^{W^a} \,=\, \frac{T^a}{\sw}\,,\qquad
\tilde I^{B} \,=\, -\frac{Y}{2\cw}\,.
\eea
Diagrams where a SU(2)$\times$U(1) gauge boson couples to a closed
quark loop give rise to traces of type

\bea
\label{eq:symmtrace}
\Tr\left(\tilde I^{\tilde V^a}
\right) &=&
\Tr\left(\tilde I_\rR^{\tilde V^a}\omega_\rR
+\tilde I_\rL^{\tilde V^a}\omega_\rL
\right)
\,=\,
2\sum_{\lambda=\rR,\rL}
\TrQ\left(\tilde I_\lambda^{\tilde V^a}\right)
\nonumber\\[3mm]
&=&
\begin{cases}
  \;\displaystyle -\frac{4}{\cw}(Q_u+Q_d) & \quad \tilde V^a= B \\[2mm]
\;0 & \quad  \tilde V^a= W^a \\
\end{cases}
\,,
\eea
where $\TrQ$ denotes the trace in the SU(2) quark-doublet space, 
the factor two in the second step arises from the traces of 
$w_\pm$,
and the expressions on the rhs follow from 
$\TrQ(T_\lambda^a)=0$ and $\TrQ(Y_\lambda)=2(Q_u+Q_d)$ for 
$\lambda=\rR, \rL$.
Traces involving a single SU(2)$\times$U(1) generator in combination with $\gamma_5$
vanish,
\bea
\label{eq:symmaxialtrace}
\Tr\left(\tilde I^{\tilde V^a}\gamma_5
\right) \,=\,
2\left[
\TrQ\left(\tilde I_\rR^{\tilde V^a}\right)-
\TrQ\left(\tilde I_\rL^{\tilde V^a}\right)
\right]\,=\,0\,.
\eea
Diagrams where two SU(2)$\times$U(1) gauge boson couple to a closed
quark loop yield traces of type
\bea
\label{eq:symmdoubletrace}
\Tr\left(\tilde I^{\tilde V^a}\tilde I^{\tilde V^b}
\right) &=&
2\sum_{\lambda=\rR,\rL}
\TrQ\left(\tilde I_\lambda^{\tilde V^a}\tilde I_\lambda^{\tilde V^b}\right)
\,=\, 
\delta_{\tilde V^a \tilde V^b}
\times\begin{cases}
\;\displaystyle\frac{1}{\cw^2}\left[4(Q_u^2+Q_d^2)-1\right] 
& \mbox{for} \quad \tilde V^a = B \\[3mm]
\;\displaystyle\frac{1}{\sw^2} 
& \mbox{for}\quad  \tilde V^a = W^a
\end{cases}
\,,
\nonumber\\
\eea
where the expressions on the rhs follow from
\bea
\TrQ\left(T^a_\lambda T^b_\lambda\right)\,=\, 
\frac{\delta_{\lambda \rL}}{2}\,,
\qquad
\TrQ\left(T^a_\lambda Y_\lambda \right)\,=\, 
0\,,\qquad
\TrQ\left(Y^2_\lambda\right)\,=\, 
4 \TrQ\left[Q^2 - \left(T^3_\lambda \right)^2\right]
\,.\qquad
\eea

In the broken phase, EW interactions are parametrised in terms of the
mass- and charge-eigenstate gauge bosons
\bea
V_\mu
 \,=\, \left(W^+_\mu,W^-_\mu,Z_\mu, A_\mu\right)^\rT\,,
\eea
which are related to the gauge-group eigenstates through the mixing
transformation
\bea
V_\mu \,=\, U(\theta_\mathrm{w})  \tilde V_\mu, \qquad
U(\theta_\mathrm{w})\,=\,
 \begin{pmatrix}
    \frac{1}{\sqrt{2}} & \frac{-\mathrm{i}}{\sqrt{2}}& 0 & 0\\
    \frac{1}{\sqrt{2}} & \frac{\mathrm{i}}{\sqrt{2}} & 0 & 0\\
    0 & 0 & \cw  & \sw \\
    0 & 0 & -\sw & \cw  \\
  \end{pmatrix}.
\eea
The EW part of the covariant derivative assumes the form
\bea
\label{eq:EWcovdev2}
g_2
T^b W^{b}_{\mu} - g_1 \frac{Y}{2} B_\mu
\,=\, e\, \sum_{V^a}
I^{V^a}\,V^a_\mu\qquad\mbox{with}\quad
V^a =W^{+},W^{-},Z,A\,,
\eea
where
\bea
I^{V} = \tilde I^{\tilde V}\,
U^{-1}(\theta_\rw)\,,
\eea
and the individual generators read
\bea
\Tgen^A \,=\, -Q, \qquad \Tgen^Z \,=\, \frac{T^3-\sw^2 Q}{\sw \cw}, \qquad
\Tgen^\pm \,=\, \frac{1}{\sw} T^\pm \,=\, \frac{1}{\sw} \frac{T^1 \pm \mathrm{i}
T^2}{\sqrt{2}}\,.
\label{eq:genmasseigenbasis}
\eea
Similarly as in~\refeq{eq:RLfundewgen}, 
the EW generators can be decomposed as
$I^{V_a}= I_\rR^{V_a}\omega_\rR + I_\rL^{V_a}\omega_\rL$, and the
components of their building blocks read
\bea
\label{eq:ewgencomp}
\left(Q_\lambda\right)_{qq'}\,=\,\delta_{qq'}Q_q\,,\quad
\left(T^3_\lambda\right)_{qq'}\,=\,\delta_{\lambda\rL}\delta_{qq'}\,
\frac{\delta_{qu}-\delta_{qd}}{2}\,,\quad
\left(T^+_\lambda\right)_{qq'}\,=\,
\left(T^-_\lambda\right)_{q'q}\,=\,
\delta_{\lambda\rL}
\frac{\delta_{qu}\delta_{q'd}}{\sqrt{2}}\,,
\nonumber\\
\eea
where $q,q'=u$ or $d$.

In the mass-eigenstate basis the 
traces~\refeq{eq:symmtrace} and~\refeq{eq:symmdoubletrace}
become
\bea
\label{eq:masstrace}
\Tr\left(I^{V^a}\right) &=& 
-4(Q_u+Q_d)\times
\begin{cases}
\;1 & \qquad V^a=A\\[1mm]
\;\displaystyle\frac{\sw}{\cw} & \qquad V^a=Z\\[1mm]
\;0 & \qquad V^a=W^\pm\\
\end{cases}\,,
\eea
and
\bea
\label{eq:massdoubletrace}
\Tr\left(I^{V^a}I^{V^b}\right) &=&
  \begin{cases}
  \; %
4 
  \left(Q_{u}^2 + Q_{d}^2\right)
  \,  & \mbox{$V^a V^b=AA$} \\[2mm]
    \; \displaystyle\frac{1}{\sw \cw} 
    \left(4\sw^2 \left(Q_{u}^2 + Q_{d}^2\right) - 1 \right)
  \,  & \mbox{$V^a V^b=AZ, ZA$} \\[2mm]
  \; \displaystyle\frac{1}{\sw^2 \cw^2} 
    \left(4\sw^4 \left(Q_{u}^2 + Q_{d}^2\right) + \cw^2-\sw^2 \right)
  \,  & \mbox{$V^a V^b=ZZ$} \\[2mm]
	\; \displaystyle \f{1}{\sw^2} &
    \mbox{$V^a V^b=W^\pm W^\mp$} \\[3mm]
    \; 0 &     \mbox{otherwise} 
\end{cases}\,.
\eea

\subsection{Treatment of $\gfive$} 
\label{se:gamma5}
As is well known, the treatment of $\gfive$ 
is delicate in dimensional
regularisation (see e.g.~\cite{Jegerlehner:2000dz}) and various schemes have been proposed 
\cite{tHooft:1972tcz,Akyeampong:1973xi,Breitenlohner:1977hr,Bonneau:1980yb,Chanowitz:1979zu,Larin:1993tq,Korner:1989is}.
For the derivation of one- and two-loop rational counterterms 
through~\refeq{eq:r1form} and \refeq{eq:r2form}
we employ the so-called KKS scheme~\cite{Korner:1989is,Kreimer:1993bh,Korner:1991sx}, 
also known as reading-point prescription. In this scheme,
the traces that arise from closed fermion loops are defined
as non-cyclic objects, starting with the same open Lorentz index associated with one of the external
vector bosons. 
In addition to the reading-point prescription,
the properties that define $\gfive$ in the KKS scheme are 
\bea
\{\gamma^\mu,\gfive\} &=& 0,
\label{eq:anticommutingg5}
\\
    \gfive^2 &=& 1 {},\\
 4\mathrm{i} \epsilon_{\mu\nu\alpha\beta} &=&
 \Tr \left[
   \gfive
   \gamma_{\bar\mu}\gamma_{\bar\nu}\gamma_{\bar\alpha}\gamma_{\bar\beta}
   \right],
  \label{eq:tracecondition}
 \\[2mm]
&&\hspace{-14mm}\text{non-cyclic trace},
  \label{eq:noncyclictrace}
\eea
where $\epsilon$ is the four-dimensional 
Levi-Civita tensor with $\epsilon_{0123} = 1 =-\epsilon^{0123}$. 
In the context of the KKS scheme, 
the identity~\eqref{eq:tracecondition}
can be replaced
by 
\bea
\gfive &=&  \f{\ri}{4!} \, \epsilon_{\mu \nu \rho \sigma} \, \gamma^{\mu} \gamma^{\nu} \gamma^{\rho} \gamma^{\sigma}
\,. 
\label{eq:gamma5eps}
\eea

In the general formulas~\refeq{eq:r1form} and \refeq{eq:r2form}
for the derivation of rational counterterms, 
the only ingredients
that require the KKS scheme are
the $D$-dimensional quantities 
$\ampbar{1}{\Gamma}{}{}$,
\mbox{$\deltaZ{1}{\gamma}{}{}\cdot\ampbar{1}{\Gamma/\gamma}{}{}$}
and $\ampbar{2}{\Gamma}{}{}$, while in all other ingredients the loop
numerator, including $\gamma_5$, is four-dimensional.
Concerning the ingredients of the master formulas for renormalised
amplitudes, \refeq{eq:masterformula1} and~\refeq{eq:masterformula2}, 
we note that all loop quantities 
are in $\numdim=4$ dimensions. 
This means that the subtleties
related to the treatment of $\gamma_5$ in dimensional regularisation 
are addressed once and for all 
through the derivation of rational counterterms.\footnote{%
We remind the reader that in this paper we restrict ourselves 
to rational terms of UV origin, while the effect of IR divergences,
including their interplay with $\gamma^5$, is deferred to future studies.
}
We also note that rational counterterms 
are independent of the $\gamma_5$-scheme employed for their derivation.
This is a consequence of the fact that renormalised amplitudes, 
as well as all $\numdim=4$ dimensional 
ingredients on the rhs of~\refeq{eq:masterformula1} 
and~\refeq{eq:masterformula2},
are independent of the $\gamma_5$-scheme.

The derivations of the rational counterterms presented in this paper 
involve at most one quark loop 
containing a $\gfive$. 
Thus in the KKS scheme we never encounter 
more than one Levi-Civita tensor in the loop amplitude.
Furthermore, in our calculations $\eps$ tensors appear 
in the contributions 
of individual up or down fermions\footnote{See e.g.~the 
one-loop rational counterterms for vertices with one 
weak vector bosons and gluons in~\cite{Draggiotis:2009yb}.} 
but cancel when summing over opposite quark flows and 
full SU(2) doublets, which is mandatory for 
UV and anomaly cancellations in the SM.
This observation~\cite{Denner:2019vbn} is consistent with the behaviour of one-loop rational terms 
in the full SM~\cite{Draggiotis:2009yb,Garzelli:2009is}, where traces
like \eqref{eq:tracecondition} drop out when combining loops with quarks and
leptons of the same generation.

To check the correctness of our results we have implemented the KKS scheme in
two different frameworks and independently.
Furthermore we have reproduced our results 
using the naive-dimensional
regularisation scheme as defined by Jegerlehner~\cite{Jegerlehner:2000dz} 
(the quasi self-chiral scheme), 
where one keeps the anti-commuting $\gfive$~\eqref{eq:anticommutingg5},
while giving up on 
\eqref{eq:tracecondition} and 
\eqref{eq:noncyclictrace}, which
are replaced by
\begin{align}
\Tr \left[
   \gfive
   \gamma_{\bar\mu}\gamma_{\bar\nu}\gamma_{\bar\alpha}\gamma_{\bar\beta}
   \right]
 \,=\,0\,,\\
\text{cyclic trace}\,.
\end{align}

\subsection{Rational counterterms}
\label{se:R2results}
In the following, we present all rational 
counterterms for the SM Lagrangian~\refeq{eq:qcdewlag}
at order $\als$ and $\als^2$.
As usual UV singularities are regularised in $\dendim=4-2\eps$ dimensions. 
The rational terms associated with a certain 1PI vertex function 
$\Gamma$ are presented in the form 
\bea
\label{eq:qed4}
\ratamp{k}{\Gamma}{\alpha_1\dots\alpha_N}{} &=& \ri
\lb\frac{\als\, t^\eps}{4\pi}\rb^k 
\sum_{a} 
\delta \hat\calR^{(a)}_{k,\Gamma}\,\,
\calT_{a,\Gamma}^{\alpha_1\dots\alpha_N}\,
\,,
\eea
where $k=1,2$ is the loop order, 
and
$\calT_{a,\Gamma}^{\alpha_1\dots\alpha_N}$ are independent tensor structures
carrying the 
indices $\alpha_1\dots\alpha_N$ of the external lines of the
vertex function at hand.
Our results are presented in the form of 
complete Feynman rules, where tree-level contributions are
supplemented by UV and rational counterterms.
For $k$-loop UV counterterms
$\deltaZ{k}{\Gamma}{\alpha_1\dots\alpha_N}{}$ 
a decomposition similar to~\refeq{eq:qed4} is used.

In the following all Dirac and colour indices associated with external quark lines are 
kept implicit. Fermion-loop contributions are either expressed as
sums over $q\in \calQ$, where the set $\calQ$ contains all 
$\nq=2N_\gen$ quarks that circulate in the loops, or are simply
proportional to $\nq$.

\subsubsection{Vertices involving SU(2)$\times$U(1) vector bosons}
\label{se:EWbosonvert}

The interaction of SU(2)$\times$U(1) gauge bosons with quarks and gluons 
gives rise to non-vanishing rational counterterms 
for the $GGVV$, $GGGV$ and $QQV$  vertices.
Rational counterterms of this kind are 
free from any quark-mass dependence, i.e.~they can be derived within the
symmetric phase of the SM using massless quarks.
The results are expressed in terms of the generic
SU(2)$\times$U(1) generators introduced in~\refse{se:EWbookkeeping}
and are applicable both to the mass-eigenstate and 
gauge-eigenstate representation.


\vskip 5mm
\subsubsection*{Four-point GGVV vertex}
The Feynman rules for the interaction of two vector bosons and two gluons
are
\bea
\vcenter{\hbox{\raisebox{0pt}{\includegraphics[width=0.22\textwidth]{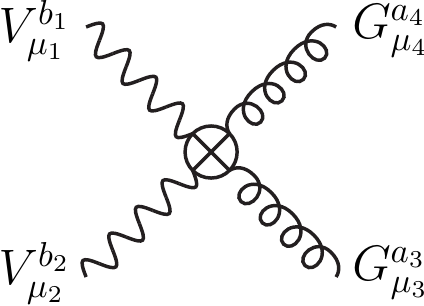} }}} 
\;  &= & \; \; {}
\ri  \,
\delta_{a_3 a_4} \, 
e^2
\,
\frac{N_{\gen}}{2}\,\Tr\left(I^{V^{b_1}}I^{V^{b_2}}\right)
\nonumber\\[-5mm]&&{}\times
  \, \Bigg\{ \,
\sum_{k=1}^2 \lb \f{\als \, t^{\eps}}{4\pi}  \rb^k  
\bigg[
\sum_{\beta=\rI,\rII} 
\calV_{\beta}^{\mu_1\mu_2\mu_3\mu_4} \, \delta \hat \calR^{(\beta)}_{k,
\mathrm{ggVV}} %
 \bigg]
\Bigg\}
\,,
\label{eq:RggVV}
\eea
with
\bea
\label{eq:ggVVstr}
\calV_{\rI}^{\mu_1\mu_2\mu_3\mu_4} &=& g^{\mu_1\mu_2}g^{\mu_3\mu_4}\,,
\qquad
\calV_{\rII}^{\mu_1\mu_2\mu_3\mu_4} \,=\, 
g^{\mu_1 \mu_3}g^{\mu_2\mu_4} + g^{\mu_1 \mu_4}g^{\mu_2\mu_3}\,.
\eea
Explicit expressions for the traces 
$\Tr(I^{V^{b_1}}I^{V^{b_2}})$
in the gauge- and mass-eigenstate basis are given 
in~\refeq{eq:symmdoubletrace} and \refeq{eq:massdoubletrace}.
For the rational counterterms we find
\begin{align}
\delta \hat \calR_{1,\mathrm{ggVV}}^{(\rI)}   \;=\; & \f{2}{3} \, \TF \,, \nonumber \\[2mm]         
\delta \hat \calR_{2,\mathrm{ggVV}}^{(\rI)}   \;=\; &
    - \TF \, \lb \frac{1}{12} \, \CA  + \frac{3}{2} \, \CF  \rb 
       + \f{2}{3}  \, \TF \, \lb    \, \dcalZ_{1,\als}  +     \, 
\dcalZ_{1,G}
\rb \,,  \\[2mm]
\delta \hat \calR_{1,\mathrm{ggVV}}^{(\rII)}   \;=\; & \f{2}{3} \, \TF \,, \nonumber \\[4mm]         
\delta \hat \calR_{2,\mathrm{ggVV}}^{(\rII)}   \;=\; &
   \left[
   \frac{1}{2} \, \TF   \, \CA    
   \right]\eps^{-1}
   - \TF \, \lb \frac{1}{2} \, \CA  + \frac{3}{2} \, \CF  \rb
    + \f{2}{3}  \, \TF \, \lb    \, \dcalZ_{1,\als}  +     \,
\dcalZ_{1,G}
\rb \,. \nonumber \\
\end{align}
We note that $\delta \hat \calR_{1,\mathrm{ggVV}}^{(\rI)}=
\delta \hat \calR_{1,\mathrm{ggVV}}^{(\rII)}$,
which implies that the one-loop part of the rational counterterm~\refeq{eq:RggVV}
is totally symmetric in the four Lorentz indices 
$\mu_1,\mu_2,\mu_3,\mu_4$. This is due to the fact that 
the four external vector bosons can be connected 
through fermionic loops in all permutations.
At two loop-level this symmetry is still present 
in the contributions proportional to $\CF$, while it is broken 
by the non-abelian contributions proportional to 
$C_A$. The latter involve virtual gluons that couple to fermionic loops and
the external gluon lines, thereby breaking the symmetry between external
gluons and electroweak bosons.

\vskip 5mm
\subsubsection*{Four-point $GGSV$ vertex}
The quartic vertex with two gluons, a vector boson and a scalar
vanishes,\\[0mm]
\bea
\label{eq:GGSVvertcanc}
\vcenter{\hbox{\raisebox{0pt}{\includegraphics[width=0.22\textwidth]{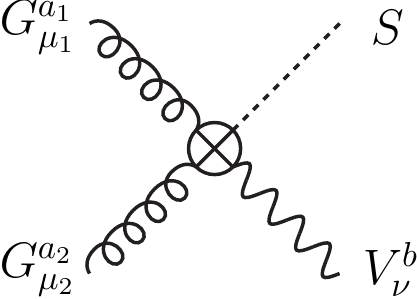}} }} \hspace{-2mm}
&\,=\,& 0\,.
\eea
This is due to the fact that the $GGSV$ vertex has a superficial degree of
divergence equal to zero. This would require 
a dimensionless object carrying three Lorentz indices, which does not exist.

\vskip 5mm
\subsubsection*{Four-point GGGV vertex}
The Feynman rules for the interaction of a vector boson with three gluons
are
\begin{align}
\vcenter{\hbox{\raisebox{0pt}{\includegraphics[width=0.22\textwidth]{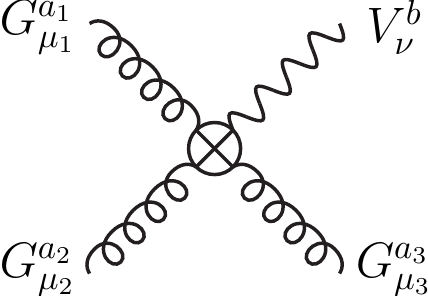} }}} 
\;  = & \; \; {}
\ri  \, \gs  \, 
e\,\frac{N_{\gen}}{2}\,\Tr\left(I^{V^b}\right)\,
d^{\, a_1 a_2 a_3} \,
\calV^{\mu_1\mu_2\mu_3\nu}
\, \Bigg\{ \,
\sum_{k=1}^2 \lb \f{\als \, t^{\eps}}{4\pi}  \rb^k   
\, \delta \hat \calR^{}_{k, \mathrm{gggV}} \Bigg\}\,,  \nonumber \\[-5mm]
%
\end{align}
with
\bea
d^{a b c } = \f{1}{\TF} \, \Tr \lb \left\{ t^{a},t^{b}\right\} \, t^{c} \rb 
\,,
\eea
and
\bea
\label{eq:gggVstra}
\calV^{\mu_1\mu_2\mu_3\nu} &=& 
g^{\mu_1\mu_2}g^{\mu_3\nu} +  g^{\mu_1\mu_3}g^{\mu_2 \nu} +
g^{\mu_2\mu_3}g^{\mu_1\nu}\,.
\eea
Explicit expressions for the traces $\Tr(I^{V^b})$ in the 
gauge- and mass-eigenstate basis are given 
in~\refeq{eq:symmtrace} and \refeq{eq:masstrace}.
For the rational terms we find 
\begin{align}
\delta \hat \calR_{1,\mathrm{gggV}}^{}   \;=\; & \f{1}{3} \, \TF \,, \nonumber \\[2mm]         
\delta \hat \calR_{2,\mathrm{gggV}}^{}   \;=\; &
 \f{1}{2} \, \TF \, \lb \,  \frac{1}{2} \, \CA  \, \eps^{-1} -  \frac{13}{24} \, \CA
- \f{3}{2} \, \CF  
        +     \, \dcalZ_{1,\als}  +     \, 
\dcalZ_{1,G}
\rb  \,.
\end{align}


\vskip 5mm
\subsubsection*{Three-point $GGV$ vertex}

The counterterm for the interaction of a vector boson with two gluons
vanishes,\\[0mm]
\bea
\vcenter{\hbox{\raisebox{0pt}{\includegraphics[width=0.22\textwidth]{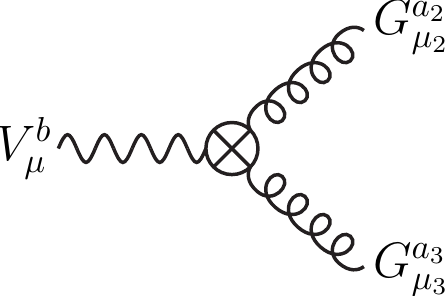}} }} \hspace{-2mm}
&\,=\,& 0\,.
\eea
This cancellation results form the vanishing fermion traces
in~\refeq{eq:symmtrace}--\refeq{eq:symmaxialtrace}, and from the fact that also the 
trace of the U(1) generator in~\refeq{eq:symmtrace}
cancels upon summation over quarks and anti-quarks in the loops.
Note that this cancellation mechanism assumes complete SU(2)
quark doublets.

\vskip 5mm
\subsubsection*{Three-point $QQV$ vertex}
The Feynman rules for the SU(2)$\times$U(1) quark interaction read
\bea
 \vcenter{\hbox{\raisebox{0pt}{\includegraphics[width=0.22\textwidth]{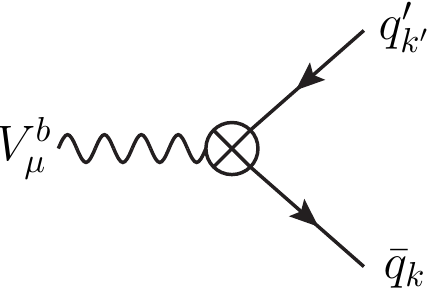}} }}
\;  &= & \; \; {}
  \ri\, \e 
 \,
\gamma^{\mu}\,  \delta_{kk'}
\sum_{\lambda=\rR,\rL}
\left(I^{V^b}_\lambda \right)_{qq'} \omega_\lambda
\, 
\nonumber\\[-5mm]
&&{}\times\bigg\{ \, 1 +
\sum_{k=1}^2 \lb \f{\als \, t^{\eps}}{4\pi}  \rb^k 
\lb \delta \hat Z^{}_{k, qq' \rV} + \delta \hat \calR^{}_{k, q q' \rV} \rb \bigg\}
\,,
\eea
where $q,q'\in\{u,d\}$ and $k,k'\in\{1,\dots, N_\gen\}$ are the generation indices.
The explicit components of the EW generators $I^{V^b}_\lambda$ in the gauge-eigenstate and 
mass-eigenstate basis can be found in~\refse{se:EWbookkeeping}.
The UV counterterms correspond to the following combinations of renormalisation
constants,
\begin{align}
\delta \hat  Z_{1,q q' \rV}^{}  \;=\; &  
\f{1}{2}  \dcalZ_{1,\als}  + \f{1}{2} \lb \delta\hat \calZ_{1,q} + \delta\hat \calZ_{1,q'} \rb
\,,    
\nonumber \\
\delta \hat  Z_{2, q q' \rV}^{}   \;=\; &
\f{1}{2}  \dcalZ_{2,\als}   + \f{1}{2} \lb  \delta\hat \calZ_{2,q}  + \delta\hat \calZ_{2,q'} \rb
 -\frac{1}{8}  \dcalZ_{1,\als}^2  
+\f{1}{4} \lb \dcalZ_{1,q}  + \delta\hat \calZ_{1,q'} \rb  \dcalZ_{1,\als}  
\,,
\end{align}
and for the rational counterterms we obtain
\begin{align}
\delta \hat \calR_{1,q q' \rV}^{}   \;=\; & - 2 \, \CF \,, \nonumber \\[2mm]         
\delta \hat \calR_{2, q q' \rV}^{}   \;=\; &
   \left[
   - \frac{29}{9}  \, \CA \, \CF + \frac{4}{3} \, \CF^2 +\frac{7}{9} \, \TF \, \nq  \, \CF    
   \right]\eps^{-1}
    -\frac{763}{108} \, \CA \, \CF +\frac{109}{18} \, \CF^2  
+\frac{55}{54} \, \TF \, \nq \, \CF  \nonumber  \\
&
   - \, \CF \, \lb  2   \, \dcalZ_{1,\als} 
          +   \frac{1}{2}    \, \dcalZ_{1,G} \nonumber  -   \frac{5}{6}   \, \dcalZ_{1,\gpar}   
         + \frac{2}{3}  \, \lb \dcalZ_{1,q} + \delta\hat \calZ_{1,q'} \rb \rb \,. \nonumber \\[-2mm]
\end{align}

\vskip5mm
\subsubsection{Vertices involving scalar bosons}
\label{se:scalarvert}

The Yukawa interactions of the scalar doublet 
with external quarks and closed quark loops
give rise to non-vanishing rational counterterms 
for the vertices of type $GGSS$, $GGS$ and $QQS$.
The $GGSS$ and $QQS$ rational counterterms 
depend only on the dimensionless Yukawa couplings
and can be derived within the symmetric phase of the SM using massless quarks.
The $GGS$ counterterm depends on the vev 
and can be determined in the broken phase, with massive quarks, 
or using vev expansions.

\vskip 5mm
\subsubsection*{Four-point $GGSS$ vertex}
The Feynman rules for the interaction of two gluons with two scalar fields
are
\begin{align}
\label{eq:GGSSverta}
&\vcenter{\hbox{\raisebox{12pt}{\includegraphics[width=0.22\textwidth]{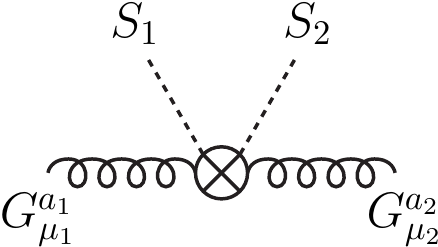}} }}  
\;=\; {}\ri \, \delta^{a_1 a_2} \, C_{\mathrm{gg}S_1 S_2}
\, g^{\mu_1\mu_2} 
\bigg\{   \, \sum_{k=1}^2 \lb \f{\als \, t^\eps}{4\pi}\rb^k  
 \delta \hat \calR^{}_{k,\mathrm{gg\rH\rH}} 
\bigg\}
\,,  
\end{align}
with 
\bea
\label{eq:GGSSvertb}
C_{\mathrm{gg}S_1 S_2} &=&
\begin{cases}
	\; {\displaystyle 1 }& \qquad S_1S_2 = HH,
\chi\chi,\phi^\pm\phi^\mp \\[2mm]
\; {\displaystyle 0} &\qquad\mbox{otherwise} 
\end{cases}\,.
\eea
For the rational counterterms we find
\begin{align}
\label{eq:GGSSvertc}
\delta \hat \calR_{1,\mathrm{gg\rH\rH}}^{} \;=\; &   
             -2 \, \TF \sum_{q\in \calQ}   \lambda_q^2
             \,,              \nonumber \\[2mm]
\delta \hat \calR_{2,\mathrm{gg\rH\rH}}^{}  \;=\;  
 &   
     -\TF \sum_{q\in \calQ}\,  \left[ 
    \left( \f{1}{2} \,  \CA  + 6 \,  \CF  \right) \eps^{-1} 
    + \frac{11}{24}  \, \CA  - \f{37}{4} \, \CF
    \right] \, \lambda_q^2
  \nonumber \\
&
        	-2 \, \TF  \sum_{q\in\calQ} 
          \left[
           \dcalZ_{1,\als} 
          +  \delta \hat \calZ_{1,G}
          + 2 \,\delta \hat \calZ_{1,\lambda_q} 
           \right] \, \lambda_q^2 \,. 
\end{align}

\vskip 5mm
\subsubsection*{Four-point $GGGS$ vertices}
The quartic vertex with three gluons and a scalar boson vanishes,\\[1mm]
\bea
\label{eq:GGGSvertcanc}
\vcenter{\hbox{\raisebox{0pt}{\includegraphics[width=0.22\textwidth]{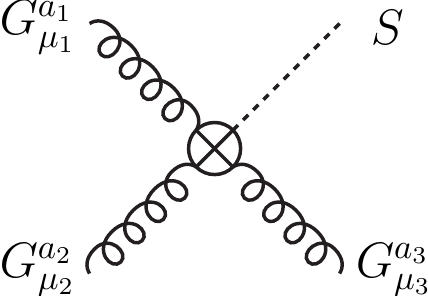}} }} \hspace{-2mm}
&\,=\,& 0\,.
\eea
The reason is the same as for~\refeq{eq:GGSVvertcanc}.


\vskip 5mm
\subsubsection*{Three-point $GGS$ vertex}
The Feynman rules for the interactions of a scalar with two gluons are
\begin{align}
\label{eq:GGSratctres}
&\vcenter{\hbox{\raisebox{12pt}{\includegraphics[width=0.22\textwidth]{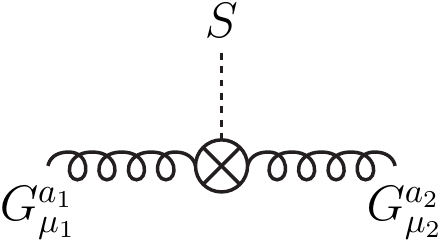}} }}   
\;=\; {}\ri \, \delta_{S H}\,
\delta^{a_1 a_2} \, v \, g^{\mu_1\mu_2} \,
\bigg\{   \, \sum_{k=1}^2 \lb \f{\als \, t^\eps}{4\pi}\rb^k  
\delta \hat \calR^{}_{k,\mathrm{gg\rH}} 
\bigg\}
\,.  
\end{align}
Note that only the 
Higgs boson, $S=H$, yields a non-zero rational counterterm.
The result is proportional to the vacuum expectation value, 
\ie it vanishes in the symmetric phase.
In the broken phase, \ie using non-zero quark masses, we find
\begin{align}
\label{eq:ggSbrokctres}
\delta \hat \calR_{1,\mathrm{gg\rH}} \;=\; &   
             -2 \, \TF \sum_{q\in \calQ}   \lambda_q^2
             \,,              \nonumber \\[2mm]
\delta \hat \calR_{2,\mathrm{gg\rH}}  \;=\;  
 &   
     -\TF \sum_{q\in \calQ}\,  \left[ 
    \left( \f{1}{2} \,  \CA  + 6 \,  \CF  \right) \eps^{-1} 
    + \frac{19}{12}  \, \CA  - 7 \, \CF
    \right] \, \lambda_q^2
  \nonumber \\
&
        	-2 \, \TF  \sum_{q\in\calQ} 
          \left[
           \dcalZ_{1,\als} 
          +  \delta \hat \calZ_{1,G}
          + 2 \,\delta \hat \calZ_{1,\lambda_q} 
           \right] \, \lambda_q^2 
           \,. 
\end{align}
These results have been cross-checked against calculations in the
symmetric phase using vev expansions. For the vertex
at hand, the vev-expansion formula~\refeq{eq:r2vevinsg} reads
\bea
\label{eq:GGSvevexpis}
\delta\calR^{}_{k,\mathrm{gg}\rH} &=& 
\delta\calR^{\ubk}_{k,\mathrm{gg}\rH} + 
\delta\calR^{\ubk}_{k,\mathrm{gg}\rH\vev} + 
\delta\calR^{\ubk}_{k,\mathrm{gg}\rH\tvev}
\,=\,
\vev\,\delta\calR^{\ubk}_{k,\mathrm{gg}\rH\rH} + 
\delta\calR^{\ubk}_{k,\mathrm{gg}\rH\tvev} \,,
\eea
where $\delta\calR^{\ubk}_{k,\mathrm{gg}\rH\rH}=\delta\calR^{}_{k,\mathrm{gg}\rH\rH}$ is given in 
\refeq{eq:GGSSverta}--\refeq{eq:GGSSvertc}. 
For the additional 
$\delta\calR^{\ubk}_{k,\mathrm{gg}\rH\tvev}$ contribution, which is 
defined by the general identities~\refeq{eq:r2vevinsone}, \refeq{eq:r2vevinsf} and 
the rules for $\tvev$ insertions~\refeq{eq:v1ct}--\refeq{eq:VDVexpD4}, we find
\bea
\label{eq:GGStvevverta}
\delta\calR^{\ubk}_{k,\mathrm{gg}S\tvev}\; &=&
\vcenter{\hbox{\raisebox{12pt}{\includegraphics[width=0.22\textwidth]{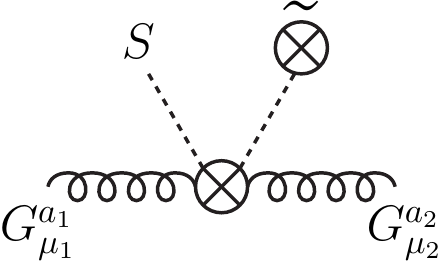}} }}   
\;=\; {}\ri \, \delta_{S H}\,
\delta^{a_1 a_2}\,\vev  
\, g^{\mu_1\mu_2} \,
\bigg\{   \, \sum_{k=1}^2 \lb \f{\als \, t^\eps}{4\pi}\rb^k  
 \delta \hat \calR^{\ubk}_{k,\mathrm{gg\rH} \tvev} 
\bigg\}
,  \qquad
\eea
with
\begin{align}
\label{eq:GGStvevvertb}
\delta \hat \calR_{1,\mathrm{gg\rH} \tvev}^{\ubk} \;=\; &   
             0
             \,,              \nonumber \\[2mm]
\delta \hat \calR_{2,\mathrm{gg\rH}  \tvev}^{\ubk}  \;=\;  
 &   
     -\TF \sum_{q\in \calQ}\,  \left[ \,
     \f{9}{8} \,  \CA  + \f{9}{4} \,  \CF   \,    \right] \, \lambda_q^2
   \,. 
\end{align}
Note that  $\delta\calR^{\ubk}_{k,\mathrm{gg}\rH\tvev}\neq 0$ in spite of the
fact that  $\delta\calR^{\ubk}_{k,\mathrm{gg}\rH}=0$ in the 
symmetric phase.
Rewriting the identity~\refeq{eq:GGSvevexpis} in terms
of the $\delta \hat\calR$  coefficients on the rhs of
\refeq{eq:GGSratctres}, \refeq{eq:GGSSverta} and~\refeq{eq:GGStvevverta}
we have
\bea
\delta\hat \calR^{}_{k,\mathrm{gg\rH} } &=& 
\delta\hat \calR^{}_{k,\mathrm{gg\rH\rH} } + 
\delta\hat \calR^{\ubk}_{k,\mathrm{gg\rH} \tvev } \,,
\eea
and combining the explicit ingredients of the vev expansion, 
\refeq{eq:GGSSvertc} and~\refeq{eq:GGStvevvertb}, 
we find agreement with the result of the 
explicit derivation in the broken phase~\refeq{eq:ggSbrokctres}.



\vskip 5mm
\subsubsection*{Three-point $QQS$ vertex}
The Feynman rules for the Yukawa interactions of quarks read
\bea
\label{eq:QQSverta}
\vcenter{\hbox{\raisebox{22pt}{\includegraphics[width=0.22\textwidth]{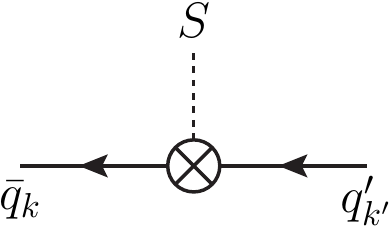}} }}   
&=& - \ri  \,  \delta_{kk'} \, \sum_{\lambda=\rR,\rL} 
C^{\lambda}_{qq' S}\,\omega_\lambda
\nonumber\\[-5mm]
&&{}\times\,
 \bigg\{ \, 1 +
  \sum_{k=1}^2 \lb\frac{\als \, t^{\eps}}{4\pi} \rb^k
  \bigg[
  \delta \hat Z^{}_{k,q q' \rH}
 +  \delta \hat\calR^{}_{k,q q' \rH }
   \bigg]
\bigg\} \,,
\eea
where $S= H,\chi, \phi^\pm$ are the components of the scalar doublet, 
while $q,q'\in\{u,d\}$, and 
$k,k'\in\{1,\dots, N_\gen\}$ are generation indices.
The tree-level coupling factors read\footnote{Alternatively, Yukawa
couplings can be written in the more generic form
\bea
\renewcommand{\arraystretch}{1.8}
\begin{array}{c@{\quad}
||@{\quad}c@{\quad}|@{\quad}c@{\quad}
}
\bar q q' S 
                 & \bar Q_i Q_j \Phi_k
                 & \bar Q_i Q_j \Phi^\dagger_k
\\\hline
C^{\rR}          
                 & \lambda_d \delta_{2i}\delta_{jk}
                 & \lambda_u \delta_{1i}\delta_{jk}
\\\hline 
C^{\rL}
                 & \lambda_u\delta_{1j}\eps_{ik}
                 & \lambda_d\delta_{2j}\eps_{ik}
\end{array}
\eea
where $i,j,k\in\{1,2\}$ are the components of the quark doublet
$Q=(u,d)^\rT$ and of the scalar doublets 
$\Phi=(\phi^+, \phi_0)^\rT$
and
$\Phi^\dagger=(\phi^-, \phi^\dagger_0)^\rT$.
Here  $\eps_{ij}=\ri\sigma^2_{ij}$
is the two-dimensional Levi-Civita symbol 
with sign convention $\epsilon^{12}=+1$.
}
\bea
\renewcommand{\arraystretch}{1.8}
\begin{array}{c@{\quad}
||@{\quad}c@{\quad}|@{\quad}c@{\quad}|@{\quad}c@{\quad}|@{\quad}c@{\quad}
}
\bar q q' S 
                 & \bar q q H                   
                 & \bar q q\chi 
                 & \bar u d \phi^+ 
                 & \bar d u \phi^-  
\\\hline
C^{\rR}          
                 & \frac{\lambda_q}{\sqrt{2}}  
                 & s_q \frac{\ri\lambda_q}{\sqrt{2}}  
                 & \lambda_d 
                 & -\lambda_u 
\\\hline 
C^{\rL}
                 & \frac{\lambda_q}{\sqrt{2}}  
                 & -s_q \frac{\ri\lambda_q}{\sqrt{2}}  
                 & -\lambda_u 
                 & \lambda_d 
\end{array}
\eea
where $s_q$ assumes the values $s_u=1$ and $s_d=-1$.
The UV counterterms read
\renewcommand{\arraystretch}{1.5}
\bea
\label{eq:QQSvertc}
\delta \hat Z_{1,q q' \rH}^{}   &=&
\frac{1}{2}\left(\delta\hat\calZ_{1,q}+\delta\hat\calZ_{1,q'}\right)
+\delta\hat\calZ_{1,qq'}^{(\rYuk)}\,,  
\nonumber\\
\delta \hat Z_{2,q q' \rH}^{}   &=&
\frac{1}{2}\left(\delta\hat\calZ_{2,q}+\delta\hat\calZ_{2,q'}\right)
+\delta\hat\calZ_{2,qq'}^{(\rYuk)}\,  
+\frac{1}{2}\left(\delta\hat\calZ_{1,q}+\delta\hat\calZ_{1,q'}\right)
\,\delta\hat\calZ_{1,qq'}^{(\rYuk)}\,,
\eea
where
\bea
\label{eq:QQSvertd}
\delta\hat\calZ_{k,qq'}^{(\rYuk)}&=&
\delta\hat\calZ_{k,\lambda_{q'}}\omega_\rR
+
\delta\hat\calZ_{k,\lambda_{q}}\omega_\rL\,,
\eea
and for the rational counterterms we obtain
\renewcommand{\arraystretch}{1.5}
\bea
\label{eq:QQSverte}
\delta \hat \calR_{1,q q' \rH }^{}   &=&
-4 \CF
  \,, \nonumber\\[2mm] 
\delta \hat \calR_{2,qq' \rH }^{}  &=&  
\left( \f{16}{3} \, \CF^2 - \frac{293}{36} \, \CA \, \CF + \frac{19}{9} \, \TF
\, \nq \, \CF  \right)  \eps^{-1} 
+ \f{20}{9} \, \CF^2 -\frac{2237}{216} \, \CA \, \CF 
\nonumber \\
&&{}+ \frac{91}{54} \, \TF \, \nq \, \CF 
-\CF \Big[ 4 \, \dcalZ_{1,\als}  + \f{2}{3} \, 
\left(\delta\hat\calZ_{1,q}+\delta\hat\calZ_{1,q'}\right)
  + 4 \, 
\delta\hat\calZ_{1,qq'}^{(\rYuk)}
\nonumber \\ &&{}
  - \f{1}{2}\dcalZ_{1,G} - \frac{5}{6} \, \dcalZ_{1,\gpar}\Big]
 \,.
\eea

\vskip 5mm
\subsubsection{Quark and gluon two-point vertices}
\label{se:massdepvert}
The QCD rational counterterms for quark and gluon
two-point vertices depend on the masses of quarks that circulate in the
loops.
Explicit derivations with massive quarks have been presented 
in~\cite{Lang:2020nnl}. In the following we show 
that such counterterms 
can be reproduced by means of vev expansions.


\vskip 5mm
\subsubsection*{Two-point $Q\bar Q$ vertex}
The Feynman rule for the two-point function of a massive quark reads
\begin{align}
& \; \vcenter{\hbox{\raisebox{-28pt}{\includegraphics[width=0.22\textwidth]{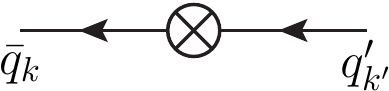}} }} 
 = \; \ri  \,  \delta_{kk'}\delta_{qq'} 
\, \bigg\{ \,\lb \slashed p - m_q \rb
\nonumber \\[2mm]
&\qquad + \,
\sum_{k=1}^2 \lb\frac{\als \, t^{\eps}}{4\pi} \rb^k 
\bigg[
\lb \delta \hat Z^{(\srp)}_{k,qq} + \delta \hat\calR^{(\srp)}_{k,qq} \rb \,
\slashed p
\,
+
\lb
  \delta \hat Z^{(\srm)}_{k,qq}
 +   \delta \hat\calR^{(\srm)}_{k,qq}
  \rb 
  m_q
\bigg] \bigg\} \,,
\label{eq:R2quark}
\end{align}
where $q,q'\in\{u,d\}$, and 
$k,k'\in\{1,\dots, N_\gen\}$ are generation indices.
The direction of the quark momentum $p$ is aligned with the fermion flow.
For the UV counterterms we have
\renewcommand{\arraystretch}{1.5}
\begin{align}
\delta \hat Z_{1,qq}^{(\srp)} &  \;=\;  
\delta\hat\calZ_{1,q}\,,                            &\quad  
\delta \hat Z_{2,qq}^{(\srp)}   \;=\; & 
\delta\hat \calZ_{2,q}\,,    \nonumber \\[2mm]
\delta \hat Z_{1,qq}^{(\srm)} &  \;=\;  
-\delta\hat\calZ_{1,q}-\delta\hat\calZ_{1,\lambda_q}\,,  &\quad  
\delta \hat Z_{2,qq}^{(\srm)}   \;=\; & 
-\delta\hat\calZ_{2,q}-\delta\hat\calZ_{2,\lambda_q}
-\delta\hat\calZ_{1,q}\,\delta\hat\calZ_{1,\lambda_q}\,,
\end{align}
where $\dcalZ_{k,\lambda_q}= \dcalZ_{k,m_q}$.
For the rational counterterms, the direct derivation with massive quarks
yields~\cite{Lang:2020nnl}
\renewcommand{\arraystretch}{1.5}
\begin{align}
\label{eq:QQR2ctP}
\delta \hat \calR_{1,qq}^{(\srp)}   \;=\; & - \CF \,, \nonumber\\[2mm] 
\delta \hat \calR_{2,qq}^{(\srp)}   \;=\; &
\left(\frac{7}{6}\, \CF^2 -\frac{61}{36} \, \CA \, \CF+\frac{5}{9} \,  \TF \, \nq \, \CF  \right)  \eps^{-1}  +\frac{43}{36}\, \CF^2 - \frac{1087}{216} \, \CA \, \CF+\frac{59}{54} \,  \TF \, \nq \, \CF   \nonumber \\
& - \CF \left( \dcalZ_{1,\als} + \frac{2}{3} \, \dcalZ_{1,q}-\frac{2}{3} \, \dcalZ_{1,\gpar}\right) \,,
\end{align}
and
\begin{align}
\label{eq:QQR2ctM}
\delta \hat \calR_{1,qq}^{(\srm)}   \;=\; &   2 \, \CF \,, \nonumber\\[2mm] 
\delta \hat \calR_{2,qq}^{(\srm)}   \;=\; &  
\left(-2 \, \CF^2+\frac{61}{12} \, \CA \, \CF - \frac{5}{3} \, \TF \, \nq \, \CF  \right)  \eps^{-1} + \CF^2+\frac{199}{24} \, \CA \, \CF-\frac{11}{6} \, \TF \, \nq \, \CF \nonumber \\
& +\CF \left( 2 \, \dcalZ_{1,\als}+ 4 \, 
\dcalZ_{1,\lambda_q}
-\frac{3}{2} \, \dcalZ_{1,G}-\frac{1}{2} \, \dcalZ_{1,\gpar}\right)\,.
\end{align}
The momentum-dependent contributions~\refeq{eq:QQR2ctP} do not depend on
quark masses and can be derived in the symmetric phase, using massless quarks.
The mass-dependent contributions~\refeq{eq:QQR2ctM} can be 
cross-checked against calculations in the
symmetric phase using vev expansions. For the vertex
at hand, the vev-expansion formula~\refeq{eq:r2vevinsg} yields the
mass-dependent contributions
\bea
\label{eq:QQvevexpis}
\delta\calR^{}_{k,qq}\bigg|_{p=0}  &=& 
\delta\calR^{\ubk}_{k,qq\vev} + 
\delta\calR^{\ubk}_{k,qq\tvev}
\,=\,
\vev\, \delta\calR^{\ubk}_{k,qq\rH} + 
\delta\calR^{\ubk}_{k,qq\tvev}\,,
\eea
where $\delta\calR^{\ubk}_{k,qq\rH}$ is given
in~\refeq{eq:QQSverta}--\refeq{eq:QQSverte}.
For the additional $\tvev$-insertion term 
$\delta\calR^{\ubk}_{k,qq\tvev}$, which is 
defined by~\refeq{eq:v1ct}--\refeq{eq:VDVexpD4}, \refeq{eq:r2vevinsone} and~\refeq{eq:r2vevinsf}, 
we find
\bea
\label{eq:ffsr2vtilde}
\delta\calR^{\ubk}_{k,qq\tvev}\;=\;
\vcenter{\hbox{\raisebox{22pt}{\includegraphics[width=0.22\textwidth]{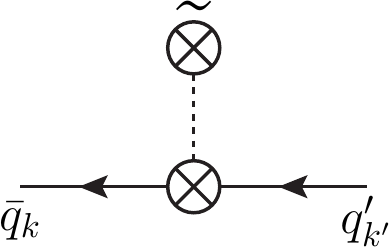}} }}   
\;=\;  - \ri  \,  \delta_{kk'}
\delta_{q q'} 
\, 
  \frac{v\, \lambda_{q}}{\sqrt{2}}\, 
  \,  \bigg\{ \, 
\sum_{k=1}^2 \lb\frac{\als \, t^{\eps}}{4\pi} \rb^k 
   \delta \hat\calR^{\ubk}_{k,qq \tvev } \,
 \bigg\} \,,
\eea
with
\renewcommand{\arraystretch}{1.5}
\begin{align}
\label{eq:QQtvevvert}
\delta \hat \calR_{1,qq \tvev }^{\ubk}   \;=\; &   2 \, \CF \,, \nonumber\\[2mm] 
\delta \hat \calR_{2,qq \tvev }^{\ubk}   \;=\; &  
-\left( \f{10}{3} \, \CF^2 - \frac{55}{18} \, \CA \, \CF + \frac{4}{9} \, \TF \, \nq \, \CF  \right)  \eps^{-1} 
- \f{29}{9} \, \CF^2 + \frac{223}{108} \, \CA \, \CF + \frac{4}{27} \, \TF \, \nq \, \CF \nonumber \\
& -\CF \left( - 2 \, \dcalZ_{1,\als}  - \frac{4}{3} \, \dcalZ_{1,q} - \dcalZ_{1,G} + \frac{1}{3} \, \dcalZ_{1,\gpar}\right) 
 \,.
\end{align}
Rewriting~\refeq{eq:QQvevexpis} in terms
of the $\delta \hat\calR$  coefficients on the rhs of~\refeq{eq:QQSverta} and~\refeq{eq:ffsr2vtilde}
we have
\bea
\label{eq:QQmvevexpis}
\delta\hat \calR^{(\srm)}_{k,{qq} } &=& 
-\left(\delta\hat \calR^{}_{k,{qq}\rH} + 
\delta\hat \calR^{\ubk}_{k,{qq} \tvev }\right) \,,
\eea
and combining the explicit ingredients of the vev expansion, 
\refeq{eq:QQSverte} and~\refeq{eq:QQtvevvert}, 
we find agreement with the result of the 
explicit derivation in the broken phase~\refeq{eq:QQR2ctM}.


\vskip 5mm
\subsubsection*{Two-point $GG$ vertex}

In the $\xi=1$ gauge, 
the Feynman rule for the gluon two-point vertex reads
\begin{align}
\vcenter{\hbox{\raisebox{-33pt}{\includegraphics[width=0.22\textwidth]{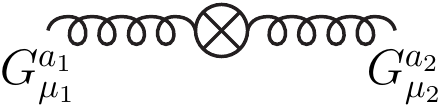}} }} 
  \;=\;& {}\ri \, \delta^{a_1 a_2} \bigg\{  -p^2 g^{\mu_1\mu_2}
+
  \, \sum_{k=1}^2 \lb \f{\als \, t^\eps}{4\pi}\rb^k  
\bigg[ \lb \delta \hat Z^{(\srp)}_{k,\mathrm{gg}} + \delta \hat \calR^{(\srp)}_{k,\mathrm{gg}}  \rb
\, p^{\mu_1}p^{\mu_2}  
 \nonumber\\[2mm]
&\qquad
+\lb \delta \hat Z^{(\srG)}_{k,\mathrm{gg}}  p^2
+ \delta \hat \calR^{(\srG)}_{k,\mathrm{gg}}\, p^2
+ \delta \tilde{Z}^{(\srG)}_{k,\mathrm{gg}} \, \tilde p^2
+ 
\vev^2
\delta \hat \calR^{(\srm)}_{k,\mathrm{gg}}
\rb \, g^{\mu_1\mu_2} 
\bigg]\bigg\}
\,. 
\label{eq:R2gluon}
\end{align}
The coefficients of the standard UV counterterms are given by
\renewcommand{\arraystretch}{1.5}
\begin{align}
\delta \hat  Z_{1,\mathrm{gg}}^{(\srp)} &  \;=\;  
\delta\hat\calZ_{1,G} - \delta\hat\calZ_{1,\gpar} \,,                            &\quad  
\delta \hat  Z_{2, \mathrm{gg}}^{(\srp)}   \;=\; & 
\delta\hat\calZ_{2,G} - \delta\hat\calZ_{2,\gpar} \,,    \nonumber \\[2mm]
\delta \hat  Z_{1,\mathrm{gg} }^{(\srG)} &  \;=\;  
{}-\delta\hat\calZ_{1,G}\,,                            &\quad  
\delta \hat  Z_{2,\mathrm{gg} }^{(\srG)}   \;=\; & 
{}-\delta\hat\calZ_{2,G}\,.
\end{align}
The additional $\tilde p^2$-dependent UV counterterm 
is required for the subtraction of
one-loop subdivergences (see~\refse{se:irredtwoloop}), and the relevant 
one-loop coefficients is
\begin{align}
\label{eq:qtildeggterm}
\delta \tilde Z^{(\srG)}_{1,\mathrm{gg}}  \;=\;  
\left( \frac{2}{3} \, \CA + \frac{2}{3} \, \TF \, \nq \right)  \eps^{-1}\,.
\end{align}
Rational counterterms in~\refeq{eq:R2gluon} are split into
momentum-dependent and vev-dependent contributions.
The former are independent of quark masses, and the corresponding
coefficients read~\cite{Lang:2020nnl}
\begin{align}
\label{eq:GGR2ctP}
\delta \hat \calR_{1,\mathrm{gg}}^{(\srp)}  \;=\; &  
- \frac{\CA}{3}\,,           \nonumber \\[2mm]
\delta \hat \calR_{2,\mathrm{gg}}^{(\srp)}   \;=\; &
   \left[\frac{19}{36} \, \CA^2 + \TF \, \nq \lb - \frac{32}{9}  \,  \CA  + 2  \, \CF \rb
\right] \eps^{-1} 
 + \TF \, \nq \lb \frac{217}{108}  \, \CA  - \frac{71}{18} \, \CF \rb
    +\frac{1211}{864} \, \CA^2 \nonumber \\
  & + \CA \lb - \frac{1}{3} \,\dcalZ_{1,\als} -\frac{35 }{12} \, \dcalZ_{1,G} +\frac{3}{4} \, \dcalZ_{1,\gpar} + \frac{1}{6} \, 
\dcalZ_{1,c}
\rb
 + \frac{4}{3} \, \TF \, \sum_{q\in\calQ} \dcalZ_{1,q}
         \,, \nonumber \\
\delta \hat \calR_{1,\mathrm{gg}}^{(\srG)} \;=\; &   
\left(\frac{\CA}{2} +\frac{2}{3} \,  \TF \, \nq \right)
             \,,              \nonumber \\[2mm]
\delta \hat \calR_{2,\mathrm{gg}}^{(\srG)}  \;=\;  &  
     \left[-\frac{4}{9} \, \CA^2+ \TF \, \nq \lb \frac{35}{9} \, \CA  - 2  \, \CF \rb
\right] \eps^{-1} 
    + \TF \, \nq \lb - \frac{193}{108} \, \CA  + \frac{109}{36} \, \CF \rb
   -\frac{541}{432} \, \CA^2  
  \nonumber \\
&        +\lb \f{\CA}{2} + \f{2}{3}\, \TF \, \nq \rb  \, \dcalZ_{1,\als}
         + \lb \f{71}{24} \, \CA + \f{2}{3} \, \TF \, \nq \rb  \, \dcalZ_{1,G} 
          - \f{7}{8} \, \CA  \, \dcalZ_{1,\gpar}
\nonumber \\
&
          + \f{\CA}{12}\, 
\dcalZ_{1,c}
        	-\f{4}{3} \, \TF  \sum_{q\in\calQ} 
          \dcalZ_{1,q} 
\,,
\end{align}
where $\dcalZ_{1,c}$ is the ghost-field renormalisation
constant.
For the vev-dependent part, 
the direct derivation of rational counterterms 
with quark masses
$m_q = v \, \lambda_q/\sqrt{2}$ yields~\cite{Lang:2020nnl}
\bea
\label{eq:GGR2ctM}
\delta \hat \calR_{1,\mathrm{gg}}^{(\srm)} &=&   
             -2 \, \TF \sum_{q\in \calQ}   \, \lambda_q^2
             \,,              \nonumber \\[2mm]
\delta \hat \calR_{2,\mathrm{gg}}^{(\srm)}  &=&  
    -\frac{\TF}{2} \sum_{q\in \calQ}\,  \Big[ 
    \left(  \CA  + 6 \,  \CF  \right) \eps^{-1} 
    + \frac{13}{6}  \, \CA  - 7 \, \CF
+4\left(
           \dcalZ_{1,\als} 
          +  \delta \hat \calZ_{1,G}
          + \delta \hat \calZ_{1,\lambda_q} 
          \right)
    \Big] \, \lambda_q^2\,.
\nonumber\\
\eea
This result can be cross-checked against 
vev expansions in the unbroken phase. According to~\refeq{eq:r2vevinsg} 
the mass-dependent contributions for the gluon--gluon vertex
are given by
\bea
\label{eq:GGvevexpis}
\delta\calR^{}_{k,\mathrm{gg}}\bigg|_{p=0}  &=& 
\delta\calR^{\ubk}_{k,\mathrm{gg}\vev\vev} + 
\delta\calR^{\ubk}_{k,\mathrm{gg}\vev\tvev} + 
\delta\calR^{\ubk}_{k,\mathrm{gg}\tvev\tvev} 
\nonumber\\
&=&
\frac{\vev^2}{2}\delta\calR^{\ubk}_{k,\mathrm{gg}\rH\rH} + 
\vev\delta\calR^{\ubk}_{k,\mathrm{gg}\rH\tvev} + 
\delta\calR^{\ubk}_{k,\mathrm{gg}\tvev\tvev}.
\eea
Note that single $\vev$ or $\tvev$ insertions
do not contribute since $\delta\calR^{}_{k,\mathrm{gg}}$ has mass dimension 
two. 
Note also that, according to~\refeq{eq:r1vevinsd} and
\refeq{eq:r2vevinsc}, double $\vev$ insertions 
give rise to a factor $1/2$ when expressed in terms of 
counterterms with external Higgs lines. This is not the case for 
double $\tvev$ insertions, since 
such insertions correspond to the 
auxiliary loop propagators~\refeq{eq:v1ct}--\refeq{eq:v2cta}, which
are not related to external Higgs lines.
The explicit expressions for 
$\delta\calR^{\ubk}_{k,\mathrm{gg}\rH\rH}=\delta\calR_{k,\mathrm{gg}\rH\rH}$
and $\delta\calR^{\ubk}_{k,\mathrm{gg}\rH\tvev}$
are given, respectively, in~\refeq{eq:GGSSverta}--\refeq{eq:GGSSvertc}
and~\refeq{eq:GGStvevverta}--\refeq{eq:GGStvevvertb}.
For the additional 
$\delta\calR^{\ubk}_{k,\mathrm{gg}\tvev\tvev}$ contribution, 
applying~\refeq{eq:v1ct}--\refeq{eq:VDVexpD4}, \refeq{eq:r2vevinsone} and
\refeq{eq:r2vevinsf}, 
we find
\bea
\label{eq:GGtvevtvevverta}
\delta\calR^{\ubk}_{k,\mathrm{gg}\tvev\tvev}
\;=\;
\vcenter{\hbox{\raisebox{12pt}{\includegraphics[width=0.22\textwidth]{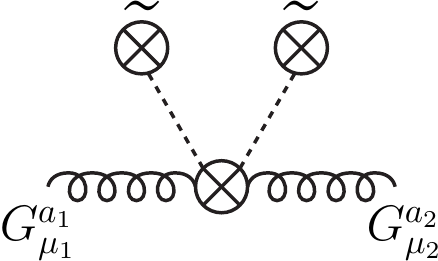}} }}   
\,=\,{}\ri \, \delta^{a_1 a_2} \, v^2 \, g^{\mu_1\mu_2} \,\bigg\{   \, \sum_{k=1}^2 \lb \f{\als \, t^\eps}{4\pi}\rb^k  
 \delta \hat \calR^{\ubk}_{k,\mathrm{gg}  \tvev \tvev}
\bigg\}
\,,  
\eea
with coefficients
\begin{align}
\label{eq:GGtvevtvevvertb}
\delta \hat \calR^{\ubk}_{1,\mathrm{gg}   \tvev \tvev} \;=\; &   
              - \, \TF \sum_{q\in \calQ}   \lambda_q^2
             \,,              \nonumber \\[2mm]
\delta \hat \calR^{\ubk}_{2,\mathrm{gg}   \tvev \tvev}  \;=\;  
 &   
      -  \TF \sum_{q\in \calQ}\,  \left[ 
    \left( \f{1}{4} \,  \CA   \right) \eps^{-1} 
    - \frac{13}{48}  \, \CA  - \f{9}{8} \, \CF
    \right] \, \lambda_q^2
        	- \, \TF  \sum_{q\in\calQ} 
          \left[
           \dcalZ_{1,\als} 
          +  \delta \hat \calZ_{1,G}
           \right] \, \lambda_q^2 \,. 
\end{align}
Finally, rewriting the vev-expansion identity~\refeq{eq:GGvevexpis} in terms
of the $\delta \hat\calR$  coefficients on the rhs 
of~\refeq{eq:GGSSverta}, \refeq{eq:GGStvevverta} and~\refeq{eq:GGtvevtvevverta}
we have
\bea
\delta \hat \calR^{(\srm)}_{k,\mathrm{gg}} &=&
  \frac{1}{2}\delta \hat \calR^{}_{k,\mathrm{gg\rH\rH}} 
 + \delta \hat \calR^{\ubk}_{k,\mathrm{gg\rH} \tvev}
 + \delta \hat \calR^{\ubk}_{k,\mathrm{gg}  \tvev \tvev }\,,
\eea
and combining the explicit ingredients of the vev expansion, 
\refeq{eq:GGSSvertc}, \refeq{eq:GGStvevvertb}
and~\refeq{eq:GGtvevtvevvertb}, 
we find full agreement with the result of the 
explicit derivation in the broken phase~\refeq{eq:GGR2ctM}.


\section{Summary}

The advent of automated numerical algorithms has
opened the door to the calculation of a vast range of 
nontrivial scattering amplitudes at one loop.
In this context, rational counterterms have proven to be a 
key ingredient for the efficient implementation of 
dimensional regularisation within numerical frameworks. The idea is that 
all process-dependent parts of the calculations can be carried out 
with numerical tools that build loop-integrand numerators in 
$\numdim=4$ dimensions,
while all contributions stemming from the 
$(\numdim-4)$-dimensional parts
of the numerators can be 
reconstructed through
process-independent rational counterterms.

The theoretical framework to extend this approach beyond one loop has been
established in~\cite{Pozzorini:2020hkx,Lang:2020nnl}.
In particular, it was shown that renormalised two-loop amplitudes can be 
constructed through a modified version of the well-known $\bfR$-operation, 
where loop amplitudes with $\numdim=4$ dimensional loop numerators 
are combined with UV counterterms and associated 
rational counterterms.

So far, the required two-loop rational counterterms were known only for U(1)
and SU(N) gauge theories~\cite{Pozzorini:2020hkx,Lang:2020nnl}, where the
relevant derivations are largely simplified by the underlying gauge
symmetry, which allows one to treat all fields in terms of generic fermion
and gauge-boson multiplets.
In the case of spontaneously broken (SB) gauge theories,
this is no longer possible since, in general,
the various states within a multiplet 
acquire different masses, mix with one another, and 
are renormalised in a different way.
For these reasons, the direct determination of 
two-loop rational counterterms for the full SM, 
including EW corrections,
can be very challenging.

In order to simplify this task, we have 
presented a general method that makes it possible to 
relate the rational counterterms 
for a SB theory
to corresponding counterterms 
in the underlying symmetric theory
via expansions in the vev parameter $\vev$.
This method is based on the fact that, in $\numdim=D$ dimensions,  
the mass dependence of loop amplitudes
in the SB phase can be generated through systematic $\vev$ expansions, where
terms of order $\vev^k$ are obtained in the symmetric phase 
via insertion of $k$ external Higgs
lines with zero momentum.
These so-called $v$ insertions are not sufficient 
in order to obtain the correct mass dependence of
loop amplitudes in $\numdim=4$ dimensions.
This is due to the fact that the
projection of loop numerators to $\numdim=4$ dimensions does not commute
with the expansion in the vev parameter.
In particular, in $\numdim=4$ dimensions
the mass dependence of fermionic loop propagators 
involves additional terms proportional to $\vev\tilq^2/\bar
q^4$ and $\vev^2\slashed q\tilq^2/\bar q^6$, 
where $\bar q$ is the loop momentum in $D$ dimensions and 
$\tilq$ its $(D-4)$-dimensional part.
As shown in~\refse{eq:tvevins}, 
such contributions can be accounted for by means of auxiliary fermion
propagators. For their systematic bookkeeping we have extended 
the Feynman rules by introducing auxiliary vertices where fermion propagators are
coupled to one or two pseudo external lines
that we have dubbed $\tvev$ insertions.

Based on this approach we have derived general vev-expansion
formulas of the form \refeq{eq:dRvexpgen}, where the one- and
two-loop rational counterterms for a generic SB theory are expressed in terms of
related counterterms with $\vev$ and $\tvev$ insertions in the symmetric
phase. 
In this way, the bulk of the derivation of 
rational counterterms is restricted to the symmetric phase, 
while all effects of SB are accounted for 
by adding (in total) at most two  $\vev$ or
$\tvev$ insertions.

This method is based on rigid invariance, i.e.~on the assumption that the
vev dependence of the Lagrangian is entirely generated through shifts of the
Higgs field, $H\to H+\vev$. Moreover, the proof presented in~\refse{se:sb} 
is based on the $\msbar$-scheme.
The extension to realistic renormalisation schemes for SB theories, such as
the on-shell scheme, has been discussed in~\refse{se:schdep}. 
In a first step we have demonstrated, still assuming rigid invariance, that 
the vev-expansion formula~\refeq{eq:dRvexpgen} 
remains valid for a wide class of renormalisation schemes for SB theories.
In particular it remains valid for any scheme that is equivalent to a 
renormalisation of the symmetric phase,
where the independent renormalisation constants 
can assume arbitrary finite parts, provided that the underlying symmetry is
preserved.
In this context we have shown that the renormalisation of the parameters of
the symmetric phase and the vev can be easily adapted such as to satisfy
typical renormalisation conditions for SB theories.  We have also discussed
how to ensure the cancellation of renormalised tadpoles, and we have shown that residual mixing
effects between on-shell mass eigenstates can be easily compensated by means
of finite LSZ factors.

Finally we have discussed possible violations of rigid invariance, which 
can arise when the gauge-fixing Lagrangian involves vev-dependent terms that 
do not arise from the symmetric phase via \mbox{$H\to H+\vev$} shifts.
Such terms are present, for instance, in the `t~Hooft gauge fixing, where
they are introduced in order to cancel the mixing between gauge bosons and
Goldstone bosons.
To keep track of vev-dependent gauge-fixing terms that violate rigid invariance
we have defined the parameter \mbox{$\hvev = \xi'\vev$}, where $\xi'$ is a
gauge-fixing parameter, and
we have introduced technical 
$\hvev$-insertion Feynman rules that make it possible 
to generate the $\hvev$-dependent parts of loop amplitudes
in a controlled way.
Gauge-fixing terms that violate rigid invariance 
give rise to $\hvev$-dependent UV divergences that are not present in the
symmetric phase. As is known, such divergences can be 
cancelled through a shift of the vev-renormalisation  
constant, which can be chosen in such a way that renormalised tadpoles
cancel exactly, while keeping all 
vev- and tadpole-independent renormalisation
constants unchanged.
As for the rational counterterms,
we have shown that the extra $\hvev$-dependence 
that results from violations of rigid invariance 
can be accommodated in the generalised vev-expansion formula~\refeq{eq:gfixampFa},
where rational counterterms for the SB phase are connected 
to the ones of the symmetric phase by means of 
combinations of 
$\vev$, $\tvev$ and $\hvev$ insertions. 

This approach can be used for the determination of 
rational counterterms in any SB gauge theory 
with wide flexibility in the choices of the 
gauge fixing and the renormalisation scheme.
As a first application, we have presented the
full set of $\ord(\alphas^2)$ rational counterterms 
for the interactions of quarks or gluons with 
EW vector bosons and scalars
in a generic renormalisation scheme. 
Together with the counterterms derived 
in~\cite{Pozzorini:2020hkx,Lang:2020nnl} these new results
provide the complete set of rational counterterms for 
two-loop QCD calculations in the full SM.

\subsection*{Acknowledgements}
This research was supported by the Swiss National Science Foundation (SNSF) 
under contract BSCGI0-157722. The work of M.Z. was supported through the
SNSF Ambizione grant PZ00P2-179877.

\appendix

\section{Commutator of mass expansion and $\numdim=4$ projection}
\label{app:DeltaV2}
In this appendix we consider the commutator 
of the mass expansion and the projection to four 
dimensions~\eqref{eq:commutator} 
for the case of generic propagators~\refeq{eq:commutid}, 
and we derive the 
identities~\refeq{eq:deltaV02gen}--\refeq{eq:gmassder}
up to second order
in the mass expansion.

At zeroth order we simply have
\bea
\Delta \bfV_{0} \, \bar G_{aa}^{\bk} (\bar q, m_a) &=&
 G_{aa}^{\bk}(\bar q, 0) - G_{aa}^{\bk}(\bar q, 0) \,=\, 0\,.
\eea
At first order we compute $\Delta \bfV_{1}  = \bfm_{1}\bfP_4 -
\bfP_4\bfm_{1} $ using~\eqref{eq:GvevexpDa}
and keeping only linear terms in $m_{a}$,
\bea
\Delta \bfV_{1} \, \bar G_{aa}^{\bk} (\bar q, m_a)
&=& m_a \lb \partial_{m_a} G_{aa}^{\bk} - G_{aa}^{\bk} \lb \partial_{m_a} \Gamma_{aa}^{\bk} \rb G_{aa}^{\bk} \rb \big|_{m_a=0} \nonumber \\
&=& m_a  \big( \partial_{m_a} G_{aa}^{\bk}  - G_{aa}^{\bk} \, \partial_{m_a} \lb
\Gamma_{aa}^{\bk}  G_{aa}^{\bk} \rb + G_{aa}^{\bk} \Gamma_{aa}^{\bk} \lb \partial_{m_a}  G_{aa}^{\bk} \rb \big) \big|_{m_a=0} \nonumber \\
&=&  m_a \, \f{\tilde q^2}{\bar q^4} \, \f{\partial g_{aa}^{\bk} (q,m_a)}{\partial m_a} \bigg|_{m_a=0} \;,
\label{eq:unifydV1}
\eea
where
in the second line we used the product rule
\bea
(\partial_{m_a} \Gamma_{aa})G_{aa}  &= & 
\partial_{m_a}( \Gamma_{aa}G_{aa}  ) -\Gamma_{aa} (\partial_{m_a} G_{aa} )\,.
\label{eq:prodrule}
\eea
To arrive to the last line we eliminate all
appearances of $G_{aa}$ and $G_{aa}\Gamma_{aa}$ in favour of 
$g_{aa}$ using
\eqref{eq:4dimprop} and \eqref{eq:4dimpropC}, respectively.

To second order 
we proceed in the same way, except that it is no longer possible to eliminate all appearances of
$\Gamma_{aa}$ without introducing inverse propagators. Instead we choose to
eliminate $G_{aa}$, as well as all $\Gamma_{aa}$ with mass derivatives
using \eqref{eq:prodrule} together with 
\bea
  G_{aa} (\partial_{m_a}^2 \Gamma_{aa})G_{aa}  &=& 
  \left[
  \partial_{m_a}^2 (G_{aa} \Gamma_{aa}) - 
  (\partial_{m_a}^2 G_{aa}) \Gamma_{aa} - 2 (\partial_{m_a} G_{aa})
  (\partial_{m_a} \Gamma_{aa} )
  \right] G_{aa}\notag\\
  &=&
  (\partial_{m_a}^2 (G_{aa} \Gamma_{aa}))G_{aa}
  -(\partial_{m_a}^2 G_{aa}) (\Gamma_{aa}G_{aa})\notag\\
  &&{}- 2 (\partial_{m_a} G_{aa}) 
  \left[
\partial_{m_a}( G_{aa} \Gamma_{aa} ) -\Gamma_{aa}(\partial_{m_a} G_{aa} )
    \right]\,.
\eea
In this way we find
\bea
  \Delta \bfV_{2} \, \bar G_{aa}^{\bk} (\bar q, m_a) &=&
\f{1}{2} \, m_a^2 \left. \lb \partial^2_{m_a} G_{aa}^{\bk} - 2 \,\lb G_{aa}^{\bk} \lb \partial_{m_a} \Gamma^{\bk}_{aa} \rb \rb^2 G^{\bk}_a  -   G_{aa}^{\bk} \lb \partial_{m_a}^2 \Gamma^{\bk}_{aa} \rb G_{aa}^{\bk} \rb \right|_{m_a=0} \nonumber \\
&=&
m_a^2 \, \left[
\f{1}{2} \, \f{\tilde q^2}{\bar q^4}  \f{\partial^2 g_{aa}^{\bk}(q,m_a) }{\partial m_a^2}  
 -  
\f{\tilde  q^2}{\bar q^6} 
\lb \f{\partial g_{aa}^{\bk} (q,m_a)}{\partial m_a} \rb^2 \Gamma_{aa}^{\bk}
    (q,m_a)
\right]_{m_a=0} 
\hspace{-2mm}
\,.\quad
\label{eq:unifydV2}
\eea
The terms~\refeq{eq:unifydV1} and~\refeq{eq:unifydV2}
correspond to the first- and second-order contributions on 
the rhs of~\refeq{eq:deltaV02gen}.
These results are applicable to any propagator of the form~\refeq{eq:4dimprop}, 
where the numerator $g_{aa}(q,m_a)$ is a polynomial in $q$ and $m_a$.

\bibliographystyle{JHEP}
\bibliography{RT_literature}

\end{document}